\documentclass[journal,twoside,web]{ieeecolor}
\usepackage{amsmath}
\usepackage{amsfonts}
\usepackage{dsfont}
 \usepackage{yfonts} 
\usepackage{mathrsfs} 
\usepackage{upgreek}
\usepackage{mathabx} 
\usepackage{amsbsy} 
\usepackage{multirow} 
\usepackage{mathtools}
\usepackage[T1]{fontenc}
\usepackage[utf8]{inputenc}
\usepackage[english]{babel}\usepackage{generic}
\usepackage{cite}
\usepackage{amsmath,amssymb,amsfonts}
\usepackage{algorithmic}
\usepackage{graphicx, subfigure}
\usepackage{textcomp} 
\usepackage{bbm}

 \newtheorem{remark}{ \textbf {Remark}}[section]
  \newtheorem{example}{\textbf {Example}}[section]

 \newtheorem{theorem}{\textbf {Theorem}}[section]
  \newtheorem{proposition}{\textbf {Proposition}}[section]
  \newtheorem{definition}{\textbf {Definition}}[section]

  \usepackage{stackengine}
\def\delequal{\mathrel{\ensurestackMath{\stackon[1pt]{=}{\scriptstyle\Delta}}}}

\begin{document}
\title{Symmetry-based observers for  ODE systems}
\author{Stefano Battilotti   
\thanks{S. Battilotti is with  Department of Computer, Control, and Management Engineering  ``Antonio Ruberti'',    Sapienza University of Rome,  Via Ariosto 25, Italy.}}

\maketitle

\begin{abstract}
In this paper we  introduce an observer design framework for  ordinary differential equation (ODE) systems based on various types of  existing or even novel one-parameter symmetries (exact, asymptotic and variational)  ending up with a certain number of semi-global  and  global observers, with bounded or unbounded system's solutions and with infinite- or finite-time convergence. We compare some of these symmetry-based observers with existing observers, recovering for instance the same performances of high-gain semiglobal observers and the finite-time convergence capabilities of sliding mode observers, while obtaining novel global observers where existing techniques are not able to provide any.   
 \end{abstract}


\section{Introduction}\label{sec:introduction}
\IEEEPARstart{O}{bserver}  design for nonlinear system  is a longstanding problem. High-gain observers (HGO: \cite{KP} and references therein),  extended state observers (ESO: \cite{FK}, \cite{GZ}) and time-varying high-gain observers  (VHGO: \cite{Gauthier_Kupka} and \cite{Besancon} with references therein) guarantee state estimation with compact error convergence domains and disturbance suppression for restricted classes of disturbances. Homogeneous observers (\cite{Andrieu_Praly_Astolfi_2008} and references therein) exploit the homogeneity  of the system for which  local and global stability properties collapse into one. 
HGOs with self-tuning gains  have been introduced to achieve
a tradeoff between a fast  transient response in a disturbance-free
setting and low sensitivity to disturbances: \cite{AK}, \cite{Andrieu_Praly_Astolfi_2009}, \cite{Battilotti_2009}, \cite{Battilotti2022}, \cite{Hammouri_Targui_Armanet}, \cite{Oueder_Farza_Abdennour_MSaad} and references therein. In particular, \cite{Andrieu_Praly_Astolfi_2009} is a first attempt of mixing  self-tuning techniques with homogeneity conditions, while performance optimization in terms of insensitivity to disturbances is considered in \cite{Battilotti_2022b}. The above observer design techniques in general are based on exploiting specific system's structures, such as for instance integrator chains,   and in many cases posing conditions  on the increments of the nonlinearities (see for instance \cite{Bernard_Praly_Andrieu} 
and references therein) rather than on  the nonlinearities themselves. Sliding-modes observers (SLO) with or without homogeneous correction terms (\!\!\cite{AFM_2017}, \cite{AMF_2018}, \cite{CM_2019}, \cite{LL_2018},  \cite{R_2022}) have revealed efficient in suppressing  certain classes of disturbances. 

Other observer design techniques  are based on transforming the system into a linear one for which a linear observer can be readily designed and obtaining a nonlinear observer for the original system by inverse transformation (the earlier works of Krener-Isidori and Krener-Respondek) or immersing the nonlinear system into a linear one (\cite{Bernard} and  references therein). Symmetries  are  multi-parameter  groups of transformations which leave  the  system's structure invariant  (\cite{Olver}). Homogeneity   is a  particular type of one-parameter symmetry (\cite{Kawski} and, more recently,  \cite{Polyakov_2020} on  functional spaces). A  symmetry-based approach  for observer design has been initiated in \cite{Bonnabel_Martin_Rouchon} (and continued in \cite{Lagerman_Trumpf_Mahony}  and \cite{Mahony_Trumpf_Hamel}) providing so-called {\sl  invariant}  observers, with    local error  convergence   and  not considering the problem of  insensitivity  to disturbances.  

The contributions of this paper and  its differences with the existing literature may be described and referenced as follows:
 
 \noindent
 - in Section  \ref{S1} we define a general framework,  based on   {\sl one-parameter groups of transformations}, \footnote{The results of this paper can be straightforwardly extended to multi-parameters groups of transformations.}  acting on the space of time and system's states,  disturbances and outputs,  {\sl infinitesimal generators  and  their prolongation}  which is a powerful tool for extending the action of a one-parameter group to the time  derivatives of the system's states. Different types of one-parameter groups are introduced and motivated with examples, according to their (local or global) action  on the {\sl time domain} (Definition \ref{time_scale}) or their (expanding or contracting) action on the {\sl  state/input domain}  (Definitions \ref{def_monotone_group},  \ref{def_IS_contracting} and \ref{def_IS_contracting_bis}). In particular,   considering time scales transformations as change of coordinates in a ``space-time'' manifold is a completely new and promising perspective in the control literature and, moreover, the expanding/contracting key feature  of the group of transformations on the state/input domain has a strategic role in achieving  global convergence results,  in contrast with the groups of transformations adopted in \cite{Bonnabel_Martin_Rouchon} and \cite{Lagerman_Trumpf_Mahony},  
 
  \noindent
 - in Section   \ref{S2}, using the notion of {\sl system map}, which is an equivalent characterization of the system as an implicit  function of time, disturbance, output, state vector and its time derivative, we recall from \cite{Olver} the notion of  one-parameter  symmetry of an ODE system with some illustrative  examples aimed to discuss the computational  aspects, the connections with classical notions (as homogeneity) and the impact on problems related to  asymptotic or prescribed finite-time convergence.  One-parameter symmetries (called {\sl orbital} symmetries: see \cite{Menini_Tornambe} for definitions and related results) for vector fields and functions,  originating from one-parameter groups of {\rm linear} dilations and much closer to the homogeneity notion,  has been considered for observer design in 
 \cite{Battilotti2022},   
 
 \noindent
 - in Sections \ref{S3} by using the symmetries introduced in Sections \ref{S1} and \ref{S2}  we prove a   number  of results on semiglobal  (Section \ref{S3.1}) and global  (Sections \ref{S3.2.1} and \ref{S3.2.2}) observer design  with {\sl infinite-time} error  convergence (Theorems \ref{TH1}, \ref{TH1.1} and \ref{TH3}). On simple benchmark examples we compare our semiglobal symmetry-based observers with classical HGOs and SLOs in terms of expected error bound and insensitivity to disturbances when the  system's solutions are   bounded,  concluding that HGOs and SLOs have better error and disturbance-insensitivity performances (discussion after Theorem \ref{TH1}). On the other hand, when the  system's solutions are not bounded, our global observers, which are obtained from the semiglobal observers coupled with a state-norm estimator (Section \ref{S3.2.1}), can be designed where  other existing techniques, including  HGOs or SLOs,  fail to apply (Examples \ref{ex_5} and  \ref{ex_6}),   
 
  \noindent
 - novel types of one-parameter symmetries, called  {\sl asymptotic}  (variational) symmetries, are introduced  in Section \ref{as_symm} and  \ref{S4}. In comparison with  exact symmetries considered in the literature (as for instance in \cite{Bonnabel_Martin_Rouchon},   just to cite one) and as demonstrated in Section \ref{S4}, Theorem \ref{TH1_AS},   asymptotic  symmetries arise as a powerful tool for improving observer's insensitivity to  disturbances but also for approximating the system's dynamics with more familiar or simplified patterns (linear, triangular, etc.) or shedding light on dominant system's dynamics. As discussed in the follow-up of Example \ref{E8} after  Remark \ref{fin_rem},  asymptotic  (variational) symmetries actually improve  observer's insensitivity, recovering the same  performances as HGOs while SLOs still perform better (with exact convergence),  

  \noindent
 - by considering symmetries with nonlinear transformation of the time scales (Section III.B and Definition \ref{time_scale}) we see how to design observers with
{\sl prescribed finite-time} error convergence  (Section \ref{S3.2.3}, Theorem \ref{TH1b}) and  compare existing homogeneous observers and SLOs, which are {\sl time-invariant  fractional-power or even discontinuous} function of the output, with our symmetry-based observers which are {\sl  time-varying continuously differentiable}  (actually smooth) function of the   output and, we should say,  more resemblant to the ones used in \cite{SWK} (discussion after Theorem \ref{TH1b}).  Moreover, symmetry-based observers with
 prescribed  finite-time  error convergence have comparable performances to SLOs  in terms of   disturbance-insensitivity.
     
For our purposes we consider ODE systems of the form \footnote{$D_t x (t) $ denotes the   derivative of the function $x(t)$ w.r.t.  $t$ and it is used in place of the more  classical notation $\dot{x}(t)$ for underlining  the actual time scale with respect to which we are differentiating the function:  if $\tau ={\mathbb T}  (t)$ is any other continuous time scale (i.e. a strictly increasing   continuous function of $t$),  $D_\tau \widetilde{x} (\tau) $  denotes the   derivative of $\widetilde{x} (\tau) \delequal x ( {\mathbb T}^{-1} (\tau) )$ (i.e. the function $x(t)$ in the new time scale $\tau$)  w.r.t.  the SNEw time scale $\tau$. }
\begin{eqnarray}
&& \hskip-1.3cm  D_t x (t) = F (t, x(t), d(t))  , \qquad  y (t) = H (t, x(t),  d(t))    \label{eq_sys}
\end{eqnarray} 
with   states  $x (t) \in {\mathbb R}^n$,     inputs $d(t) \in {\mathbb R}^m$ (exogenous disturbances or time-varying parameters) and measured outputs  $y (t) \in {\mathbb R}^p$.  We assume $C^\infty$  functions $F:{\mathbb R}\times {\mathbb R}^n \times {\mathbb R}^m \to {\mathbb R}^n$ and $H:{\mathbb R}\times {\mathbb R}^n \times {\mathbb R}^m \to {\mathbb R}^p$ and forward completeness  of \eqref{eq_sys}. For the sake of simplicity, we are not including {\sl control} inputs $u(t)$ in \eqref{eq_sys}, recalling that it is  straightforward to extend the results of this paper when considering also control variables. 

\section{Notation I}\label{note}

 \noindent $  \blacktriangleright$ ({\sl vector spaces}). ${\mathbb R}^n$ (resp. ${\mathbb R}^{n \times s}$)  is the set of $n$-dimensional real   vectors (resp. $n\times s$ matrices).  ${\mathbb R}_+^0$ (resp. ${\mathbb R}_+$)  denotes the set of   non-negative (resp.     positive) real numbers. For any vector ${\mathsf v}  \in {\mathbb R}^s$ and matrix $A \in {\mathbb R}^{n \times s}$  we denote by ${\mathsf v}_i  \in {\mathbb R}$, resp. $( A{\mathsf v} )_i  \in {\mathbb R}$,     the $i$-th   component   of ${\mathsf v}$,  resp.  $  A{\mathsf v}  $.   Moreover,  ${\textbf 0}_{n \times s}$, resp.  ${\textbf I}_n$,  denote  the zero matrix in  $ {\mathbb R}^{n \times s}$, resp. the identity matrix in $ {\mathbb R}^{n \times n}$, and   ${\rm diag} \{{\mathsf v}_1, \dots , {\mathsf v}_n \}$ or simply  ${\rm diag}_i \{ {\mathsf v}_i \}$  is the diagonal matrix with diagonal entries ${\mathsf v}_1, \dots , {\mathsf v}_n \in {\mathbb R}$.  ${\mathbb S}_+ (n)$ (resp. ${\mathbb S}_- (n)$)  is the set of symmetric positive (resp. negative) definite  matrices $S \in {\mathbb R}^{n \times n}$ and ${\mathbb {GL}} (n)$  is the set of nonsingular matrices $S \in {\mathbb R}^{n \times n}$. 

\noindent $  \blacktriangleright$ ({\sl norms}).  $\vert {\mathsf v} \vert$ denotes the absolute value of ${\mathsf v} \in {\mathbb R}$, $\Vert {\mathsf v} \Vert \delequal \sqrt{{\mathsf v}^\top {\mathsf v} }$  denotes the euclidean norm of ${\mathsf v} \in {\mathbb R}^n$ and the induced  norm   of $S \in {\mathbb R}^{m \times n}$ is $  \Vert S \Vert \delequal  \sup_{{\mathsf v} \in {\mathbb R}^n }  (  \Vert S{\mathsf v} \Vert / \Vert {\mathsf v} \Vert  ) $.  We  identify a vector  ${\mathsf v}  \in {\mathbb R}^r$  with the $r$-uple  $ ({\mathsf v}_1 ,  \dots , {\mathsf v}_r ) $ and by $\Vert ({\mathsf v}_1 ,  \dots , {\mathsf v}_r )  \Vert $ we mean the norm of  ${\mathsf v}  \in {\mathbb R}^r$. 
   
\noindent $  \blacktriangleright$ ({\sl monotone functions}). Let $ {\mathcal K}_+$ (resp. $ {\mathcal K}$,   resp. $ {\mathcal K}_\infty$) be the set of  continuous  strictly increasing functions $f : {\mathbb R}_+^0  \to {\mathbb R}_+^0$ such that $f(0) > 0$ (resp.  $f(0) = 0$,  resp.  $f(0) = 0$ and $\lim_{s \to+\infty} f(s) =+\infty$). Let  $ {\mathcal L}$ be the set of  continuous   decreasing   functions $f : {\mathbb R}_+^0  \to {\mathbb R}_+$  such that $\lim_{s \to+\infty} f(s) = 0$.     
Finally,   let  $ {\mathcal {KL}}$ (resp. $ {\mathcal {K_+ L}}$) be the set of  continuous  functions $f : {\mathbb R}_+^0 \times {\mathbb R}_+^0   \to {\mathbb R}_+^0$ such that $ f ( \cdot ,  s ) \in {\mathcal K} $ (resp. $ f ( \cdot ,  s ) \in {\mathcal K}_+$)  for each $ s \ge  0$ and  $ f ( r , \cdot ) \in {\mathcal L}$  for each $r  \ge  0$. 
 
\noindent $  \blacktriangleright$ ({\sl saturation functions}). A locally Lipschitz function $ ({\mathsf c}  , {\mathsf v}) \in {\mathbb R}_+ \times {\mathbb R}^n   \mapsto {\rm sat}_{\mathsf c} ({\mathsf v}) \delequal   \begin{pmatrix} {\rm sat}_{\mathsf c} ({\mathsf v}_1 )  &  \cdots &   {\rm sat}_{\mathsf c} ({\mathsf v}_n ) \end{pmatrix}  \in {\mathbb R}^n$ 
  is a {\sl  saturation function}   if ${\rm sat}_{\mathsf c} ({\mathsf v}_i ) = {\mathsf v}_i $ for all ${\mathsf v}_i \in [-{\mathsf c}  , {\mathsf c} ]$, $i=1,  \dots , n$,   and  $\vert {\rm sat}_{\mathsf c} ({\mathsf v}_i) \vert \le   {\mathsf c} $ and  $\vert {\rm sat}_{\mathsf c} ({\mathsf v}_i) - {\rm sat}_{\mathsf c} ({\mathsf w}_i ) \vert \le \vert {\mathsf v}_i -{\mathsf w}_i \vert$ for all  ${\mathsf v}_i ,  {\mathsf w}_i  \in {\mathbb R}$, $i=1,  \dots , n$.  

{\bf Notational remark}.   We adopt the following notation with regard to \eqref{eq_sys} and its variables: $t \in  {\mathbb R}$ is the independent (time) variable, $ z(t) \delequal ( x(t) , d(t) , y(t)) $ are the dependent (state, disturbance, output) variables and  $\mathsf{z} \delequal  (\mathsf{x}   , \mathsf{d}  , \mathsf{y} )  $ represents  the  {\sl values} of the function  $ z(\cdot) = ( x(\cdot) , d(\cdot) , y(\cdot)) $ at (time) $t$. \footnote{Using different notations for the  dependent variables  $f (\cdot) $ and their {\sl values} ${\mathsf f}= f(t)$  at  $t$  is instrumental  to distinguish between  {\sl point transformations} (such as symmetries) and  {\sl functional transformations}: see also \cite{Olver}.} Moreover,  ${\mathcal Z}:= {\mathbb R}^n \times {\mathbb R}^m \times {\mathbb R}^p$ (resp. ${\mathcal M} \delequal {\mathbb R} \times {\mathcal Z} $) represents the space of points   $  \mathsf{z}  $ (resp. $(t , \mathsf{z} )$).  $\hfill \blacktriangleleft$ 

\section{Group of transformations and prolongations}\label{S1}

In this paper we consider  families of (local)  nonlinear transformations $\Psi ({\mathfrak p},  \cdot ) $ of the time, state, input and output space ${\mathcal M} $, parametrized by ${\mathfrak p} \in {\mathbb R}$ which conveys the structure of {\sl group} to the family itself (i.e. one-parameter groups of transformations). Moreover, specifically for observer design applications,  we consider nonlinear transformations   of the form 
 \begin{eqnarray}
  &&\hskip-1.2cm    \Psi ({\mathfrak p} ,  t, \mathsf{z}) \delequal   (  \Psi^t ( {\mathfrak p} , t) ,   \Psi^{\mathsf{x}}  ( {\mathfrak p} ,  t, \mathsf{x})  ,  \Psi^{\mathsf{d}}  ( {\mathfrak p}   ,  t, \mathsf{x}, \mathsf{d})  ,   \Psi^{\mathsf{y}}  ({\mathfrak p} ,     t , \mathsf{y})    )    \label{symm2}
  \end{eqnarray}
  in which $ \Psi^t $ transforms the time scales,  $\Psi^{\mathsf{x}}$ transforms the state variables,  $\Psi^{\mathsf{d}}$ transforms the  disturbance  variables and   $  \Psi^{\mathsf{y}} $ transforms the output variables. \footnote{An important  generalization   follows from considering in \eqref{symm2} time scales transformations of the form  $\Psi^t ( {\mathfrak p} ,  t, \mathsf{x}) $ (rather than simply  $\Psi^t ( {\mathfrak p} ,  t  ) $)   which include for example Lorentz transformations and, more generally, Poincar\'e groups of transformations on the time/space domain.}
  
\subsection{One-parameter group of transformations}

For characterizing precisely the above families of transformations  we recall below the   definition of one-parameter group of transformations. 
\begin{definition}\label{onep}
 (\textit{One-parameter group  of transformations}).  A {\sl  local one-parameter  group of $C^\infty$-transformations} (LGT) $ \Psi  $  acting  on $ {\mathcal M}$  consists of  an open subset ${\mathcal V}$    with  $\{ 0 \}     \times  {\mathcal M}  \subset {\mathcal V} \subset  {\mathbb R}   \times  {\mathcal M} $  and a  $C^\infty$ mapping  $  \Psi   :  {\mathcal V}  \rightarrow   {\mathcal M}$ such that 

 \noindent  (i) ({\sl identity}) $ \Psi (0 ,t, \mathsf{z}) = (t, \mathsf{z})$ for all $(t, \mathsf{z}) \in {\mathcal M}$,  

\noindent  (ii)  ({\sl inversion})  if $( {\mathfrak p}  , t, \mathsf{z})  \in {\mathcal V}$ then $(  -{\mathfrak p}  ,  \Psi ({\mathfrak p}  , t, \mathsf{z}) ) \in {\mathcal V}$  and   $\Psi  (  -{\mathfrak p}  ,  \Psi ({\mathfrak p}  , t, \mathsf{z}) ) = ( t, \mathsf{z})$, 

\noindent  (iii) ({\sl composition})  if $(  {\mathfrak p}_1  ,  \Psi ({\mathfrak p}_2 , t, \mathsf{z}) ) \in {\mathcal V}$, $( {\mathfrak p}_2 , t, \mathsf{z})  \in {\mathcal V}$   and also $ (    {\mathfrak p}_1 + {\mathfrak p}_2  ,  t, \mathsf{z})   \in {\mathcal V}$ then $ \Psi ({\mathfrak p}_1 ,  \Psi ( {\mathfrak p}_2  ,t, \mathsf{z})  )  = \Psi (  {\mathfrak p}_1 + {\mathfrak p}_2   ,t, \mathsf{z})$.   $ \hfill \blacktriangleleft $ 
\end{definition}  
The set ${\mathcal V}$,  the domain of the mapping $\Psi$, is  denoted by ${\rm Dom} ( \Psi )$,   and when $ {\rm Dom} ( \Psi ) = {\mathbb R}    \times {\mathcal M}$   the group of transformations is  {\sl global} (GGT).  The underlying group of a  LGT or GGT  is always $( {\mathbb R}, + )$, i.e. the additive   group ${\mathbb R}$. \footnote{In general we can replace $( {\mathbb R}, + )$ with  a (local)  {\sl Lie group} $  {\mathbb G}  $ endowed with an identity element ${\mathfrak e} \in  {\mathbb G}$, a smooth composition law $m : {\mathbb G} \times {\mathbb G} \to {\mathbb G}$ and a smooth inversion law $i : {\mathbb G} \to {\mathbb G} $.}  A   simple example of LGT is   the group of {\sl nonlinear} (time) transformations  $ \Psi ({\mathfrak p} ,t, \mathsf{z}) = (  \frac{t}{1-t {\mathfrak p}}   ,     \mathsf{z}  )$ 
  with   $  {\rm Dom} ( \Psi )   = \{  ({\mathfrak p} ,  t,\mathsf{z})  \in {\mathbb R} \times {\mathcal M} :    t  {\mathfrak p} \ne 1   \}$, while an example of GGT  is   the group of {\sl linear} transformations $
  \Psi  ({\mathfrak p} ,t, \mathsf{z}) = (    t  ,  e^{{\mathfrak p} A^{\mathsf x}  } \mathsf{x} ,e^{{\mathfrak p} A^{\mathsf d}  } \mathsf{d} , e^{{\mathfrak p} A^{\mathsf y}  } \mathsf{y} ) $ 
 for  $A^{\mathsf x}  \in {\mathbb R}^{n \times n} $,  $A^{\mathsf d}  \in {\mathbb R}^{m \times m} $ and  $A^{\mathsf y}  \in {\mathbb R}^{p \times p} $.  
 
 For each ${\mathfrak p}$, $\Psi ({\mathfrak p} , \cdot) $ represents a (local) diffeomorphism. Indeed, for each ${\mathfrak p}$ let ${\mathcal V} ({\mathfrak p}) \delequal \{ (t, \mathsf{z}) \in  {\mathcal M} :  (  {\mathfrak p} , t, \mathsf{z}  ) \in {\rm Dom}(\Psi )  \}$. 
 \begin{proposition}
 The set  ${\mathcal V} ({\mathfrak p})$  is open and  the mapping $ ( t, \mathsf{z}) \in {\mathcal V} ({\mathfrak p})   \xmapsto{ \Psi_{\mathfrak p} }   \Psi_{\mathfrak p}  (  t, \mathsf{z}) \delequal \Psi ({\mathfrak p} , t, \mathsf{z}) \in {\mathcal V} (-{\mathfrak p})   $  is a $C^\infty$ transformation  
  with $C^\infty$ inverse  $  \Psi_{\mathfrak p}^{-1} = \Psi_{-\mathfrak p}$ (i.e. a diffeomorphism).  $ \hfill \blacktriangleleft $ 
  \end{proposition} 
  The set ${\mathcal V} ({\mathfrak p})$, the domain of the mapping $\Psi_{\mathfrak p}$,  is  denoted by ${\rm Dom} ( \Psi_{\mathfrak p} ) $ for similarities with the notation ${\rm Dom} ( \Psi  ) $. 
 
 {\bf Notational remark}. In what follows we  use without distinction  the notations  $\Psi ({\mathfrak p} , \cdot  )$ and   $\Psi_{\mathfrak p} (\cdot)$   to stress the action of $\Psi ({\mathfrak p} , \cdot  )$ on ${\mathcal M}$ for each $\mathfrak p$ and when no ambiguity may arise we even drop the letter $ {\mathfrak p} $ in  $\Psi_{\mathfrak p} (\cdot)$.  $\hfill \blacktriangleleft$ 
  
 For a vector field ${\textbf g}$ on ${\mathcal M}$ we denote by ${\textbf g} \vert_{(t, \mathsf{z})}$ the tangent vector assigned by ${\textbf g}$ at $(t, \mathsf{z}) \in  {\mathcal M}$, i.e.  ${\textbf g} \vert_{(t, \mathsf{z})} \in  T_{(t, \mathsf{z})} {\mathcal M} \cong  {\mathcal M}$.  The tangent vector  ${\textbf g} \vert_{(t, \mathsf{z})}$   is   represented   in the canonical basis $\{   \frac{\partial}{\partial t} \vert_{(t, \mathsf{z})}  ,  \frac{\partial}{\partial \mathsf{z}_1} \vert_{(t, \mathsf{z})}  , \cdots , \frac{\partial}{\partial \mathsf{z}_{n+m+p}} \vert_{(t, \mathsf{z})}     \} $ of the tangent space $T_{(t, \mathsf{z})} {\mathcal M}  $ as $  {\textbf g} \vert_{  (t, \mathsf{z}) }  \delequal  g^t  (t, \mathsf{z}) \frac{\partial}{\partial t} \vert_{  (t, \mathsf{z}) }  +  \sum_{j=1}^{n+m+p} g^{\mathsf{z}_j}  (t, \mathsf{z}) \frac{\partial}{\partial \mathsf{z}_j} \vert_{  (t, \mathsf{z}) } $.\footnote{Throughout the paper  we  omit the application point $ (t, \mathsf{z})$ in each component  of the canonical basis $\{   \frac{\partial}{\partial t} \vert_{(t, \mathsf{z})}  ,  \frac{\partial}{\partial \mathsf{z}_1} \vert_{(t, \mathsf{z})}  , \cdots , \frac{\partial}{\partial \mathsf{z}_{n+m+p}}  \vert_{(t, \mathsf{z})}    \} \in T_{(t, \mathsf{z})} {\mathcal M} $.}   A LGT     on $ {\mathcal M}$ is the local flow generated by a suitable vector field  on $ {\mathcal M}$, the {\sl  infinitesimal generator} of the group.
 
\begin{definition}
 (\textit{Infinitesimal generator}).  The \emph{infinitesimal generator} of  a $C^\infty$ LGT $  \Psi_{ {\mathfrak p}  } $ on $ {\mathcal M}$    is  the  $C^\infty$ vector field ${\textbf g}$ on ${\mathcal M}$ such that ${\textbf g} \vert_{(t, \mathsf{z})}= \frac{\partial}{\partial {\mathfrak p}}     \Psi_{ {\mathfrak p}  } (  t, \mathsf{z})   \vert_{ {\mathfrak p} 
 = 0}  $ at each point $(t, \mathsf{z}) \in{\mathcal M}$. Moreover, for all $({\mathfrak p}, t, \mathsf{z} ) \in {\rm Dom} (\Psi ) $  
 \begin{eqnarray}
  \frac{\partial}{\partial {\mathfrak p}}     \Psi_{ {\mathfrak p}  } (  t, \mathsf{z})  =  {\textbf g} \vert_{(t, \mathsf{z})=  \Psi_{ {\mathfrak p}  } (  t, \mathsf{z}) } , \;   \Psi_0 (  t, \mathsf{z})   = (  t, \mathsf{z})   .  \hfill \blacktriangleleft \label{inf_gen}
  \end{eqnarray}  
 \end{definition}
 The corresponding structure of the infinitesimal generator of  a one-parameter group \eqref{symm2}  is  $ {\textbf g} \vert_{ (t, \mathsf{z}) } \delequal g^t (t) \frac{\partial}{\partial t}  +   \sum_{j=1}^n  g^{\mathsf{x}_j} (t, \mathsf{x}) \frac{\partial}{\partial \mathsf{x}_j}   + \sum_{j=1}^m g^{\mathsf{d}_j}   (t,\mathsf{x}, \mathsf{d}) \frac{\partial}{\partial \mathsf{d}_j}    +   \sum_{j=1}^p g^{\mathsf{y}_j}   (t,\mathsf{y}) \frac{\partial}{\partial {\mathsf y}_j} $. Notice that, with the claimed  structure  of $\Psi_{\mathfrak p}$ and $ {\textbf g}$,  $\Psi_{\mathfrak p}^t  $ (resp.  $\Psi_{\mathfrak p}^{t,\mathsf{x}} \delequal (  \Psi_{\mathfrak p}^t , \Psi_{\mathfrak p}^{\mathsf{x}} )$, resp.  $\Psi_{\mathfrak p}^{t,\mathsf{x},\mathsf{d}} \delequal (  \Psi_{\mathfrak p}^t , \Psi_{\mathfrak p}^{\mathsf{x}} , \Psi_{\mathfrak p}^{\mathsf{d}}  )$)  is  a  LGT  with infinitesimal generator $ {\textbf g}^t \vert_{t} \delequal g^t (t ) \frac{\partial}{\partial t}  $ (resp. $ {\textbf g}^{t,\mathsf{x}} \vert_{ (t, \mathsf{x}) } \delequal g^t (t ) \frac{\partial}{\partial t}  +   \sum_{j=1}^n  g^{\mathsf{x}_j} (t, \mathsf{x}) \frac{\partial}{\partial \mathsf{x}_j}$,  resp. $ {\textbf g}^{t,\mathsf{x},\mathsf{d}} \vert_{ (t, \mathsf{x},\mathsf{d}) } \delequal g^t (t ) \frac{\partial}{\partial t}  +   \sum_{j=1}^n  g^{\mathsf{x}_j} (t, \mathsf{x}) \frac{\partial}{\partial \mathsf{x}_j} +   \sum_{j=1}^m  g^{\mathsf{d}_j} (t, \mathsf{x}, \mathsf{d}) \frac{\partial}{\partial \mathsf{d}_j}$).
   
   \subsection{Time scales transformation}
   
One    important feature of our groups of transformations (also in relation to other types of groups of transformations considered in the literature, for instance \cite{Bonnabel_Martin_Rouchon} or \cite{Lagerman_Trumpf_Mahony})  is that also time scales are transformed and each variable  transformation is the result of a nonlinear intertwined transformation of  time, state, input and output variables. For this reason   the  groups of {\sl  time  transformations} (or some of their generalizations) we use in this paper represent a  novelty in the control literature in comparison with classical time scalings, more properly originating from  transformations such as {\sl  Lorentz} groups of   transformations, more familiar in physical  contexts,  where each point $ (t,{\mathsf x})$ of the  space-time is transformed into the point  $  ( \frac{ t - \frac{\mathfrak p}{c}{\mathsf x}  }{\sqrt{1-{\mathfrak p}^2}}  , \frac{ {\mathsf x}    -  {\mathfrak p}ct  }{\sqrt{1-{\mathfrak p}^2}} )$ of the same space ($c$ is the light velocity).  

In this section we classify and discuss  for later use a certain number of ways by which $\Psi_{\mathfrak p} $  transforms the time  variable $t$ or, more generally, a continuous time scale.\footnote{
 A  continuous  {\sl time scale} ${\mathbb T} $   is a  strictly  increasing continuous  function ${\mathbb T} : {\mathcal I} \to {\mathcal  J}$,   with ${\mathcal I} ,  {\mathcal J} \subseteq {\mathbb R}  $   intervals of the form $[a,b)$ and $[c,d)$, with $b,d \le +\infty  $. The trivial time scale is the identity function $  \mathbbm{1}   $ on  ${\mathbb R}  $ (i.e. $  \mathbbm{1} (t ) = t $). A time scale ${\mathbb T} : [a,b) \to  [c,d) $ is {\sl  non-negative} if $c= 0$, {\sl  positive} if $c > 0$, it is  {\sl  bounded} if $\lim_{t \nearrow b }{\mathbb T} (t)  < + \infty$ (i.e. $d < +\infty$)  and {\sl 
 unbounded} if $\lim_{t \nearrow b }{\mathbb T} (t)  =  + \infty$  (i.e. $d = +\infty$) .}   In order to proceed rigorously,  for each  ${\mathfrak p} > 0$ define $   {\rm Dom}^+  \Psi^t_{\mathfrak p}  = \sup \{ a \ge  0  :  [0 , a ) \subset {\rm Dom}   ( \Psi^t_{\mathfrak p} ) \}$.  Clearly, $   {\rm Dom}^+  \Psi^t_{\mathfrak p}   \ge 0$ for all  ${\mathfrak p} > 0$,  $   {\rm Dom}^+  \Psi^t_{\mathfrak p}   =0$ when  $[0,a) \not\subset {\rm Dom}   ( \Psi^t_{\mathfrak p} )  $ for any $a > 0$ and $  {\rm Dom}^+  \Psi^t_{\mathfrak p}   = +\infty $ when ${\mathbb R}_+^0 \subset {\rm Dom}   ( \Psi^t_{\mathfrak p} )$.   
 It is easy to see from  \eqref{inf_gen} that 
 $\frac{\partial \Psi_{\mathfrak p}^t}{\partial t}  (t ) 
 > 0$ for each  ${\mathfrak p} >  0 $  and $t \in [ 0 ,   {\rm Dom}^+  \Psi^t_{\mathfrak p}   )$. 
Hence, $\Psi_{\mathfrak p}^t  $ is for each ${\mathfrak p} > 0 $  a monotonically increasing function of $t \in [ 0 ,   {\rm Dom}^+  \Psi^t_{\mathfrak p}   )$ or, restated in more appealing terms, $\Psi_{\mathfrak p}^t  $   for each   ${\mathfrak p} > 0$  {\sl transforms  non-negative  continuous time scales   into continuous time scales}.  However,   when  $  {\rm Dom}^+  \Psi^t_{\mathfrak p}  \le  +\infty$ for     ${\mathfrak p} > 0$  it may not happen that $\lim_{t \nearrow   {\rm Dom}^+  \Psi^t_{\mathfrak p} } \Psi_{\mathfrak p}^t  (t) =+\infty$: consider for instance the case $\Psi_{\mathfrak p}^t  (t ) =   \frac{  t  }{ 1  +  {\mathfrak p}   t    }  $. In other words, unbounded (or bounded) continuous  non-negative  time scales  are not necessarily transformed  into continuous unbounded time scales.  
 \begin{definition}\label{time_scale}
 (\textit{Time scale transformations}). A LGT  \eqref{symm2}    transforms (continuous)  non-negative   time scales into  (continuous)  time scales for each ${\mathfrak p} > 0 $. 
 
  \noindent $\blacktriangleright$ If   $  {\rm Dom}^+  \Psi^t_{\mathfrak p}  = +\infty$  and  $ \lim_{t \to +\infty} \Psi_{\mathfrak p}^t  (t) =+\infty$ for each    $ {\mathfrak p} > 0$,   we say that the LGT  \emph{transforms    unbounded   non-negative  time scales into unbounded time scales} or it is $UU$-transforming  for ${\mathfrak p} > 0$.  

 \noindent $\blacktriangleright$ If $0 <   {\rm Dom}^+  \Psi^t_{\mathfrak p}  < + \infty$   and  $ \lim_{t \nearrow   {\rm Dom}^+  \Psi^t_{\mathfrak p}  } \Psi_{\mathfrak p}^t  (t) =+\infty$ for each    $ {\mathfrak p}> 0$, we say that \eqref{symm2}  \emph{transforms bounded  non-negative   time scales into unbounded time scales} or it is BU-transforming  for  
$ {\mathfrak p}> 0$.  If, in addition,   the sequence $  {\rm Dom}^+  \Psi^t_{\mathfrak p}  \searrow 0 $ as $ {\mathfrak p}   \to +\infty$  we say that  \eqref{symm2}  \emph{transforms    contractively  bounded  non-negative  time scales into unbounded time}  scales or it is CBU-transforming  for   ${\mathfrak p} > 0$.  
  $\hfill \blacktriangleleft$ 
 \end{definition}
In what follows some illustrative examples.   If $\Psi_{\mathfrak p}^t (t) = e^{   {\mathfrak p} }  t$  (or $\Psi_{\mathfrak p}^t (t) = e^{- {\mathfrak p} }  t$) then \eqref{symm2} is  UU-transforming for  ${\mathfrak p} > 0$. 
On the other hand,  if $\Psi_{\mathfrak p}^t  (t ) =  \frac{ t  }{ 1 -  {\mathfrak p}   t    }  $  
 then   $  {\rm Dom}^+  \Psi^t_{\mathfrak p}   =\frac{1}{\mathfrak p}$  if $   {\mathfrak p} >  0$ 
and the LGT  is  CBU-transforming for   ${\mathfrak p} >0$. A final example is $\Psi_{\mathfrak p}^t  (t ) =  \frac{ e^{2{\mathfrak p}}(t-1)+t+1  }{t+1- e^{2{\mathfrak p}} (t -1)   }  $ (with generator ${\textbf g}^t \vert_t = ( t^2-1) \frac{\partial }{\partial t}$) for which $  {\rm Dom}^+  \Psi^t_{\mathfrak p}  = \frac{ e^{2{\mathfrak p}}+1}{ e^{2{\mathfrak p}} - 1}$  for $   {\mathfrak p} > 0$, hence the LGT  is  BU-transforming (but not CBU!) for   ${\mathfrak p} >0$. 
 
We remark that either UU- or  BU-transforming LGT's  may not guarantee  that  the transformed time scale  $\Psi^t_{\mathfrak p} (t) $ is  non-negative  for all ${\mathfrak p} > 0$ and  $t \in [0,    {\rm Dom}^+  \Psi^t_{\mathfrak p}   )$: for instance, consider the case  $\Psi^t_{\mathfrak p} (t) = t - {\mathfrak p}$ with infinitesimal generator ${\textbf g}^t \vert_t = - \frac{\partial}{\partial t}$. For  the sake of simplicity,  we assume   $ \Psi_{\mathfrak p}^t   (0) = 0$ for all ${\mathfrak p} > 0$
:  for instance $ \Psi_{\mathfrak p}^t   (t) =   \frac{t}{ 1-{\mathfrak p} t } $ and  $ \Psi_{\mathfrak p}^t   (t) =  e^{a{\mathfrak p}} t$.  \footnote{ Needless to say, our case list is not exhaustive: for instance,  if   $\Psi_{\mathfrak p}^t (t) = \frac{t}{ 1+{\mathfrak p} t } $  than $ \lim_{t \to +\infty} \Psi_{\mathfrak p}^t (t) = \frac{1}{ {\mathfrak p}   } < +\infty $ for each ${\mathfrak p}>0$ and  $\Psi_{\mathfrak p}$ transforms  \emph{unbounded positive time scales into  bounded  time scales} for  ${\mathfrak p} >  0$.  }

\subsection{Prolongation of  groups and generators}\label{prolong}
Consider the dependent variable  $ x(t)$  with its first-order derivative $D_t x (t)$ and let $\mathsf{x}  \in {\mathcal X} \delequal {\mathbb R}^n$ represent the  value of $x(t) $ at $t$, while let $\mathsf{x}^{(1)} \in {\mathcal X}$ represent  the  value  of  $D_t x(t)$  at time $t$. Moreover, $ \mathsf{x}^{[1]} \delequal (\mathsf{x} ,\mathsf{x}^{(1)}) \in  {\mathcal X}^{[1]} \delequal {\mathcal X}  \times {\mathcal X}$ (the \emph{$1$-st order jet space} of ${\mathcal X}$)   and ${\rm pr}^{[1]} x (t) \delequal (x(t) , D_t x(t))$ (\emph{the $1$-st order prolongation}  of the variable $x (t) $), hence $ \mathsf{x}^{[1]}  $ represents the value of  ${\rm pr}^{[1]} x (t)  $ at $t$.  
 There is an induced local action of  $ \Psi_{\mathfrak p}^{t,\mathsf{x}}$   on   $ {\mathbb R}  \times  {\mathcal X}^{[1]}$ called the   \emph{$1$-st order prolongation}   of $ \Psi_{\mathfrak p}^{t,\mathsf{x}}$, denoted by ${\rm pr}^{[1]} \Psi_{\mathfrak p}^{t,\mathsf{x} }$  and  
 %
 computed as follows.
\begin{proposition}\label{P1}
 (\textit{Prolongation formula for $\Psi_{\mathfrak p}^{t,\mathsf{x}}$}): 
\begin{eqnarray}
 && \hskip-.8cm {\rm pr}^{[1]} \Psi_{\mathfrak p}^{t,\mathsf{x}}  ( t,  \mathsf{x}^{[1]}) = ( \Psi^t_{\mathfrak p} (t ), \Psi^{  \mathsf{x}^{[1]} }_{\mathfrak p} (t ,  \mathsf{x}^{[1]})  ) , \nonumber \\
 && \hskip-.8cm \Psi^{  \mathsf{x}^{[1]} }_{\mathfrak p} (t ,  \mathsf{x}^{[1]})  = ( \Psi^{ \mathsf{x} }_{\mathfrak p} (t , \mathsf{x}), \Psi^{ \mathsf{x}^{(1)} }_{\mathfrak p} (t ,  \mathsf{x}^{[1]}) ) ,
 \label{EPRW2}   \\
  && \hskip-.8cm \Psi^{ \mathsf{x}^{(1)} }_{\mathfrak p} (t ,  \mathsf{x}^{[1]})  =    ( \frac{\partial  \Psi^t_{\mathfrak p} (t ) }{\partial t}   )^{-1} \!  \left (    \frac{\partial  \Psi^{\mathsf{x}}_{\mathfrak p} (t,\mathsf{x}) }{\partial t} \!  +   \! \frac{\partial  \Psi^{\mathsf{x}}_{\mathfrak p} (t,\mathsf{x}) }{\partial \mathsf{x}}   \mathsf{x}^{(1)} \right )   .   \hfill \blacktriangleleft    \nonumber
 \end{eqnarray}
 \end{proposition}
Similarly,   the  {\sl  $1$-order prolongation} ${\rm pr}^{[1]}  {\textbf g}^{t,\mathsf{x}}  $ of   $  {\textbf g}^{t,\mathsf{x}}  $   is  defined  as   the infinitesimal generator of   ${\rm pr}^{[1]} \Psi_{\mathfrak p}^{t,\mathsf{x}}$  (see \cite{Olver} for technical details).
\begin{proposition}\label{P0}
 (\textit{Prolongation formula for ${\textbf g}^{t,\mathsf{x}}$}):   
 \begin{eqnarray}
&& \hskip-1cm {\rm pr}^{[1]}  {\textbf g}^{t,\mathsf{x}} \vert_{  (t,    \mathsf{x}^{[1]} )}  =  g^t   (t ) \frac{\partial  }{\partial t }   \nonumber  \\
&& \hskip-1cm  +\sum_{j=1}^n   g^{ \mathsf{x}_j  }   (t,   \mathsf{x}  ) \frac{\partial  }{\partial  \mathsf{x}_j } +  \sum_{j=1}^n   g^{ \mathsf{x}_j^{(1)} }   (t,    \mathsf{x}^{[1]} ) \frac{\partial  }{\partial  \mathsf{x}_j^{(1)}}  , \label{EPRW1}    \\
&& \hskip-1cm g^{ \mathsf{x}_j^{(1)} }   (t,    \mathsf{x}^{[1]} ) =    \frac{\partial g^{\mathsf{x}_j}}{\partial \mathsf{x}}   (t, \mathsf{x})  \mathsf{x}^{(1)}   +  \frac{\partial  g^{\mathsf{x}_j}}{\partial t}  ( t, \mathsf{x})   -  \frac{\partial g^t}{\partial t}   (t ) \mathsf{x}_j^{(1)}    .   \nonumber  \hfill \blacktriangleleft 
\end{eqnarray}  
\end{proposition}
 
\section{Symmetries  of ODE systems}\label{S2}
     
One-parameter symmetries  of a system are one-parameter  groups of transformations which preserve the system's structure. Symmetries are meant more generally  as {\sl  multi-parameter (or Lie) groups} of transformations but in this paper for the sake of simplicity we limit ourselves to one-parameter symmetries.   Symmetries of  differential  systems and their applications to nonlinear   system analysis (system reduction, integration, etc)  have been  extensively studied in \cite{Olver}. More recently, symmetries have been introduced in the feedback linearization problem (\!\!\cite{Menini_Tornambe}) while particular classes of symmetries, leaving time scales unchanged and using a simplified decoupled structure, have been used  for designing  observers  with local asymptotic convergence (\!\!\cite{Bonnabel_Martin_Rouchon}). In this paper we want to illustrate on different examples the impact of symmetries on semiglobal and global  observer design with either  asymptotic or   prescribed finite-time convergence. The way a  symmetry transforms a differential system passes through the notion of {\sl system  map} and its transformation under the action of a LGT. 
For each pair of $C^\infty$  functions $(t, \mathsf{x} ,  \mathsf{d}  )    \xmapsto{F}  F (t, \mathsf{x} ,  \mathsf{d}  )   \in  {\mathbb R}^n$ and $(t, \mathsf{x} ,  \mathsf{d}  )    \xmapsto{H} H (t, \mathsf{x} ,  \mathsf{d}  )   \in  {\mathbb R}^p$   define  the {\sl system map} $\Sigma \delequal   \mathsf{(F,H)} (t ,   \mathsf{x}^{[1]} , \mathsf{d} , \mathsf{y} ) $ (associated to the pair of functions  $(F  ,  H  )$)  as  \footnote{We remark that the generality of the system map approach allows to tackle the observer problem not only for {\sl explicit } differential  systems  \eqref{eq_sys}  but  also for {\sl implicit  } differential  systems   like  $( \Phi (t ,  x (t) , D_t x (t) , d (t)    )  , \Psi   (t ,  x (t) , y(t)   ,  d (t)     ) )  \equiv  (0,0)$ and in perspective for even more challenging types of differential equations.}
\begin{eqnarray}
&&\hskip-1cm (t ,   \mathsf{x}^{[1]} , \mathsf{d} , \mathsf{y} ) \in {\mathbb R}  \times {\mathcal X}^{[1]}   \times {\mathbb R}^m \times  {\mathbb R}^p  \xmapsto{  \mathsf{(F,H)} }   \nonumber \\
 &&\hskip-1cm   \mathsf{(F,H)} (t ,   \mathsf{x}^{[1]} , \mathsf{d} , \mathsf{y} ) \delequal \begin{pmatrix} \mathsf{x}^{(1)}  -F (t, \mathsf{x} ,  \mathsf{d}  )  \cr \mathsf{y} -H(t, \mathsf{x} ,  \mathsf{d}  )    \end{pmatrix}  \in {\mathbb  R}^{n+p} .\label{syst_map}
 \end{eqnarray} 

\subsection{Notation II}

 In this paragraph, for later use in the main  definitions  and proofs of the main results,  we collect all the notations relative to points,   functions, time scales and system maps transformed under the action of a  LGT.  We recall that  $ \mathsf{x}^{[1]} \delequal ( \mathsf{x} , \mathsf{x}^{(1)})$ and ${\rm pr}^{[1]}  x (t) \delequal (x(t) ,D_t x(t))$.  
    \begin{eqnarray}
  && \hskip-.8cm  \blacktriangleright  \hskip.2cm  \text{\sl  transformed points $(t ,   \mathsf{x}^{[1]} , \mathsf{d} , \mathsf{y} ) $} :  \mathsf{t}_{\mathfrak p} \delequal \Psi_{\mathfrak p}^t  (t )  ,       \nonumber \\
  &&\hskip-.8cm  \mathsf{z}_{\mathfrak p} \delequal (   \mathsf{x}_{\mathfrak p}  ,  \mathsf{d}_{\mathfrak p}   ,  \mathsf{y}_{\mathfrak p}   )   \delequal  (   \Psi_{\mathfrak p}^{\mathsf{x}}  (t,\mathsf{x})    ,  \Psi_{\mathfrak p}^{ \mathsf{d}}  (t, \mathsf{x},\mathsf{d})   , \Psi_{\mathfrak p}^{\mathsf{y}}  (t, \mathsf{y})   )  ,   \label{zp} \\
   &&\hskip-.8cm  
    \mathsf{x}^{(1)}_{\mathfrak p} \delequal  \Psi_{\mathfrak p}^{\mathsf{x}^{(1)}}(t,   \mathsf{x}^{[1]}) ,   \mathsf{x}^{[1]}_{\mathfrak p} \delequal ( \mathsf{x}_{\mathfrak p}  , \mathsf{x}^{(1)}_{\mathfrak p})  ,   \nonumber \\
  && \hskip-.8cm  \blacktriangleright \hskip.1cm \text{\sl  transformed   variables $ {\rm pr}^{[1]} x (t)   , d (t)    , y  (t)     $ in  time-scale $t$}: \nonumber \\
    &&\hskip-.8cm   z_{\mathfrak p} (t)   \delequal (  x_{\mathfrak p} (t)   , d_{\mathfrak p} (t)    , y_{\mathfrak p} (t)    )   \label{zpp1}   \\
  &&\hskip-.8cm   
  \delequal  (  \Psi_{\mathfrak p}^{\mathsf{x}}  (t,x(t))    , \Psi_{\mathfrak p}^{\mathsf{d}}  (t,x(t),d(t))   , \Psi_{\mathfrak p}^{\mathsf{y}}  (t,y(t))   )  ,  \nonumber \\
  &&\hskip-.8cm ( D_t x  )_{\mathfrak p} (t) \delequal  \Psi_{\mathfrak p}^{\mathsf{x}^{(1)}}(t , {\rm pr}^{[1]} x (t)  )  ,   ( {\rm pr}^{[1]}  x  )_{\mathfrak p} (t) \delequal (  x_{\mathfrak p} (t) , ( D_t x  )_{\mathfrak p} (t) )  ,     \nonumber \\
  && \hskip-.8cm \blacktriangleright \hskip.2cm   \text{\sl  transformed   variables $ {\rm pr}^{[1]} x (t)   , d (t)    , y  (t)     $  in  time-scale $\mathsf{t}_{\mathfrak p}$} :  \nonumber \\
  &&\hskip-.8cm   \widetilde{z}_{\mathfrak p} (\mathsf{t}_{\mathfrak p}) \delequal (\widetilde{x}_{\mathfrak p}   (\mathsf{t}_{\mathfrak p}),\widetilde{u}_{\mathfrak p}   (\mathsf{t}_{\mathfrak p}),\widetilde{y}_{\mathfrak p} (\mathsf{t}_{\mathfrak p}))      \delequal    z_{\mathfrak p}(t)  \vert_{ [ \Psi_{\mathfrak p}^t     ]^{-1}  (\mathsf{t}_{\mathfrak p})  }   ,      
  \label{transf_z_tp}  \\  
   && \hskip-.8cm  {\rm pr}^{[1]} \widetilde{x}_{\mathfrak p}  (\mathsf{t}_{\mathfrak p} ) \delequal ( {\rm pr}^{[1]} x )_{\mathfrak p}  (t)   \vert_{ t = [ \Psi_{\mathfrak p}^t     ]^{-1}  (\mathsf{t}_{\mathfrak p})  }  , \nonumber    \\
  &&\hskip-.8cm \blacktriangleright  \hskip.2cm  \text{\sl  system  map  in  time-scale $t$} :  \nonumber \\ 
 &&\hskip-.8cm   \Sigma  (t)  \delequal   \mathsf{(F,H)}  (  t    ,{\rm pr}^{[1]} x (t)   , d (t)    , y  (t)  )  ,     \label{syst_map_t}  \\ 
  &&\hskip-.8cm \blacktriangleright \hskip.2cm  \text{\sl  transformed  system  map  in  time-scale $t$} :   \nonumber \\
 &&\hskip-.8cm   \Sigma_{\mathfrak p} \delequal   \mathsf{(F,H)} (\mathsf{t}_{\mathfrak p}  ,  \mathsf{x}_{\mathfrak p}^{[1]} , \mathsf{d}_{\mathfrak p} , \mathsf{y}_{\mathfrak p}  )  , \label{point_transf_syst_map_t}     \\
  &&\hskip-.8cm  \Sigma_{\mathfrak p} (t) \delequal \mathsf{(F,H)}   (  \Psi_{\mathfrak p}^t  (t)    ,  ( {\rm pr}^{[1]} x )_{\mathfrak p} (t), d_{\mathfrak p} (t) , y_{\mathfrak p} (t) ) ,     
  \nonumber \\
  && \hskip-.8cm \blacktriangleright \hskip.2cm     \text{\sl  transformed  system map   in   time-scale $\mathsf{t}_{\mathfrak p}$} :    \nonumber \\
  &&\hskip-.8cm  \widetilde{  \Sigma}_{\mathfrak p}  (\mathsf{t}_{\mathfrak p}) \delequal \mathsf{(F,H)}   (\mathsf{t}_{\mathfrak p}  ,  {\rm pr}^{[1]} \widetilde{x}_{\mathfrak p}  (\mathsf{t}_{\mathfrak p} )   , \widetilde{u}_{\mathfrak p}   (\mathsf{t}_{\mathfrak p}),\widetilde{y}_{\mathfrak p} (\mathsf{t}_{\mathfrak p})  )  .  \label{transf_syst_map_tp} 
\end{eqnarray} 
With the system map \eqref{syst_map} and notation \eqref{syst_map_t}, the  system \eqref{eq_sys}  can be   characterized in the following three equivalent ways 
\begin{eqnarray}
&&\hskip-1.2cm \eqref{eq_sys} \Leftrightarrow  \Sigma (t  ) = 0  \Leftrightarrow  \mathsf{(F,H)} (t   , {\rm pr}^{[1]} x (t)  , d(t) , y(t) ) = 0 . \label{syst_char}
\end{eqnarray}
For our discussion, we introduce also 
 the vector field  ${\rm pr}^{[1]} {\textbf g}$ on ${\mathbb R} \times {\mathcal X}^{[1]} \times {\mathbb R}^m \times {\mathbb R}^p$ (representing the infinitesimal generator of the prolonged group of transformations)  defined as
 \begin{eqnarray}
&&\hskip-.6cm {\rm pr}^{[1]} {\textbf g} \vert_{(t ,   \mathsf{x}^{[1]} , \mathsf{d} , \mathsf{y} )} :={\rm pr}^{[1]} {\textbf g}^{t,\mathsf{x}} \vert_{(t,  \mathsf{x}^{[1]}) }  + 
 \sum_{j=1}^m g^{\mathsf{d}_j}  (t, \mathsf{x} , \mathsf{d}  ) \frac{\partial  }{\partial \mathsf{d}_j } \nonumber \\
 &&\hskip-.6cm  + 
 \sum_{j=1}^p  g^{\mathsf{y}_j}  ( t, \mathsf{y}) \frac{\partial }{\partial \mathsf{y}_j }   \label{pr_g}
 \end{eqnarray}  
 with ${\rm pr}^{[1]} {\textbf g}^{t,\mathsf{x}} $  as in \eqref{EPRW1}  and by  $ L_{   {\rm pr}^{[1]} {\textbf g}    }\Sigma $ we denote  the Lie derivative  of  $ \mathsf{(F,H)}  $ along the vector field $ {\rm pr}^{[1]} {\textbf g} $ (\!\!\cite{Boothby}), i.e. $   ( L_{   {\rm pr}^{[1]} {\textbf g}    }\Sigma ) (q)   = 
 {\rm pr}^{[1]} {\textbf g}  \vert_{ q }  \mathsf{(F,H)}  = \frac{\partial \mathsf{(F,H)}}{\partial t} (q) g^t (t) + 
 \sum_{j=1}^n \frac{\partial \mathsf{(F,H)}}{\partial \mathsf{x}_j } (q)  g^{\mathsf{x}_j}  ( t, \mathsf{x})   + 
 \sum_{j=1}^n \frac{\partial \mathsf{(F,H)}}{\partial \mathsf{x}^{(1)}_j } (q)  g^{\mathsf{x}^{(1)}_j} ( t, \mathsf{x}^{[1]} ) + 
 \sum_{j=1}^m \frac{\partial \mathsf{(F,H)}}{\partial \mathsf{d}_j } (q)   g^{\mathsf{d}_j}  (t, \mathsf{x} , \mathsf{d}  )  + 
 \sum_{j=1}^p \frac{\partial \mathsf{(F,H) }}{\partial \mathsf{y}_j } (q)   g^{\mathsf{y}_j}  ( t, \mathsf{y}) $, where for brevity we denoted the point  $( t ,   \mathsf{x}^{[1]} , \mathsf{d} , \mathsf{y} )$ by $q$. 
 
\subsection{Definitions, examples and practical aspects}
  
A LGT   transforms the system map $\Sigma$  into $\Sigma_{\mathfrak p} $ (and also $\Sigma(t) $ into $\Sigma_{\mathfrak p} (t)$)  and any solution  $ z(t) $ of the system $ \Sigma (t )  = 0$   into a function   $      \widetilde{z}_{\mathfrak p} (\mathsf{t}_{\mathfrak p})  $ (in time scale $\mathsf{t}_{\mathfrak p}$) which, however, is not always a solution of the system $ \widetilde{ \Sigma}_{\mathfrak p}  (\mathsf{t}_{\mathfrak p})  = 0$. A LGT   is a symmetry of the system $ \Sigma (t )  =  0$  if $    \widetilde{z}_{\mathfrak p} (\mathsf{t}_{\mathfrak p})   $ is a solution of  $ \widetilde{ \Sigma}_{\mathfrak p}  (\mathsf{t}_{\mathfrak p})  =  0$.
 \begin{definition}\label{1d}
 \emph{(Symmetries)}. A LGT (resp. GGT)  $  \Psi_{\mathfrak p}  $ on $ {\mathcal M}$  (with infinitesimal generator $\textbf g$) is a \emph{local  (resp. global) $C^\infty$   symmetry (LS, resp. GS)} of the   system  $   \Sigma (t) = 0$    if  either one of the following  equivalent  conditions  holds:
    \begin{eqnarray}
&&\hskip-1.5cm  {\rm  (i)} \; \forall {\mathfrak p} \in {\mathbb R} ,  \forall ( t , \mathsf{z} )  \in {\rm Dom}(\Psi_{\mathfrak p}  )   :   \Sigma  =  0     \Rightarrow      \Sigma_{\mathfrak p}  = 0   ,    \label{1a}    \\
&&\hskip-1.5cm  {\rm  (ii)}  \; \forall (  t ,\mathsf{z} ) \in  {\mathcal M} : \Sigma  =  0     \Rightarrow       L_{{\rm pr}^{[1]} {\textbf g}}   \Sigma  =  0   .  \hfill \blacktriangleleft \label{1b}  
\end{eqnarray}  
\end{definition}
 
 \begin{remark}\label{geometric}
 {\sl  (A geometric interpretation of conditions \eqref{1a} and \eqref{1b}}).  Condition \eqref{1b} is given for an ODE (more generally, for  a partial differential  equations) system   in   \cite{Olver} and it is a condition based on the invariance of the system map evaluated at  the integral curves of ${\rm pr}^{[1]} {\textbf g}$, i.e. 
 \begin{eqnarray}
&&\hskip-1.4cm \frac{\partial}{\partial {\mathfrak p}}    \mathsf{(F,H)}   (\mathsf{t}_{\mathfrak p}  ,  {\rm pr}^{[1]} \widetilde{x}_{\mathfrak p}  (\mathsf{t}_{\mathfrak p} )   , \widetilde{u}_{\mathfrak p}   (\mathsf{t}_{\mathfrak p}),\widetilde{y}_{\mathfrak p} (\mathsf{t}_{\mathfrak p})  ) \equiv \frac{\partial}{\partial {\mathfrak p}}  \widetilde\Sigma (\mathsf{t}_{\mathfrak p}   ) \equiv 0 . \label{1bbis}
\end{eqnarray}  
In particular, by setting ${\mathfrak p} = 0$ in \eqref{1bbis} and using the (prolonged) generator equation we get  condition \eqref{1b}.    
   
 Condition \eqref{1a} is an alternative (equivalent)  characterization of a LS and, as it will be seen in Section \ref{as_symm},  lends itself to interesting and useful generalizations  and recalls in a general framework the notion of orbital symmetry of a vector field  (or  map): see  \cite{Menini_Tornambe}. The  geometric rationale behind condition \eqref{1a} is the following. By the prolongation formula \eqref{EPRW2} the vector field 
 $F(t,\mathsf{x},\mathsf{d})$ is mapped by   the group of transformations  into
 \begin{eqnarray}
 && \hskip-.8cm  \Psi_{\mathfrak p}^{  \mathsf{x}^{(1)}  } (t, \mathsf{x} ,  F(t,\mathsf{x},\mathsf{d} ) )  \nonumber \\
  && \hskip-.8cm \delequal  ( \frac{\partial  \Psi^t_{\mathfrak p} (t ) }{\partial t}   )^{-1} \left (    \frac{\partial  \Psi^{\mathsf{x}}_{\mathfrak p} (t,\mathsf{x}) }{\partial t}   +    \frac{\partial  \Psi^{\mathsf{x}}_{\mathfrak p} (t,\mathsf{x}) }{\partial \mathsf{x}}   F(t,\mathsf{x},\mathsf{d}) \right )     \nonumber
 \end{eqnarray}
 while the function $ H(t,\mathsf{x},\mathsf{d})$ is mapped  into $   \Psi^{\mathsf{y}}_{\mathfrak p} (t,H(t,\mathsf{x},\mathsf{d})) $. Condition \eqref{1a} amounts to  
   \begin{eqnarray}
 && \hskip-.8cm  \Psi_{\mathfrak p}^{  \mathsf{x}^{(1)}  } (t, \mathsf{x} ,  F(t,\mathsf{x},\mathsf{d} ) )   \equiv  F(\Psi^t_{\mathfrak p} (t ) ,\Psi^\mathsf{x}_{\mathfrak p} (t , \mathsf{x})  ,\Psi^\mathsf{d}_{\mathfrak p} (t , \mathsf{x} , \mathsf{d}) ) ,    \nonumber \\
 && \hskip-.8cm     \Psi^{\mathsf{y}}_{\mathfrak p} (t,H(t,\mathsf{x},\mathsf{d}))  \equiv   H(\Psi^t_{\mathfrak p} (t ) ,\Psi^\mathsf{x}_{\mathfrak p} (t , \mathsf{x})  ,\Psi^\mathsf{d}_{\mathfrak p} (t , \mathsf{x} , \mathsf{d}) )  \label{hom}
 \end{eqnarray}
i.e. $F$ (resp.  $H$) is mapped into itself or it is {\sl invariant under $\Psi_{\mathfrak p}$}. Hence, solutions of $\Sigma(t)=0$ are mapped by the symmetry into other solutions. From  this geometric perspective we recover the well-known notion of {\sl homogeneity} of vector field and maps (see also \cite{Kawski}).  Consider the system map  $\Sigma  $ associated to  the pair $( F( t, \mathsf{x},\mathsf{d} ) , H ( t, \mathsf{x},\mathsf{d} ) )  \delequal  ( A (\mathsf{x})   , C (\mathsf{x})   )$,     and a GGT  of the form  $ \Psi_{\mathfrak p}  (t,   \mathsf{z} ) \delequal    (  e^{ {\mathfrak p}a^t  }  t ,  {\rm diag}_i \{  e^{  {\mathfrak p} (   a^{\mathsf{x}_i} -  a^t ) }   \} \mathsf{x}   ,   \mathsf{d} ,   {\rm diag}_i \{  e^{ {\mathfrak p}  a^{\mathsf{y}_i}   }  \}  \mathsf{y} )$, where  $a^{\mathsf{x}_i} < 0$  and $a^{\mathsf{y}_j} , a^t    \in {\mathbb R}$.  The map   $ \Phi_{ \mathfrak p}  ({\mathsf x}) \delequal {\rm diag}_i \{  e^{ - {\mathfrak p}   a^{\mathsf{x}_i}  }   \}  \mathsf{x}$ is known as {\sl one-parameter group of standard (linear)  dilations} on ${\mathbb R}^n$ (\cite{Pol}).  On account of the geometric interpretation given above  $ \Psi_{\mathfrak p} $ is a GS  of  $\Sigma (t) = 0$ if (and only if) the vector field $A $,  resp.  the mapping $C $,  is {\sl  homogeneous  with weights  $(  - a^{\mathsf{x}_1}   , \cdots , - a^{\mathsf{x}_n}  )$ and degree  $g^t$, resp. degrees $( -a^{\mathsf{y}_1} , \cdots , -a^{\mathsf{y}_p} )$}:  in formulas from \eqref{hom} we have 
 $ {\rm diag}_i \{  e^{ {\mathfrak p} (  a^{\mathsf{x}_i} -  a^t ) }  \}A  (  \mathsf{x} )  \equiv  A ( {\rm diag}_i \{  e^{   {\mathfrak p} a^{\mathsf{x}_i}  } \} \mathsf{x}  ) $ and  ${\rm diag}_i \{  e^{   {\mathfrak p} a^{\mathsf{y}_i}  } \}  C ( \mathsf{x} ) \equiv  C ( {\rm diag}_i \{  e^{   {\mathfrak p} a^{\mathsf{x}_i}  } \}  \mathsf{x}  )$.  $\hfill\blacktriangleleft   $
  \end{remark}
  
In what follows we give some illustrative and motivating examples together with direct procedures for computing symmetries. 
 \begin{example}\label{ex_0}
 \emph{(Computing symmetries from PDE's)}. A first way of computing a LS of a system $\Sigma (t) = 0$ is by solving the PDE resulting from \eqref{1b} with unknown    infinitesimal generator  ${\textbf g}$.  Consider the system map $\Sigma $ associated to  
   \begin{eqnarray}
 (F  , H ) \delequal \left ( \begin{pmatrix} \mathsf{x}_2  \cr  \mathsf{x}_2^2 + \mathsf{d} \end{pmatrix} ,   \mathsf{x}_1  \right  ) , \label{sys_0b}
  \end{eqnarray}
 $\mathsf{x}=(\mathsf{x}_1,\mathsf{x}_2) \in {\mathbb R}^2$. Assuming an infinitesimal generator $\textbf g$  of the form  $\textbf g  \vert_{(t, \mathsf{z} )}  \delequal  g^t (t ) \frac{\partial}{\partial t}  +  g^{\mathsf{x}_1} ( \mathsf{x}_1) \frac{\partial}{\partial \mathsf{x}_1}  +  g^{\mathsf{x}_2} ( \mathsf{x} )   \frac{\partial}{\partial \mathsf{x}_2}+  g^{\mathsf{d}} (\mathsf{x} , \mathsf{d} )   \frac{\partial}{\partial \mathsf{d} }+  g^{\mathsf{y}} ( \mathsf{y} )   \frac{\partial}{\partial \mathsf{y}}$ and using the  prolongation formula \eqref{EPRW1} for computing ${\rm pr}^{[1]} {\textbf g} $ in \eqref{pr_g}, the symmetry condition \eqref{1b} is
   \begin{eqnarray}
 && \hskip-.8cm   g^{\mathsf{y}} \vert_{ \mathsf{y}  =   \mathsf{x}_1 }  \equiv   g^{\mathsf{x}_1}   ,  \left ( \frac{\partial g^{\mathsf{x}_1} }{\partial \mathsf{x}_1}   -    \frac{\partial g^t }{\partial t} \right ) \mathsf{x}_2 \equiv g^{\mathsf{x}_2} , \nonumber \\
 && \hskip-.8cm  \frac{\partial g^{\mathsf{x}_2} }{\partial \mathsf{x}_1} \mathsf{x}_2 +  \left ( \frac{\partial g^{\mathsf{x}_2} }{\partial \mathsf{x}_2}  -   \frac{\partial g^t }{\partial t}    \right ) ( \mathsf{x}_2^2 + \mathsf{d} ) \equiv  2g^{\mathsf{x}_2} \mathsf{x}_2 + g^{\mathsf{d}} ,  \label{gsol} 
 \end{eqnarray}
 from which we get a solution for $\textbf g $ as $ \textbf g  \vert_{(t, \mathsf{z} )}  =  t  \frac{\partial}{\partial t}  -(1+ e^{ \mathsf{x}_1} ) \frac{\partial}{\partial \mathsf{x}_1} - ( 1+  e^{ \mathsf{x}_1}   ) \mathsf{x}_2 \frac{\partial}{\partial \mathsf{x}_2}   - ( 2+  e^{ \mathsf{x}_1}   ) {\mathsf d}  \frac{\partial}{\partial {\mathsf d}}   -(1+ e^{ \mathsf{y}} )     \frac{\partial}{\partial \mathsf{y}}  $.   
 By solving the group generator equation \eqref{inf_gen}  we obtain the local symmetry  $\Psi_{\mathfrak p}({\mathsf t} , {\mathsf z}  )=({\mathsf t}_{\mathfrak p} , {\mathsf x}_{\mathfrak p} , {\mathsf d}_{\mathfrak p}, {\mathsf y}_{\mathfrak p}) $ where 
 \begin{eqnarray}
   &&\hskip-.8cm  {\mathsf t}_{\mathfrak p} \delequal e^{\mathfrak p} t , {\mathsf x}_{\mathfrak p} \delequal \left ( \ln \frac{ e^{  - {\mathfrak p} + {\mathsf x}_1  }   }{ 1+  e^{   {\mathsf x}_1  } -  e^{  - {\mathfrak p} + {\mathsf x}_1  } } ,   \frac{ e^{  - {\mathfrak p}  }   }{ 1+  e^{   {\mathsf x}_1  } -  e^{  - {\mathfrak p} + {\mathsf x}_1  } }  {\mathsf x}_2 \right )  , \nonumber \\
 &&\hskip-.8cm  {\mathsf d}_{\mathfrak p} \delequal  \frac{ e^{  -2 {\mathfrak p}  }   }{ 1+  e^{   {\mathsf x}_1  } -  e^{  -   {\mathfrak p} + {\mathsf x}_1  } }  {\mathsf d} ,  {\mathsf y}_{\mathfrak p} \delequal \ln \frac{ e^{  - {\mathfrak p} + {\mathsf y}   }   }{ 1+  e^{   {\mathsf y}   } -  e^{  - {\mathfrak p} + {\mathsf y}   } }       . \label{Psi_0}
 \end{eqnarray} 
 Notice that ${\rm Dom} (\Psi_{\mathfrak p}) = {\mathcal M}$ if ${\mathfrak p} > 0$ and  $ = \{   (t,{\mathsf z} ) \in  {\mathcal M} : {\mathsf x}_1 , {\mathsf y}  < -\ln (   e^{-\mathfrak p} - 1) \}$   if ${\mathfrak p} < 0$. Moreover, since $\Psi_{\mathfrak p}^t (t ) =   e^{\mathfrak p} t $,  $\Psi_{\mathfrak p}$ is UU-transforming for    ${\mathfrak p} > 0$. A  simpler alternative  solution of \eqref{gsol} leads to $\textbf g  \vert_{(t, \mathsf{z} )}  =  t  \frac{\partial}{\partial t}   -  \mathsf{x}_2 \frac{\partial}{\partial \mathsf{x}_2}  - 2 {\mathsf d}  \frac{\partial}{\partial {\mathsf d}}  + \frac{\partial}{\partial \mathsf{y}}    $ from which we get the global symmetry  $\Psi_{\mathfrak p}({\mathsf t} , {\mathsf z}  ) =({\mathsf t}_{\mathfrak p}  , {\mathsf x}_{\mathfrak p} , {\mathsf d}_{\mathfrak p}, {\mathsf y}_{\mathfrak p} ) $    with  
 \begin{eqnarray}
 &&\hskip-1.2cm {\mathsf t}_{\mathfrak p} \delequal e^{\mathfrak p} t , {\mathsf x}_{\mathfrak p} \delequal  (    {\mathsf x}_1   ,    e^{  - {\mathfrak p}  }     {\mathsf x}_2 )  ,   {\mathsf d}_{\mathfrak p} \delequal e^{  -  2{\mathfrak p}    }     {\mathsf d} ,  {\mathsf y}_{\mathfrak p} \delequal   {\mathsf y}    . \label{Psi_0_alt}
 \end{eqnarray}
 \end{example}
 \begin{example}\label{ex_12}
 {\sl (Symmetries with nonlinear time transformations)}.   Consider the system map $\Sigma $ associated to the pair $(F  , H  )  \delequal  (A\mathsf{x} + B \mathsf{d} , C\mathsf{x}  + D \mathsf{d})$  and assume a candidate infinitesimal generator  $ \textbf g$ of the form  
 $  \textbf g  \vert_{(t, \mathsf{z} )}  =   t^2  \frac{\partial}{\partial t} +  \sum_{j=1}^n ( G^{\mathsf{x}} (t) \mathsf{x} )_j \frac{\partial}{\partial \mathsf{x}_j}   +  \sum_{j=1}^m ( G^{\mathsf{d}} (t) \mathsf{d} )_j \frac{\partial}{\partial \mathsf{d}_j}     +  \sum_{j=1}^p ( G^{\mathsf{y}} (t ) \mathsf{y} )_j     \frac{\partial}{\partial \mathsf{y}_j}  $,  where $G^{\mathsf{x}} (t)  \in {\mathbb R}^{n\times n}$, $G^{\mathsf{d}} (t)  \in {\mathbb R}^{m\times m}$ and $G^{\mathsf{y}} (t)  \in {\mathbb R}^{p\times p}$. Since $\textbf g^t  \vert_{t}  =   t^2  \frac{\partial}{\partial t}$ then $\Psi_{\mathfrak p}^t  (t  ) =    \frac{  t   }{1- {\mathfrak p} t}$ with $ {\rm Dom} (\Psi_{\mathfrak p}^t ) = \{ t \in {\mathbb R} : t \ne  \frac{1}{\mathfrak p} \}$,  $   {\rm Dom}^+  \Psi^t_{\mathfrak p}   =  \frac{1}{\mathfrak p} > 0 $ for ${\mathfrak p} >0$ and  $   {\rm Dom}^+  \Psi^t_{\mathfrak p}    \searrow 0$ for ${\mathfrak p} \to +\infty$, hence the LGT with infinitesimal generator $\textbf g$ is CBU-transforming for   ${\mathfrak p} > 0$ and acts locally  on the  time domain while  globally on the  space ${\mathcal Z}$ of points ${\mathsf z }$. Using  the prolongation formula \eqref{EPRW1}  for computing ${\rm pr}^{[1]} {\textbf g} $ in \eqref{pr_g},  the symmetry condition \eqref{1b} for $\Sigma (t) = 0$ boils down to
  \begin{eqnarray}
 && \hskip-1cm \frac{\partial G^{\mathsf{x}} }{\partial t} (t) \equiv A G^{\mathsf{x}} (t) -  G^{\mathsf{x}} (t) A  + 2t A  ,  \label{ric1} \\
 && \hskip-1cm  0 \equiv G^{\mathsf{x}} (t ) B -  B  (  2t {\textbf I}_m  +    G^{\mathsf{d}} (t) )  ,    \label{ric2} \\
  && \hskip-1cm 0 \equiv G^{\mathsf{y}} (t ) C -  C G^{\mathsf{x}} (t)  , 0 \equiv   G^{\mathsf{y}} (t ) D -  D   G^{\mathsf{d}} (t ) .  \label{ric3}
 \end{eqnarray}
As well-known, the unique solution of the  \emph{differential Sylvester equation} \eqref{ric1}    is $G^{\mathsf{x}} (t) = e^{At} G^{\mathsf{x}} (0) e^{-At}  + A t^2$,  $G^{\mathsf{x}} (0)  \in {\mathbb R}^{n \times n}$. The remaining \eqref{ric2}-\eqref{ric3} are (algebraic)  \emph{Sylvester equations}. 
Once the matrices $G^{\mathsf{x}} (t)$,  $G^{\mathsf{d}} (t)$ and $G^{\mathsf{y}} (t)$ satisfying \eqref{ric1}-\eqref{ric3} are found, we solve  the group generator equation \eqref{inf_gen}   and   get the LS of $\Sigma(t)=0$ in the form  
  \begin{eqnarray}
&&\hskip-1cm  \Psi_{\mathfrak p}  (t,  \mathsf{z} )  \delequal   (  \frac{  t   }{1- {\mathfrak p} t} , \Phi^{\mathsf{x}} (  {\mathfrak p}, t  ) \mathsf{x}, \Phi^{\mathsf{d}} (  {\mathfrak p}, t  )  \mathsf{d} ,  \Phi^{\mathsf{y}} (  {\mathfrak p}, t  ) \mathsf{y}  ) , \label{sym2}
 \end{eqnarray}
where $\Phi^{\mathsf{x}} (  {\mathfrak p}, t  ) \in {\mathbb R}^{n \times n}$, $\Phi^{\mathsf{d}} (  {\mathfrak p}, t  )  \in {\mathbb R}^{m \times m}$ and  $\Phi^{\mathsf{y}} (  {\mathfrak p}, t  ) \in {\mathbb R}^{p \times p}$. 
For instance,  if  
\begin{eqnarray}
(F  , H ) \delequal \left  (\begin{pmatrix} \mathsf{x}_2 \cr   \mathsf{d}  \end{pmatrix}  , \mathsf{x}_1 \right ) , \;  \mathsf{x} = ( \mathsf{x}_1, \mathsf{x}_2 )  , \label{infinite}
\end{eqnarray}
as solution of  \eqref{ric1}-\eqref{ric3} we get  
\[
 G^{\mathsf{x}} (  t  ) = \begin{pmatrix} t &0 \cr 1 &-t  \end{pmatrix} ,  G^{\mathsf{d}} (  t  ) = - 3t   , G^{\mathsf{y}} (  t  ) = t .
 \]
  From these matrices,  we compute the matrices $\Phi_{\mathfrak p}^{\mathsf  x}  ( t , {\mathsf  x} )  $,  $\Phi_{\mathfrak p}^{\mathsf{d}} (  t , {\mathsf  u} ) $ and $ \Phi_{\mathfrak p}^{\mathsf{y}} ( t,  \mathsf{y} )  $ in \eqref{sym2} resulting in the local symmetry
  \begin{eqnarray}
&&\hskip-.8cm  \Psi_{\mathfrak p}  (t,  \mathsf{z} )  \delequal  \left  (  \frac{  t   }{1- {\mathfrak p} t} , \begin{pmatrix} \frac{ 1  }{1- {\mathfrak p} t}  &0 \cr    {\mathfrak p}   & 1-{\mathfrak p}t  \end{pmatrix}   \mathsf{x}, ( 1-{\mathfrak p}t )^3 \mathsf{d} ,  \frac{\mathsf{y}}{1- {\mathfrak p} t}  \right  )   \nonumber \\
&&\hskip-.8cm  \label{locgr}
 \end{eqnarray}
 \end{example}
\begin{example}\label{ex_3}
 {\sl (Triangular systems maps)}. Symmetries can be computed as well by exploiting the system's structure.  Consider the {\sl lower triangular} system map $\Sigma $ associated to  
    \begin{eqnarray} 
 &&\hskip-1.3cm (F( t,\mathsf{x},\mathsf{d} ) , H( t,\mathsf{x},\mathsf{d} )) \! = \!  \left ( \begin{pmatrix} A_1 (t,  \mathsf{x}_1 ) +  B_1 ( t, 
 \mathsf{x}_1) \mathsf{x}_2    \cr    A_2 ( t, \mathsf{x})  +  B_2 ( t, \mathsf{x})  \mathsf{d}    \end{pmatrix}  \! ,  \mathsf{x}_1 \!  \right )   \label{triang}
      \end{eqnarray}
with  $\mathsf{x} \delequal (\mathsf{x}_1, \mathsf{x}_2 )$,   $\mathsf{x}_1 , \mathsf{x}_2 ,   \mathsf{d} \in {\mathbb R}^n$,  $ B_1 ( t,\mathsf{x}_1), B_2( t,\mathsf{x}) \in {\mathbb {GL}} (n)$. We compute a symmetry of  $\Sigma (t) = 0$ according to a two-steps procedure.   
 The first step consists of computing, with the help   of   the prolongation formula \eqref{EPRW1} and the symmetry condition \eqref{1b}, the infinitesimal generator  ${\textbf g}_1$ of a LS of the  system   $\Sigma_1 (t) =  0$, with system map  $\Sigma_1  $ associated to the pair   $ ( F_1      , H_1    )  \delequal
  ( A_1 (t,  \mathsf{x}_1 ) +  B_1 ( t, 
 \mathsf{x}_1 )  \mathsf{d}_1  ,   \mathsf{x}_1 )$.  The system   $\Sigma_1 (t) =  0$  is obtained  from   $\Sigma  (t) =  0$ by considering the first block of equations    $D_t x_1 (t) =    A_1 (t, x_1(t) ) +  B_1 ( t, 
x_1(t) )x_2 (t)$, $y (t) = x_1 (t)$ in  $\Sigma  (t)  = 0$  and taking   $u_1 (t) = x_2 (t)$ as input and $y_1 (t) = y (t)$ as output.  Relying once more on  the prolongation formula \eqref{EPRW1} and the symmetry condition \eqref{1b}, the second step consists of   extending the infinitesimal generator ${\textbf g}_1$ (on the $(t,{\mathsf x}_1,{\mathsf d}_1,{\mathsf y}_1)$-space)  to the infinitesimal generator ${\textbf g}$ (on the $(t,{\mathsf z})$-space) of  a LS  of $\Sigma  (t)  = 0$ by setting $\mathsf{d}_1= \mathsf{x}_2  $ and $\mathsf{y} = \mathsf{y}_1  $. For instance, the generator    ${\textbf g}_1 \vert_{( t, \mathsf{x}_1 ,  \mathsf{d}_1 ,  \mathsf{y}_1 ) } \delequal  g^t (t ) \frac{\partial}{\partial t}  +  g^{\mathsf{x}_1} ( \mathsf{x}_1) \frac{\partial}{\partial \mathsf{x}_1}  +  g^{\mathsf{d}_1} ( t,  \mathsf{x}_1 ,  \mathsf{d}_1 )   \frac{\partial}{\partial \mathsf{d}_1}+   g^{\mathsf{y}_1} ( \mathsf{y}_1 )   \frac{\partial}{\partial \mathsf{y}_1}$   of a LS of $\Sigma_1 (t)  = 0$ must satisfy on account of \eqref{1b}
  \begin{eqnarray}
 && \hskip-.8cm   g^{\mathsf{y}_1} \vert_{\mathsf{y}_1 = \mathsf{x}_1}  \equiv  g^{\mathsf{x}_1}   , \;  g^{\mathsf{d}_1}  \equiv  B_1^{-1}   \Big [   \left ( \frac{\partial g^{\mathsf{x}_1} }{\partial \mathsf{x}_1}   -    \frac{\partial g^t }{\partial t} \right )(A_1  +  B_1    \mathsf{d}_1 ) \nonumber \\
  && \hskip-.8cm - (\frac{\partial A_1 }{\partial t}    + \frac{\partial  B_1}{\partial t}     \mathsf{d}_1 )  g^t  - ( \frac{\partial A_1 }{\partial \mathsf{x}_1}      +  \frac{\partial  B_1}{\partial \mathsf{x}_1}      \mathsf{d}_1 )  g^{\mathsf{x}_1} \Big  ] . \nonumber 
 \end{eqnarray}
 The infinitesimal generator    ${\textbf g}  \vert_{( t, \mathsf{x}  ,  \mathsf{d}  ,  \mathsf{y} ) }  $   of a LS of $\Sigma  (t)  = 0$ is   obtained by setting $\mathsf{d}_1= \mathsf{x}_2  $ and $\mathsf{y}_1 = \mathsf{y}  $  in ${\textbf g}_1 \vert_{( t, \mathsf{x}_1 ,  \mathsf{d}_1 ,  \mathsf{y}_1 ) }$:
   \begin{eqnarray}
&&\hskip-1cm   {\textbf g}  \vert_{( t, \mathsf{z} ) } \delequal {\textbf g}_1  \vert_{( t, \mathsf{x}_1  ,  \mathsf{x}_2  ,  \mathsf{y} ) }  +   g^{\mathsf{d}} ( t,\mathsf{x} , \mathsf{d}  )   \frac{\partial}{\partial \mathsf{d} }  = g^t (t ) \frac{\partial}{\partial t} \nonumber \\
&&\hskip-1cm  +  g^{\mathsf{x}_1} ( \mathsf{x}_1) \frac{\partial}{\partial \mathsf{x}_1}  +  g^{\mathsf{d}_1} ( t, \mathsf{x}  )   \frac{\partial}{\partial \mathsf{x}_2} +   g^{\mathsf{d}} ( t,\mathsf{x} , \mathsf{d}  )   \frac{\partial}{\partial \mathsf{d} } +   g^{\mathsf{y}_1} ( \mathsf{y}  )   \frac{\partial}{\partial \mathsf{y} }   \nonumber 
\end{eqnarray} 
with $ g^{\mathsf{d}} $ computed as satisfying the  symmetry condition \eqref{1b}  for $\Sigma(t) =0$.  
For instance, for the system map $\Sigma  $ associated to  the pair
    \begin{eqnarray} 
 &&\hskip-1cm (F  , H )  \delequal  \left ( \begin{pmatrix}  \mathsf{x}_2    \cr   \mathsf{x}_2 \mathsf{x}_1^k +  (1+\mathsf{x}_1^2)  \mathsf{d}    \end{pmatrix}   ,  \mathsf{x}_1  \right )  ,\label{triang_ex}
      \end{eqnarray}
 $k \ge 2$,    we obtain the infinitesimal generator 
\begin{eqnarray}
&&\hskip-1cm   {\textbf g}  \vert_{( t, \mathsf{z} ) } =  k t \frac{\partial}{\partial t}  
- \mathsf{x}_1  \frac{\partial}{\partial \mathsf{x}_1}   -(1+k) \mathsf{x}_2    \frac{\partial}{\partial \mathsf{x}_2} \nonumber \\
&&\hskip-1cm + \left ( -( 1+2k ) + \frac{2  \mathsf{x}_1^2 }{1+\mathsf{x}_1^2} \right ) \mathsf{d}    \frac{\partial}{\partial \mathsf{d} }  - \mathsf{y} \frac{\partial}{\partial \mathsf{y} }  \label{triang_gen}
\end{eqnarray}
and, by integration of the group generator equation \eqref{inf_gen} we obtain the GS $ \Psi_{\mathfrak p}  ( t, \mathsf{z} ) =  ( {\mathsf t}_{\mathfrak p} ,  {\mathsf x}_{\mathfrak p} , {\mathsf d}_{\mathfrak p}, {\mathsf y}_{\mathfrak p} )$ of $\Sigma (t) = 0$ with 
\begin{eqnarray}
&&\hskip-1cm  {\mathsf t}_{\mathfrak p}  \delequal  e^{k{\mathfrak  p}}t , \; {\mathsf x}_{\mathfrak p} \delequal {\rm diag} \{  e^{-{\mathfrak  p}}    ,   e^{-(1+k){\mathfrak  p}} \} \mathsf{x}  ,  \nonumber \\
&&\hskip-1cm   {\mathsf d}_{\mathfrak p} \delequal    e^{-( 2k - 1 ){\mathfrak  p}} \frac{1+\mathsf{x}_1^2}{ e^{ 2{\mathfrak  p}}  +\mathsf{x}_1^2}  \mathsf{d}  , \; {\mathsf y}_{\mathfrak p} \delequal   e^{-{\mathfrak  p}}\mathsf{y}  .  \label{triang_symm}
\end{eqnarray}
Notice that the GGT \eqref{triang_symm} is  UU-transforming for   ${\mathfrak p} > 0$.  The reader is left with patiently  extracting from  above   a {\sl top-down} step-by-step symmetry-computing method    for higher-dimensional  {\sl lower  triangular} systems as well as with similar technicalities  a  {\sl bottom-up} step-by-step method  for    {\sl upper  triangular} systems.    
 \end{example}

\subsection{Contracting symmetries}\label{S2.2}

The symmetries are one-parameter groups of transformations mapping solutions into other (parameterized) solutions.   A useful property of a symmetry, which opens completely new perspectives in control design, is its capability of  contracting  solutions into other   solutions of arbitrarily small magnitude, where the magnitude is parameterized by the group parameter itself.  This motivates the following couple of related definitions.    
\begin{definition}\label{def_monotone_group}
(\textit{Diagonal non-contracting transformations}). 
We  say that  a   one-parameter group of linear transformations $ \Gamma_{\mathfrak p}^\mathsf{x} \times  \Gamma_{\mathfrak p}^\mathsf{d}    :  {\mathbb R}^n \times {\mathbb R}^m \to {\mathbb R}^n \times {\mathbb R}^m$,   $(\Gamma_{\mathfrak p}^\mathsf{x}\times \Gamma_{\mathfrak p}^\mathsf{d} ) ( \mathsf{x} ,\mathsf{d} ) \delequal  (\Gamma_{\mathfrak p}^\mathsf{x}  \mathsf{x}  , \Gamma_{\mathfrak p}^\mathsf{d}  \mathsf{d}  )$,  is  a {\sl one-parameter group of diagonal non-contracting transformations} (GGDT) on $ {\mathbb R}^n \times {\mathbb R}^m$ if    $\Gamma_{\mathfrak p}^\mathsf{x} = {\rm diag}_{i=1, \dots , n} \{  \gamma^{\mathsf{x}_i}_{\mathfrak p} \}$ and  $\Gamma_{\mathfrak p}^\mathsf{d}  = {\rm diag}_{j=1, \dots , m} \{  \gamma^{\mathsf{d}_j}_{\mathfrak p} \}$  and for all $i=1, \dots , n$ and $j=1, \dots , m$
\begin{eqnarray}
&&\mathfrak{p}_1 \le \mathfrak{p}_2 \Rightarrow \gamma^{\mathsf{x}_i}_{\mathfrak{p}_1}    \le \gamma^{\mathsf{x}_i}_{\mathfrak{p}_2}  ,  \gamma^{\mathsf{d}_j}_{\mathfrak{p}_1}  \le \gamma^{\mathsf{d}_j}_{\mathfrak{p}_2}  .\hfill   \blacktriangleleft  \label{mono} 
\end{eqnarray}  
\end{definition}
A one-parameter group of identity  transformations $\mathbbm{1}_{\mathfrak p}^n \times \mathbbm{1}_{\mathfrak p}^m$ on ${\mathbb R}^n \times {\mathbb R}^m$ \footnote{$   \mathbbm{1}_{\mathfrak p}^n \times \mathbbm{1}_{\mathfrak p}^m   $ denotes for each ${\mathfrak p}$ the identity transformation $ \mathbbm{1}^n \times \mathbbm{1}^m   $ on ${\mathbb R}^n \times {\mathbb R}^m$, i.e. $(\mathbbm{1}^n \times \mathbbm{1}^m   ) ( \mathsf{x} ,\mathsf{d} )  =  ( \mathsf{x} ,\mathsf{d} ) $.}
and a one-parameter group of linear  dilations  on ${\mathbb R}^n \times {\mathbb R}^m$  (\!\!  \cite{Kawski})  are examples of  GGDT's on ${\mathbb R}^n \times {\mathbb R}^m$. Notice that     
  $ ( \Gamma_0^\mathsf{x} \times  \Gamma_0^\mathsf{d}  ) ( \mathsf{x} , \mathsf{d} )   =   ( \mathsf{x}  , \mathsf{d}  )    $ by the group property  and $  \left \Vert   ( \mathsf{x} , \mathsf{d}  )    \right  \Vert \le  \left \Vert ( \Gamma_{\mathfrak p}^\mathsf{x} \times  \Gamma_{\mathfrak p}^\mathsf{d}  ) ( \mathsf{x}  , \mathsf{d}  )    \right  \Vert  $ for all $\mathfrak{p}\ge 0$ and $ ( \mathsf{x}  , \mathsf{d} )\in {\mathbb R}^n \times {\mathbb R}^m $ by the non-contracting condition  \eqref{mono}. 
\begin{definition}\label{def_IS_contracting}
(\emph{SI contraction}).  A  LGT  (resp. GGT)   $  \Psi_{\mathfrak p}  $ on ${\mathcal M}$  is \emph{state-input  contracting (SI-contracting)} with contracting maps $(\sigma, \mu,  \Gamma_{\mathfrak p}^\mathsf{x} \times  \Gamma_{\mathfrak p}^\mathsf{d}   )$ if  there exist $\sigma \in {\mathcal K}_\infty$, $\mu   \in {\mathcal L}$  and a  GGDT  $ \Gamma_{\mathfrak p}^\mathsf{x} \times  \Gamma_{\mathfrak p}^\mathsf{d}  $  on $ {\mathbb R}^n \times {\mathbb R}^m$ such that for all   $  {\mathfrak p} \ge  \sigma   (   \Vert    ( \mathsf{x} , \mathsf{d}  )  \Vert $ and $ ( t , \mathsf{x}  , \mathsf{d} ) \in {\rm Dom} ( \Psi_{\mathfrak p}^{t , \mathsf{x}  , \mathsf{d}} )$ it holds that
\begin{eqnarray}
\left \Vert ( \Gamma_{\mathfrak p}^\mathsf{x} \times  \Gamma_{\mathfrak p}^\mathsf{d}  ) ( \mathsf{x}_{\mathfrak p} , \mathsf{d}_{\mathfrak p} )    \right  \Vert  \le \mu  ( {\mathfrak p} )        .  \label{SM2}  \hfill  
 \blacktriangleleft 
\end{eqnarray}
\end{definition}
For a group of transformations  $\Psi_{\mathfrak p}$  with ${\rm Dom} ( \Psi_{\mathfrak p}  ) = {\mathcal M}$ for each ${\mathfrak p} > 0$  and  SI-contracting  with contracting maps $(\sigma, \mu,  \Gamma_{\mathfrak p}^\mathsf{x} \times  \Gamma_{\mathfrak p}^\mathsf{d}   )$, if we pick any $(  t , \mathsf{x} , \mathsf{d} ) \in {\mathbb R} \times {\mathbb R}^n \times {\mathbb R}^m$ then there always  exists ${\mathfrak p}_0 > 0$ such that $  \sigma   (   \Vert    ( \mathsf{x} , \mathsf{d}  )  \Vert   )   \le    {\mathfrak p}$ for  ${\mathfrak p} \ge {\mathfrak p}_0 $, hence the right-hand side inequality in \eqref{SM2} holds  for all ${\mathfrak p}  \ge {\mathfrak p}_0 $. Since  $\mu \in {\mathcal L}$ we get  $      \Vert ( \Gamma_{\mathfrak p}^\mathsf{x} \times  \Gamma_{\mathfrak p}^\mathsf{d}  ) ( \mathsf{x}_{\mathfrak p} , \mathsf{d}_{\mathfrak p} )   \Vert \to 0$ as ${\mathfrak p} \to +\infty$. In conclusion,  $ \Psi_{\mathfrak p}  $  acts contractively  on the  $( \mathsf{x} ,\mathsf{d} )$-space  with guaranteed contraction rates $\Gamma_{-\mathfrak p}^\mathsf{x} \times  \Gamma_{-\mathfrak p}^\mathsf{d}$  as  ${\mathfrak p} \to +\infty$. 

\textbf{\emph{Example}} \emph{\ref{ex_3}} \textbf{\emph{(cont'ed)}}.  
For   \eqref{triang_symm} we have   $ \Vert  (  \mathsf{x}_{\mathfrak p} ,  \mathsf{d}_{\mathfrak p} ) \Vert  \le e^{{- {\mathfrak p} }}  \vert  \mathsf{x}_1 \vert  + e^{{- (1+k){\mathfrak p} }}  \vert  \mathsf{x}_2 \vert  +  e^{ - (2k-1) {\mathfrak p} }  \frac{1+ \mathsf{x}_1^2}{e^{ 2 {\mathfrak p} }  +  \mathsf{x}_1^2}    \vert  \mathsf{d}  \vert \le e^{{- {\mathfrak p} }}  \Vert  (  \mathsf{x}  ,  \mathsf{d}  ) \Vert  $ 
 for all $ {\mathfrak p} \ge 0$ and $(t , \mathsf{x}  ,\mathsf{d})\in {\mathbb R} \times {\mathbb R}^n \times {\mathbb R}^m$  . Hence, \eqref{triang_symm}  is  SI-contracting with contracting maps  
     \begin{eqnarray}
&&\hskip-1cm ( \sigma , \mu , \Gamma_{\mathfrak p}^\mathsf{x} \times  \Gamma_{\mathfrak p}^\mathsf{d}  )  = ( \frac{1}{1 - \delta}   \ln (\varphi (s) +1),  e^{-\delta  s }   ,   \mathbbm{1}_{\mathfrak p}^2 \times  \mathbbm{1}_{\mathfrak p}  )  \label{ex5.3.2.b0}
\end{eqnarray}           
for any $\delta \in (0, 1)$ and  $\varphi \in {\mathcal K}_\infty$ such that $\varphi (s) \ge s$ for all $s \ge 0$ \footnote{The additional degrees of freedom in the choice of $\sigma$ represented  by the function $\varphi  $ and the parameter $\delta$ turns in advantage  for other different purposes: see Section \ref{S3.2.2}}. Different contracting maps $\Gamma_{\mathfrak p}^\mathsf{x} \times  \Gamma_{\mathfrak p}^\mathsf{d} $ with the same pair of functions $(\sigma, \mu)$  can be considered as well,  for instance  $ \Gamma_{\mathfrak p}^\mathsf{x} \times  \Gamma_{\mathfrak p}^\mathsf{d}  = {\rm diag} \{ 
1, e^{ {\mathfrak p} k }  
 \} \times   e^{ 2(k-1){\mathfrak p}  }   $.

\textbf{\emph{Example}} \emph{\ref{ex_0}} \textbf{\emph{(cont'ed)}}.  
For  \eqref{Psi_0} we have ${\rm Dom} ( \Psi_{\mathfrak p} ) = {\mathcal M}$   for all $ {\mathfrak p} > 0 $ and, furthermore,  for all $(  \mathsf{x}  ,  \mathsf{d}  ) \in {\mathbb R}^n \times {\mathbb R}^m $
     \begin{eqnarray}
     &&\hskip-.8cm \frac{1}{\delta} \Vert (  \mathsf{x}  ,  \mathsf{d}  ) \Vert  \le  {\mathfrak p}    \Rightarrow   \label{ex5.3.2}  \\
      &&\hskip-.8cm \Vert  (  \mathsf{x}_{\mathfrak p} )_2 ,  \mathsf{d}_{\mathfrak p} ) \Vert \le  \frac{e^{-  {\mathfrak p}  } }{ 1+e^{ \mathsf{x}_1}-e^{ -{\mathfrak p} 
 + \mathsf{x}_1}} ( \vert \mathsf{x}_2 \vert +e^{-\mathfrak p}  \vert  \mathsf{d} \vert  )   \le  
 e^{ - {\mathfrak p} (1-\delta)  }      \nonumber
     \end{eqnarray}
 with any $\delta \in (0,1) $.  However,  $ \lim_{{\mathfrak p} \to +\infty} \vert ( \mathsf{x}_{\mathfrak p} )_1 \vert = \lim_{{\mathfrak p} \to +\infty} \vert  \ln \frac{ e^{  - {\mathfrak p} + {\mathsf x}_1  }   }{ 1+  e^{   {\mathsf x}_1  } -  e^{  - {\mathfrak p} + {\mathsf x}_1  } } \vert = +\infty $ for each $\mathsf{x}_1$, hence \eqref{Psi_0}  is not SI-contracting. On the other hand,  for \eqref{Psi_0_alt}   we have  
     \begin{eqnarray}
     &&\hskip-.8cm     \frac{1}{\delta} \ln ( 1 + \Vert (  \mathsf{x}  ,  \mathsf{d}  )  \Vert  )  \le  {\mathfrak p}   \Rightarrow   \label{ex5.3.2_alt}  \\
      &&\hskip-.8cm \Vert  (  \mathsf{x}_{\mathfrak p} )_2 ,  \mathsf{d}_{\mathfrak p} ) \Vert \le   e^{-  {\mathfrak p}  }  ( \vert \mathsf{x}_2 \vert +e^{-\mathfrak p}  \vert  \mathsf{d} \vert  )   \le   
 e^{ - {\mathfrak p} (1-\delta)  }    \nonumber
     \end{eqnarray}
for any $\delta \in (0,1 ) $   and for all $ (\mathsf{x}  , \mathsf{d} ) \in  {\mathbb R}^n \times {\mathbb R}^m$. However,    \eqref{Psi_0_alt} is not SI-contracting since $  \vert ( \mathsf{x}_{\mathfrak p} )_1 \vert \equiv \vert  \mathsf{x}_1    \vert   $. 

The fact that there exist symmetries like \eqref{Psi_0} or   \eqref{Psi_0_alt} which are contractive only partially  on the $(\mathsf{x},\mathsf{d})$-space motivate  the following relaxed notion of SI-contraction, which will be worthwhile for some of the forthcoming results.

\begin{definition}\label{def_IS_contracting_bis}
(\textit{Partial SI-contraction}).  A  LGT    $  \Psi_{\mathfrak p}  $   is \emph{partially state-input  contracting (PSI-contracting)} with contracting maps $(\sigma, \mu,  \Gamma_{\mathfrak p}^{\mathsf{x}_{j_1},\dots,\mathsf{x}_{j_r}} \times  \Gamma_{\mathfrak p}^{\mathsf{d}_{i_1},\dots,\mathsf{d}_{i_s}}   )$, where $ \{  j_1 ,\dots, j_r \} \subseteq  \{1, \dots , n \}  \ $ and $\{ i_1 ,\dots, i_s \} \subseteq  \{1, \dots , m \}   $ are (possibly empty) subsets of indexes,  if  there exist $\sigma \in {\mathcal K}_\infty$, $\mu   \in {\mathcal L}$  and a  GGDT   $ \Gamma_{\mathfrak p}^{\mathsf{x}_{j_1},\dots,\mathsf{x}_{j_r}} \times  \Gamma_{\mathfrak p}^{\mathsf{d}_{i_1},\dots,\mathsf{d}_{i_s}} $  on $ {\mathbb R}^r \times {\mathbb R}^s$ such that for all ${\mathfrak p} \ge   \sigma   (   \Vert    ( \mathsf{x} , \mathsf{d}  )  \Vert   ) $ and $ ( t , \mathsf{x}  , \mathsf{d} ) \in {\rm Dom} ( \Psi_{\mathfrak p}^{t , \mathsf{x}  , \mathsf{d}} )  )  $ it holds that 
    \begin{eqnarray}    
&&\hskip-1.2cm     \Vert ( \Gamma_{\mathfrak p}^{\mathsf{x}_{j_1},\dots,\mathsf{x}_{j_r}} \times  \Gamma_{\mathfrak p}^{\mathsf{d}_{i_1},\dots,\mathsf{d}_{i_s}}  ) (\mathsf{x}_{\mathfrak p}^{ j_1  ,\dots, j_r }   , \mathsf{d}_{\mathfrak p}^{ i_1  ,\dots, i_s }   )     \Vert   \le \mu  ( {\mathfrak p} )    \label{SM2_bis}
\end{eqnarray}
where $ \mathsf{x}_{\mathfrak p}^{ j_1  ,\dots, j_r }  \delequal ( ( \mathsf{x}_{\mathfrak p} )_{j_1}  ,\dots, ( \mathsf{x}_{\mathfrak p} )_{j_r} )$ and $ \mathsf{d}_{\mathfrak p}^{ i_1  ,\dots, i_s}  \delequal ( ( \mathsf{d}_{\mathfrak p} )_{ i_1}  ,\dots, ( \mathsf{d}_{\mathfrak p} )_{i_s} )$.     $\hfill \blacktriangleleft$
\end{definition}

\textbf{\emph{Example}} \emph{\ref{ex_0}} \textbf{\emph{(cont'ed)}}.  
The local group of transformations  \eqref{Psi_0},   on account of \eqref{ex5.3.2},     is PSI-contracting with contracting maps $( \sigma , \mu , \Gamma_{\mathfrak p}^{\mathsf{x}_2} \times  \Gamma_{\mathfrak p}^\mathsf{d}  )  = (  \delta s ,  e^{-  s (1-\delta) }   ,   \mathbbm{1}_{\mathfrak p}  \times  \mathbbm{1}_{\mathfrak p}  )$, any $\delta \in (0,1 )$. The  global group of transformations    \eqref{Psi_0_alt},  on account of \eqref{ex5.3.2_alt},   is PSI-contracting with contracting maps  
\begin{eqnarray}
( \sigma , \mu , \Gamma_{\mathfrak p}^{\mathsf{x}_2} \times  \Gamma_{\mathfrak p}^\mathsf{d}  )  = ( \frac{1}{\delta} \ln (1+\varphi(s)) ,  e^{-  s (1-\delta) }   ,   \mathbbm{1}_{\mathfrak p}  \times  \mathbbm{1}_{\mathfrak p}  ) \label{contr_part}
\end{eqnarray}
for any $\delta \in (0,1)$ and $\varphi \in {\mathcal K}_\infty$ such that $\varphi (s) \ge s$ for all $s \ge 0$. 

\section{Symmetry-based observers}\label{S3}
   
The idea behind the design of symmetry-based observers is the following. Let $ z (t) $ be a solution of $\Sigma (t) = 0$ and $\Psi_{\mathfrak p}$ be  a  symmetry  of $\Sigma (t) = 0$ with ${\rm Dom} (  \Psi_{\mathfrak p} ) = {\mathcal M}$ for ${\mathfrak p} > 0$,   SI-contracting     with contracting maps $(\sigma, \mu ,  \Gamma_{\mathfrak p}^\mathsf{x} \times  \Gamma_{\mathfrak p}^\mathsf{d}  )$ and UU-transforming  for ${\mathfrak p} > 0$ (similar considerations can be repeated  with proper modifications for symmetries which are BU-transforming). Pick any $\varrho > 0$.  If the group parameter ${\mathfrak p} $ (possibly time-varying)  is  such   that 
\begin{eqnarray}
{\mathfrak p}    \ge   \sigma    (   \Vert    (  x  (t) , d(t)  )    \Vert + \omega       )  \label{widesigma}
\end{eqnarray}
for all $t \ge T $ \footnote{$T\ge 0$ may vary with the solution  $ z(t)$.}  and for some $\omega  \ge   0$ such that $ ( \mu \circ \sigma ) ( \omega  )  \le \varrho$  \footnote{The existence of such $\omega$ follows from $\sigma \in {\mathcal K}_\infty$ and $ \mu  \in {\mathcal L} $, hence $ \mu \circ \sigma \in {\mathcal L}  $.},  by the SI-contraction property \eqref{SM2} of  $\Psi_{\mathfrak p}$   it follows that   $  \Vert ( \Gamma_{\mathfrak p}^\mathsf{x} \times  \Gamma_{\mathfrak p}^\mathsf{d}  ) (  x_{{\mathfrak p}} (t) , u_{{\mathfrak p}} (t)  )   \Vert    \le  ( \mu \circ \sigma ) ( \omega  )     \le \varrho $  for  all $t \ge T$. 
 Hence,   $ \Sigma_{\mathfrak p}   (\mathsf{t} ) = 0 $ (and $   \widetilde\Sigma_{\mathfrak p}    (\mathsf{t}_{\mathfrak p}) = 0$ in time scale ${\mathsf t}_{\mathfrak p}$) can be approximated to any   degree  $\varrho$ by its linear approximation  around the origin.   
Under  some detectability assumptions on  the linear approximation of  $ \Sigma   (t) = 0 $,  we design    an exponential  observer for $ \widetilde\Sigma  (\mathsf{t}_{\mathfrak p}) = 0$  with exponentially converging estimate $\widehat\chi   ( \mathsf{t}_{\mathfrak p} ) \to  \widetilde{x}_{{\mathfrak p}} (\mathsf{t}_{\mathfrak p})$ as $\mathsf{t}_{\mathfrak p} \to +\infty  $ (asymptotic convergence in time scale $ \mathsf{t}_{\mathfrak p} $ is allowed since $ \Psi_{\mathfrak p}$ is   UU-transforming). By inverse group transformation $[ \Psi_{ {\mathfrak p}}^{ t, \mathsf x} ]^{-1}$   we get an asymptotically converging estimate $\widehat{x} (t) \to x  (t) $ as $t \to +\infty$ (asymptotic convergence in time scale $ t $ is implied by asymptotic convergence in time scale $ \mathsf{t}_{\mathfrak p} $  since $ \Psi_{\mathfrak p}$ is   UU-transforming). 
\begin{remark}\label{detect}
Assuming detectability of the linearization of $  \Sigma     (t) = 0$  represents a  simple way of guaranteeing the existence of a local $C^\infty$ observer for the transformed system $ \widetilde\Sigma  (\mathsf{t}_{\mathfrak p}) = 0$ but any other local  observer can be used, including  non-Lipschitz observers such as SLOs. Mixing up symmetry-based techniques with non-Lipschitz observers may remarkably enlarge the   range of applications of both SLOs and symmetry-based observers.  $\hfill \blacktriangleleft$
\end{remark}

There are two ways of satisfying \eqref{widesigma}: if we consider all the   solutions $ ( x  (t) ,  d (t)  ) $ bounded   for all $t \ge 0$ by a known number $N > 0$ then  we   use a {\sl constant} ${\mathfrak p} \ge \sigma (N) $ ({\sl semiglobal} observers) or   if we consider also unbounded   solutions $ ( x  (t) ,  d (t)  ) $   but still bounded for all $t \ge T \ge 0$ by a {\sl time-varying} variable  $\widehat{V} (t)$, filtered from the system's observations $y(t)$,   then  we   use a {\sl time-varying} ${\mathfrak p} (t) \ge \sigma (\widehat{V} (t)) $  ({\sl global} observers).

  \subsection{Semiglobal symmetry-based observers}\label{S3.1}
  
 As discussed above, when $\sup_{t \ge 0}  \Vert     ( x  (t) ,  d (t)  )     \Vert  < +\infty$  we set  ${\mathfrak p}$ constant satisfying    \eqref{widesigma}. We sum  up below (with remarks) the main  assumptions   behind our semiglobal observer design. First of all, we assume the existence of a SI-contracting and  UU-transforming symmetry. 

{\textbf{\emph{Assumption S.1}}:  
 $ \Psi_{\mathfrak p}$   is a LS  of  $\Sigma (t)  =  0$ with ${\rm Dom} ( \Psi_{\mathfrak p} ) = {\mathcal M}$ for ${\mathfrak p} > 0$, SI-contracting with  contracting maps $(\sigma , \mu , \Gamma_{\mathfrak p}^\mathsf{x} \times  \Gamma_{\mathfrak p}^\mathsf{d}  )$ and     UU-transforming for   ${\mathfrak p} > 0$. $\hfill \blacktriangleleft$ 
 
 Some (mild) conditions are required on the inverse group of transformation $[ \Psi_{ {\mathfrak p}}^{ t, \mathsf x} ]^{-1}$ and the input transformation map  $\Psi_{ {\mathfrak p}}^{\mathsf d} $ for obtaining a converging estimate $\widehat{x} (t)$ of $ x  (t) $ (in time scale $t$) from a converging estimate $\widehat\chi (\mathsf{t}_{\mathfrak p} ) $  of  the transformed state $ \widetilde{x}_{{\mathfrak p}}  (\mathsf{t}_{\mathfrak p} ) $  (in time scale $\mathsf{t}_{\mathfrak p}  $). We remind the reader the notational meaning of  $\mathsf{t}_{\mathfrak p} ,  \mathsf{x}_{\mathfrak p}  ,  \mathsf{d}_{\mathfrak p}   ,  \mathsf{y}_{\mathfrak p}$  (see Notation \eqref{zp}).

{\textbf{\emph{Assumption S.2}}:  There exist ${\mathfrak p}_0 > 0$,   $\pi_1 \in {\mathcal K}_+$ and $\pi_2 \in {\mathcal L}$ such that for all ${\mathfrak p}  \ge {\mathfrak p}_0 $ and  $ ( t ,  \overline{\mathsf{x}}   ) \in   {\rm Dom} ( \Psi^t_{ \mathfrak p} ) \times {\mathbb R}  $ for which  $  \Vert  \overline{\mathsf{x}}  \Vert  \le n \mu ({\mathfrak p} )$ we have 
$    (  \mathsf{t}_{\mathfrak p}  ,{\mathsf x} ) \in {\rm Dom} ( \Psi^{ t,{\mathsf x} }_{-\mathfrak p} )  $ and $\Vert \frac{\partial 
\Psi^{\mathsf{x}}_{-\mathfrak p} ( \mathsf{t}_{\mathfrak p}   ,  \overline{\mathsf{x}}  ) }{\partial  \overline{\mathsf{x}} }   \Vert  \le \frac{ \pi_1  (  \Vert  \overline{\mathsf{x}}   \Vert ) }{ \pi_2 ({\mathfrak p})   }$.$\hfill \blacktriangleleft
$
      
The way we present  Assumption S.2 is motivated by the fact that, although ${\rm Dom} ( \Psi_{\mathfrak p} ) = {\mathcal M}$ for ${\mathfrak p} > 0$ by Assumption S.1,  we may as well  have $  (\mathsf{t}_{\mathfrak p} ,  \overline{\mathsf{x}}  )  \notin {\rm Dom} ( \Psi^t_{-\mathfrak p} ) $ for some ${\mathfrak p} > 0$ and  $( t ,  \overline{\mathsf{x}}  ) \in   {\rm Dom} ( \Psi^t_{ \mathfrak p} ) \times   {\mathbb R}^n$, hence  $\frac{\partial 
\Psi^{\mathsf{x}}_{-\mathfrak p} (\mathsf{t}_{\mathfrak p}  ,   \overline{\mathsf{x}}  ) }{\partial   \overline{\mathsf{x}}  } $ not defined at such point $( \mathfrak p , t,  \overline{\mathsf{x}} )$. Recall that,  by (ii) of Definition \ref{onep},  we have  $ (\mathsf{t}_{\mathfrak p}  ,\mathsf{x}_{\mathfrak p}  ) \in {\rm Dom} ( \Psi^{ t,\mathsf{x} }_{-\mathfrak p} ) $ {\sl  but not} $ (\mathsf{t}_{\mathfrak p}  , \overline{\mathsf{x}}    ) \in {\rm Dom} ( \Psi^{ t,\mathsf{x} }_{-\mathfrak p} ) $ for all ${\mathfrak p} > 0$ and  $( t ,  \mathsf{x}, \overline{\mathsf{x}} ) \in {\mathbb R}  \times  {\mathbb R}^n   \times  {\mathbb R}^n $.

{\textbf{\emph{Assumption S.3}}:
 There exist  $\pi_3 \in {\mathcal K}$   such that $  \Vert \mathsf{d}_{ \mathfrak p}  \Vert   \le \pi_2 (  {\mathfrak p})   \pi_3  (   \Vert  \mathsf{d}  \Vert )$  for all $ {\mathfrak p} > 0$,  $( t ,  \mathsf{x} ,  \mathsf{d} ) \in  {\mathbb R} \times  {\mathbb R}^n \times {\mathbb R}^m$. $\hfill \blacktriangleleft$ 

Let   
     \begin{eqnarray} 
 &&\hskip-1.3cm  A  (t) \delequal   \frac{\partial    F (t,\mathsf{x},\mathsf{d})  }{\partial \mathsf{x} } \Big \vert_{(\mathsf{x},\mathsf{d})=0}  ,  B  (t) \delequal   \frac{\partial    F (t,\mathsf{x},\mathsf{d})  }{\partial  \mathsf{d}} \Big \vert_{(\mathsf{x},\mathsf{d})=0}  ,  \nonumber \\
 &&\hskip-1.3cm C  (t) \delequal   \frac{\partial  H (t,\mathsf{x},\mathsf{d})  }{\partial \mathsf{x} } \Big \vert_{(\mathsf{x},\mathsf{d})=0}  ,  D  (t) \delequal   \frac{\partial   H (t,\mathsf{x},\mathsf{d})  }{\partial  \mathsf{d}} \Big \vert_{(\mathsf{x},\mathsf{d})=0}  ,  \label{abcd} \\
    &&\hskip-1.3cm
 ( \Delta \Sigma_L  ) ( t, \mathsf{x} , \mathsf{d})  \delequal  \begin{pmatrix}  F   ( t,  \mathsf{x} , \mathsf{d}) \cr  H   ( t,   \mathsf{x} , \mathsf{d})  \end{pmatrix}   -   \begin{pmatrix} A  (t) &  B  (t) \cr  C  (t)& D  (t)\end{pmatrix} \begin{pmatrix}  \mathsf{x}   \cr   \mathsf{d} \end{pmatrix}   . \label{Delta_2}
     \end{eqnarray}   
As announced in Remark \ref{detect}, we introduce some detectability property on the linearization of $\Sigma (t) = 0$.  
           
{\textbf{\emph{Assumption S.4}}:  
  $ \sup_{t \ge 0}  \Vert B (t) \Vert 
 < +\infty$, $ \sup_{t \ge 0} \Vert D (t) \Vert 
 < +\infty$  and there exist  $K   : {\mathbb R}_+^0  \to  {\mathbb R}^{n \times p} $,  $P  : {\mathbb R}_+^0  \to    {\mathbb S}_+ (n)$ and  $ {\bf \underline{p}}  ,  {\bf \overline{p}}    > 0$ such that  $ \sup_{t \ge 0}  \Vert K (t) \Vert  
 < +\infty$ and   $ D_t P (t) + P (t) (A (t)   -K  (t)  C   (t))  + (A (t)   -K  (t)  C   (t))^\top (t)  P (t)  + 2{\textbf I}_n   \in {\mathbb  S}_- (n)  $ with  
${\bf {\bf \underline{p}}}  {\textbf I}_n \le P (t) \le {\bf {\bf \overline{p}}} {\textbf I}_n $ for all $t \ge 0$.$\hfill \blacktriangleleft$  
  
{\textbf{\emph{Assumption S.5}}:   There exist   $\xi   \in {\mathcal K}$ such that  $   \Vert   \frac{\partial  ( \Delta \Sigma_L  )  (t , {\mathsf x} , {\mathsf d}  )  }{\partial ( {\mathsf x} , {\mathsf d}   )   }   \Vert \le  \xi  (   \Vert  ( {\mathsf x} , {\mathsf d}  ) \Vert )       $ for all $t \ge 0$ and $( {\mathsf x} , {\mathsf d} ) \in {\mathbb R}^n \times {\mathbb R}^m$. $\hfill \blacktriangleleft$ 
  
Notice that \emph{for each $t \ge 0$},  by continuity of $ \frac{\partial  ( \Delta \Sigma_L  )  (t , {\mathsf x} , {\mathsf d}  )  }{\partial ( {\mathsf x} , {\mathsf d}   )   } $  and since  $\Vert \frac{\partial ( \Delta \Sigma_L )   }{\partial ( {\mathsf x} , {\mathsf d}   )   } ( t, \mathsf{x} , \mathsf{d}) \Vert_{( \mathsf{x} , \mathsf{d})=0}  = 0$, there always exists such $\xi  \in {\mathcal K}$. 
Define the dynamic output filter
   \begin{eqnarray}
 &&\hskip-.8cm\begin{pmatrix}  D_t \chi (t) \cr \zeta (t)   \end{pmatrix}  = \begin{pmatrix} \frac{\partial \Psi^t_{\mathfrak p} (t)  }{\partial t} {\textbf I}_n & {\textbf 0}_{n \times p} \cr {\textbf 0}_{p \times n} & {\textbf I}_p \end{pmatrix} \Bigg (   \begin{pmatrix} A   ( \Psi^t_{\mathfrak p} (t)  )     \cr  C ( \Psi^t_{\mathfrak p} (t)  )     \end{pmatrix}    \chi (t)    \nonumber     \\
  &&\hskip-.8cm   +  ( \Delta \Sigma_L  ) (\Psi^t_{\mathfrak p} (t)   ,    \chi^{\rm sat} (t) , 0 )   \label{observer_1}    \\
  &&\hskip-.8cm  +    \begin{pmatrix}        K  (\Psi^t_{\mathfrak p} (t)  )   ( \Psi^{\mathsf y}_{\mathfrak p} (t, y(t))       - \zeta(t)) \cr   {\textbf 0}_{p \times 1}   \end{pmatrix}   \Bigg )  \nonumber 
   \end{eqnarray}
with    $  \chi^{\rm sat} (t)  \delequal  
 \Gamma^{\mathsf x}_{-\mathfrak p}    {\rm sat}_{\mu ( {\mathfrak p}  )}   (  \Gamma^{\mathsf x}_{\mathfrak p}  (  \chi  (t)  ) )$. 
      \begin{theorem}\label{TH1}
 Under Assumptions S.1-S.5 and for each $ N > 0$  consider the solutions   $z(t)$ of $\Sigma (t) = 0 $ for which $\sup_{t \ge 0} \Vert    (   x (t)  , d (t)  )  \Vert  \le N$ together with the filter   \eqref{observer_1}, its design parameter ${\mathfrak p} > 0$ and its output $ \widehat{x} (t)  \delequal   \Psi_{ -{\mathfrak p} }^{ \mathsf{x} }  ( \Psi^t_{\mathfrak p} (t)  , \chi^{\rm sat} (t) )$. For each $\varepsilon > 0$ there exists  $ {\mathfrak p} >0$   such that   
  \begin{eqnarray}
 && \hskip-.7cm  \limsup_{t \rightarrow +\infty}  \Vert x (t) - \widehat{x}  (t)   \Vert   \le \{  \varepsilon +  8{\bf \overline{p}}  \sqrt\frac{2{\bf \overline{p}}}{{\bf \underline{p}}}     (1+ \sup_{ t \ge 0 }   \Vert K ( t ) \Vert  ) \times \nonumber \\
  && \hskip-.7cm \times (  \sup_{ t \ge 0  }   \Vert B (  t ) \Vert  +   \sup_{t \ge  0 }   \Vert D ( t ) \Vert \}  \pi_1 (0) ) \pi_3 (\sup_{t \ge 0} \Vert d(t) \Vert  )  .  \label{error} \hfill \blacktriangleleft
   \end{eqnarray}   
 \end{theorem} 
\begin{remark}\label{tuning}
 The proof of Theorem \ref{TH1}, reported in the Appendix,  points out the procedure for tuning the parameter  $ {\mathfrak p} >0$ of the filter \eqref{observer_1} (inequalities  \eqref{fine1.0}-\eqref{fine1}). Notice that  if $B (t) \equiv 0$ and  $D  (t)  \equiv 0$ (i.e. the linearization of  $\Sigma (t)= 0$ is input-insensitive) it follows from 	\eqref{error}   that  $\widehat{x}  (t) $ is an asymptotic estimate of any solution $x(t)$ of $\Sigma (t) = 0 $ for which $\sup_{t \ge 0} \Vert (x(t), d(t) ) \Vert  \le N$ with  arbitrarily small asymptotic  error bound  $\varepsilon   \pi_3 (N) $. If, in addition, $d(t) \equiv 0$ we have exact convergence  $ x (t) \to \widehat{x}  (t) $ as $t \to +\infty$. 
\end{remark}  
           
\textbf{\emph{Example}} \emph{\ref{ex_3}} \textbf{\emph{(cont'ed)}}.   
 Consider  the system map $\Sigma $ associated to   \eqref{triang_ex} and  the GS of $\Sigma(t)=0$ in \eqref{triang_symm}. As already seen, \eqref{triang_symm}  is SI-contracting with contracting maps \eqref{ex5.3.2.b0}  and UU-transforming (Assumption  S.1). Let's see how it is possible to meet the remaining Assumptions  S.2-S.5. From \eqref{triang_symm},  $ \left \Vert    \frac{\partial \Psi_{-{\mathfrak p}}^{\mathsf{x}} (   {\mathsf t}_{\mathfrak p}     ,\mathsf{x})  }{\partial     \mathsf{x}   } \right \Vert  \le  \frac{\pi_1  (\Vert {\mathsf x} \Vert ) }{ \pi_2 ( {\mathfrak p} )} $ for all $ {\mathfrak p} \ge 0$ and $( t ,   {\mathsf x} ) \in {\mathbb R} \times {\mathbb R}^n$,  with  
 \begin{eqnarray}
     \pi_1 (s) \delequal 1+s  \in {\mathcal K}_+  ,   \pi_2 (s ) \delequal   e^{ - (1+k) s    } \in {\mathcal L}     \label{pi2.1}
 \end{eqnarray}
 (Assumption  S.2 with ${\mathfrak p}_0=0$). Since $k \ge 2$,     $ \Vert  \mathsf{d}_{ \mathfrak p} \Vert   \le  \pi_2 ({\mathfrak p})  \pi_3 ( \Vert   \mathsf{d}  \Vert    ) $  for all $ {\mathfrak p} \ge 0$ and $( t ,   {\mathsf x}  ,   {\mathsf d}) \in {\mathbb R} \times {\mathbb R}^n\times {\mathbb R}^m$,   with $\pi_3 (s) \delequal  s $ (Assumption  S.3). Also, Assumption  S.4 is satisfied since $ ( C(t) , A(t)  ) =  \left (  \begin{pmatrix} 1 & 0\end{pmatrix}  ,   \begin{pmatrix} 0 & 1 \cr 0  &0 \end{pmatrix}   \right ) \delequal ( C,A )
$  is observable, while  Assumption  S.5 trivially follows since the functions  $F$ and $H$ are time-invariant.     Theorem \ref{TH1} applies to  $\Sigma(t)=0$  providing the asymptotic error bound   \eqref{error} with the symmetry-based observer \eqref{observer_1} 
    \begin{eqnarray}
 &&\hskip-.8cm D_t \chi (t)  = e^{ k {\mathfrak p} }     \begin{pmatrix}  \chi_2   (t)     + k_1    Y(t)  \cr   {\rm sat}^k_{   e^{-  {\mathfrak p}  \delta }  }   (    \chi_1   (t)    )   {\rm sat}_{   e^{-  {\mathfrak p}  \delta }  }   (    \chi_2 (t)    )      + k_2   Y(t)  \end{pmatrix}    , \nonumber \\
  &&\hskip-.8cm \widehat{x} (t) =e^{   {\mathfrak p }  } \begin{pmatrix}  
  {\rm sat}_{   e^{-  {\mathfrak p}  \delta }   }   (   \chi_1 (t)    )  \cr e^{   {\mathfrak p }k } {\rm sat}_{   e^{-  {\mathfrak p}  \delta }   }   (   \chi_2 (t)    )    \end{pmatrix} ,  \label{obs1}    
   \end{eqnarray} 
where $Y(t):=  e^{- {\mathfrak p}  } y (t)        -    \chi_1    (t)$,  the gain matrix $K = \begin{pmatrix}     k_1  &  k_2  \end{pmatrix}^\top$ is such that $A - KC$ is Hurwitz and the parameter ${\mathfrak p}>0$ is sufficiently large (Remark \ref{tuning}).   It might be useful to compare our  observer \eqref{obs1}  with  a classical  semi-global HGO (see for instance \cite{KP}, \cite{Bernard_Praly_Andrieu}) 
      \begin{eqnarray}
 &&\hskip-1.4cm D_t \widehat{x} (t)  =  \begin{pmatrix}    \widehat{x}_2 (t)  + \gamma  k_1 Y(t) \cr    {\rm sat}^k_N  (   \widehat{x}_1   (t)    )  {\rm sat}_N   (  \widehat{x}_2 (t)    )  +  \gamma^2    k_2 Y(t)  \end{pmatrix}     ,  \label{due} 
 \end{eqnarray}
 where   $Y(t):= y (t)        -  \widehat{x}_1    (t)$ and   $\gamma>0$ is sufficiently large,  or   a SLO (see for instance \cite{CM_2019}, \cite{Bernard_Praly_Andrieu} with references therein) 
   \begin{eqnarray}
 &&\hskip-1.2cm D_t \widehat{x} (t)  =  \begin{pmatrix}    \widehat{x}_2 (t)  +  \gamma  k_1  \lfloor Y(t)  \rceil^{\frac{1}{2}}  \cr    {\rm sat}^k_N  (    \widehat{x}_1   (t)    ))  {\rm sat}_N   (  \widehat{x}_2 (t)    )   + \gamma^2  k_2  \lfloor Y(t)  \rceil    \end{pmatrix}     , \label{tre} 
 \end{eqnarray}
where $Y(t):= y (t)        -  \widehat{x}_1    (t)$,  $ \lfloor s  \rceil^k  =  {\rm sgn} (s) \vert s \vert^k $  and $\gamma>0$ is sufficiently large.   While with our symmetry-based observer \eqref{obs1}   we are guaranteed from  \eqref{error} with an asymptotic estimation  error bound $ \{  \varepsilon  +  8{\bf \overline{p}}  \sqrt\frac{2{\bf \overline{p}}}{{\bf \underline{p}}}    (1+\Vert K \Vert   )  \}  N $, with the HGO \eqref{due}   we  achieve the arbitrarily small  asymptotic estimation  error bound $    \varepsilon    N $  (see for instance Proposition 2, \cite{Bernard_Praly_Andrieu}) and with the SLO \eqref{tre} we have the best performance with  asymptotic  convergence of $  \widehat{x} (t) $ to $x(t)$  (see for instance Proposition 4, \cite{Bernard_Praly_Andrieu}).  By introducing  relaxed notions of symmetry  in  Section \ref{as_symm}  we will be able to recover the same 
 performances of  semiglobal HGOs when $\sup_{t \ge 0} \Vert (x(t) , d(t) \Vert < +\infty$ (Theorem \ref{TH1_AS}),  not yet the exact error convergence observed in the   SLOs. On the other hand,  global symmetry-based observers may be still designed (Theorem \ref{TH3}) where SLOs or homogeneous  observers do not when  $\sup_{t \ge 0}  \Vert (x(t) , d(t) ) \Vert = +\infty$ (Examples \ref{ex_5} and \ref{ex_6} later on). 
$\hfill \blacktriangleleft$
 
We can state an analogous result to Theorem \ref{TH1} for PSI-contracting symmetries such as \eqref{Psi_0_alt}. To this aim we refer to a system map  $\Sigma $ with $( F ,  H  ) \delequal ( F_0 (t,{\mathsf x} ,{\mathsf d})+F_1 (t,{\mathsf x}_1)   ,  {\mathsf x}_1 )$  
and ${\mathsf x} = ( {\mathsf x}_1 , {\mathsf x}_2 )$, ${\mathsf x}_1 \in {\mathbb R}^p$ and ${\mathsf x}_2 \in {\mathbb R}^{n-p}$, where $H$ and  $F_1$  depend only  on the  points $ t$ and ${\mathsf x}_1 $ which may not be subject to the SI-contracting action of the group. The following Assumptions will replace Assumptions  S.1, S.2 and S.5:

{\textbf{\emph{Assumption PS.1}}:  $ \Psi_{\mathfrak p}$   is a LS  of  $\Sigma (t)  =  0$ with ${\rm Dom} ( \Psi_{\mathfrak p} ) = {\mathcal M}$ for ${\mathfrak p} > 0$, $ \Psi^{\mathsf{x}_1}_{\mathfrak p} \equiv \Psi^{\mathsf{y}}_{\mathfrak p}$,   PSI-contracting with  contracting maps $(\sigma , \mu , \Gamma_{\mathfrak p}^{\mathsf{x}_2} \times  \Gamma_{\mathfrak p}^\mathsf{d}  )$ and UU-transforming. $\hfill \blacktriangleleft$
}

{\textbf{\emph{Assumption PS.2}}:  There exist ${\mathfrak p}_0 > 0$,   $\pi_1 \in {\mathcal K}_+$ and $\pi_2 \in {\mathcal L}$ such that  for all ${\mathfrak p}  \ge {\mathfrak p}_0$,  $( t ,  \mathsf{x}  ) \in  {\rm Dom} ( \Psi^{ t,{\mathsf x}  }_{ \mathfrak p} )  $ and   $\overline{{\mathsf x}}_2 \in {\mathbb R}^n$ for which ${\mathfrak p} \ge \sigma ( \Vert  \mathsf{x}   \Vert )$ and   $  \Vert   \overline{{\mathsf x}}_2  \Vert  \le  (n-p ) \mu ( {\mathfrak p} )$ we have  
$    ({\mathsf t}_{\mathfrak p}  , ( {\mathsf x}_{\mathfrak p} )_1 ,\overline{{\mathsf x}}_2 ) \in {\rm Dom} ( \Psi^{ t,{\mathsf x} }_{-\mathfrak p} ) ) $  and  
$  \Vert \frac{\partial 
\Psi^{\mathsf{x}_2}_{-\mathfrak p}   ({\mathsf t}_{\mathfrak p}  , ( {\mathsf x}_{\mathfrak p} )_1 ,  \overline{{\mathsf x}}_2 ) }{\partial \overline{{\mathsf x}}_2 }   \Vert   \le \frac{ \pi_1  (  \Vert \overline{{\mathsf x}}_2 \Vert ) }{ \pi_2 ({\mathfrak p})   }$. $\hfill \blacktriangleleft $

Let 
\begin{eqnarray}
&&A  (t) \delequal   \frac{\partial  F_0 ( t,  \mathsf{x}  , \mathsf{d})   }{\partial \mathsf{x}  }  \vert_{(\mathsf{x} ,\mathsf{d})=0} , 
 B  (t) \delequal   \frac{\partial  F_0  ( t,  \mathsf{x}  , \mathsf{d})   }{\partial \mathsf{d}  }  \vert_{(\mathsf{x},\mathsf{d})=0}  \nonumber \\
 && ( \Delta \Sigma_L  )  (t , {\mathsf x}   , {\mathsf d}  )  \delequal   F_0 ( t,  \mathsf{x}  , \mathsf{d} ) -   A  (t) \mathsf{x} -  B  (t)   \mathsf{d} .
\end{eqnarray}

{\textbf{\emph{Assumption PS.5}}:  There exist   $\xi   \in {\mathcal K}$ such that $ \left \Vert   \frac{\partial  ( \Delta \Sigma_L  )  (t , {\mathsf x}   , {\mathsf d}  )  }{\partial ( {\mathsf x}_2 , {\mathsf d}   )   }   \right  \Vert \le  \xi  (   \Vert  ( {\mathsf x}_2 , {\mathsf d}  ) \Vert )      $ 
  for all $t \ge 0$ and $ ( {\mathsf x} , {\mathsf d} ) \in {\mathbb R}^n \times {\mathbb R}^m$.  $ \hfill \blacktriangleleft$
}

Define the dynamic output  filter 
   \begin{eqnarray}
 &&\hskip-.8cm  D_t \chi (t)  = \frac{\partial \Psi^t_{\mathfrak p} (t)  }{\partial t}   \Big  (    A   ( \Psi^t_{\mathfrak p} (t)  )    \chi  (t)  +  F_1 (\Psi^t_{\mathfrak p} (t)  , \Psi^{\mathsf y}_{\mathfrak p} (t, y(t))  )     \nonumber   \\
  &&\hskip-.8cm   +  ( \Delta \Sigma_L  ) (\Psi^t_{\mathfrak p} (t)   ,  \Psi^{\mathsf{y}}_{\mathfrak p} (t, y(t))   ,   \chi_2^{\rm sat} (t) , 0 )    \label{observer_1.1}   \\
  &&\hskip-.8cm  +     K  (\Psi^t_{\mathfrak p} (t)  )   ( \Psi^{\mathsf y}_{\mathfrak p} (t, y(t))       - \chi_1 (t))    \Big  ) \nonumber 
   \end{eqnarray}
with  $\chi = ( \chi_1 ,  \chi_2)$, $\chi_1 \in {\mathbb R}^p$,  $\chi_2 \in {\mathbb R}^{n-p}$,  and   $\chi_2^{\rm sat} (t)  \delequal  
 \Gamma^{\mathsf{x}_2}_{-\mathfrak p}    {\rm sat}_{\mu ( {\mathfrak p}  )}   (  \Gamma^{\mathsf{x}_2}_{\mathfrak p}  (  \chi_2 (t)  ) )$.   
      \begin{theorem}\label{TH1.1}
 Under Assumptions PS.1, PS.2, S.3, S.4,  and PS.5 and  for each $ N > 0$  consider the solutions   $z(t)$ of $\Sigma (t) = 0 $ for which $\sup_{t \ge 0} \Vert    (   x (t)  , d (t)  )  \Vert  \le N$ together with the filter   \eqref{observer_1.1}, its design parameter ${\mathfrak p}>0$ and its output   $\widehat{x} (t)  \delequal ( y(t) ,   \Psi_{ -{\mathfrak p} }^{ \mathsf{x}_ 2 }  ( \Psi^t_{\mathfrak p} (t)  , \Psi^{\mathsf{y}}_{\mathfrak p} (t, y(t))  
 , \chi_2^{\rm sat} (t) )$. For each $\varepsilon > 0$ there exists  $ {\mathfrak p} >0$   such that
  \begin{eqnarray}
 && \hskip-.7cm  \limsup_{t \rightarrow +\infty}  \Vert x (t) - \widehat{x}  (t)   \Vert   \le \{ \varepsilon +  8{\bf \overline{p}}  \sqrt\frac{2{\bf \overline{p}}}{{\bf \underline{p}}}     (1+ \sup_{ t \ge 0 }   \Vert K ( t ) \Vert  ) \times \nonumber \\
  && \hskip-.7cm \times (  \sup_{ t \ge 0  }   \Vert B (  t ) \Vert   \pi_1 (0)  \}  \pi_3 (\sup_{t \ge 0} \Vert d(t) \Vert  )  .  \label{error.1} \hfill \blacktriangleleft 
   \end{eqnarray}  
 \end{theorem} 
   The proof of Theorem \ref{TH1.1} is similar to the proof of Theorem \ref{TH1} and it is omitted. 
   
\textbf{\emph{Example}} \emph{\ref{ex_0}} \textbf{\emph{(cont'ed)}}.  
Consider  the system map $\Sigma $ associated to  \eqref{sys_0b}. As already seen,    its global symmetry  \eqref{Psi_0_alt}  is  PSI-contracting with contracting maps \eqref{contr_part}  for  any $\delta \in (0,1)$ and $\varphi \in {\mathcal K}_\infty$ such that $\varphi (s) \ge s$ for all $s \ge 0$. It is also  UU-transforming for   ${\mathfrak p} > 0$ and $ \Psi^{\mathsf{x}_1}_{\mathfrak p} \equiv \Psi^{\mathsf{y}}_{\mathfrak p}$ (Assumption  PS.1). Furthermore,  $\Vert  \mathsf{d}_{ \mathfrak p}   \Vert \le  \pi_2 ({\mathfrak p})  \pi_3 ( \Vert   \mathsf{d}  \Vert    )$  for all ${\mathfrak p} > 0$ and $(t,  {\mathsf x},{\mathsf d})  \in {\mathbb R} \times {\mathbb R}^2 \times {\mathbb R} $,   with  $\pi_3 (s) \delequal s \in {\mathcal K}$ and  
 $\pi_2 (s) \delequal e^{-s} \in {\mathcal L}$,  and  $ \Vert \frac{\partial}{\partial  \overline{\mathsf{x}}_2}  \Psi^{\mathsf{x}_2}_{- \mathfrak p} ({\mathsf t}_{\mathfrak p}  , ( {\mathsf x}_{\mathfrak p} )_1  ,  \overline{\mathsf{x}}_2 )   \Vert  =  e^{\mathfrak p} \le    \frac{  \pi_1 ( \Vert   \overline{\mathsf{x}}_2  \Vert    ) }{ \pi_2 ({\mathfrak p}) }  $ for all ${  \mathfrak p} \ge 0 $ and $( t,{\mathsf x} ) \in {\mathbb R} \times {\mathbb R}^2$ and $\overline{\mathsf{x}}_2 \in {\mathbb R} $,  
  with  $\pi_1 (s) \delequal 1+s \in {\mathcal K}_+$ (Assumptions PS.2 and  S.3). Moreover, $(C(t),A(t)) = \left (  \begin{pmatrix} 1 & 0\end{pmatrix} ,    \begin{pmatrix} 0 & 1 \cr 0  &0 \end{pmatrix}  \right ) \delequal (C,A)$  is observable (Assumption S.4)  and the functions $F $ and $H$ are time-invariant (Assumption  S.5). Our  symmetry-based observer is 
\begin{eqnarray}
&&D_t \chi (t)  = e^{  {\mathfrak p} }      \begin{pmatrix}  \chi_2   (t)  +  k_1 Y(t)   \cr {\rm sat}^2_{  e^{-{\mathfrak p} (1-\delta) }  } (\chi_2 (t))  +  k_2 Y(t)  \end{pmatrix} ,  \nonumber \\
&&   \widehat{x} (t) = \begin{pmatrix}  
y (t)  \cr  e^{ {\mathfrak p} }  {\rm sat}_{  e^{-{\mathfrak p} (1-\delta) }  } (\chi_2 (t))   \nonumber\end{pmatrix}  , 
\end{eqnarray}
  with  $Y(t):=y(t)  -    \chi_1    (t)$,  the gain matrix $K:= (k_1, k_2)^\top$ such that  $A-KC 
 $ is Hurwitz and ${\mathfrak p}>0$ sufficiently large. A different symmetry-based observer can be designed  for the same   system  $\Sigma(t) = 0$ by using the  LS \eqref{Psi_0} which,   on account of \eqref{ex5.3.2},     is PSI-contracting as well.  

\subsection{Global symmetry-based observers}\label{S3.2}  
 
 By considering only the solutions $z(t) $ of $ \Sigma (t) = 0 $ for which  $\sup_t \Vert  ( x(t) ,  d(t) )  \Vert \le N < +\infty$, we restrict as well the set of   initial conditions $x(0) $ for which the observer   convergence to $x(t)$  is guaranteed.  A powerful option we have for overcoming this problem and let the initial  condition $x(0)$  be any   point in ${\mathbb R}^n$ is to design a \emph{time-varying} ${\mathfrak p} = p (t)$ estimating the norm $\Vert   ( x(t) , d(t) )  \Vert  $ in the sense of \eqref{widesigma}. These type of estimates  are computed by suitable 
 dynamic output filters ({\sl  norm estimators}) originally introduced in  \cite{SS} and, more recently,   used for global  observer design in  \cite{Battilotti2022} ({\sl state-norm estimators}). 
 
 \subsubsection{SI-norm estimators}\label{S3.2.1}
 
For the design of a norm-estimator of $  ( x(t) , d(t) ) $ (called \emph{SI-norm estimator}) we introduce the following assumption.
  
\noindent
\noindent{\textbf{\emph{Assumption SINE}}:  There exist  $\alpha \in {\mathcal K}$, $ \beta_1 \in {\mathcal K}_\infty$, $\eta , \Phi \in {\mathcal K}_+$, $\beta_0 \ge 0$ and a $C^1$ function $ V : {\mathbb R}_+^0  \times {\mathbb R}^n \to {\mathbb R}_+^0 $  such that   for  all $t \ge  0$  
\begin{eqnarray}
 &&\hskip-1.4cm  D_t V (t,x(t)) \vert_{ \Sigma (t)  = 0  } \!  \le \!  -\alpha  ( V  (t,x(t)) ) \! + \! \Phi (   \Vert ( y  (t)  , d (t)  )       \Vert    )     \label{SN1}  \\
&&\hskip-1.4cm   \Vert   x (t) \Vert    \le    \beta_1 (  V (t,x(t)) ) + \beta_0    \label{SN2}   \\
&&\hskip-1.4cm \Phi (   \Vert ( y  (t)  , d (t)  )       \Vert    )  \vert_{ \Sigma (t)  = 0  }   \le 
 \eta (  \Vert ( x  (t)  , d (t)  )       \Vert   )   .  \label{SN3}  \hfill \blacktriangleleft
 \end{eqnarray}   
Assumption SINE takes inspiration from the corresponding Assumption SNE introduced in  \cite{Battilotti2022} for state-norm estimator design but it marks a significant advancement with $\alpha \in {\mathcal K}$ substituting $\alpha(s)= s$ in \cite{Battilotti2022}. 

\textbf{\emph{Example}} \emph{\ref{ex_3}} \textbf{\emph{(cont'ed)}}. 
  Consider  the system map $\Sigma $ associated to   \eqref{triang} where, in addition,    
  \begin{eqnarray}
&&\hskip-1.2cm A_2 ( t, \mathsf{x} ) \delequal \sum_{j=1}^2  A^\ast_{2,j} (t,  \mathsf{x}  ) \mathsf{x}_j   , \label{F0} \\
&&\hskip-1.2cm  B_1  ( t,   \mathsf{x}_1) - B_1 ( 0 ,  0) \delequal  B^\ast_1 ( t, \mathsf{x}_1  ) \mathsf{x}_1 , \label{F1} \\
&&\hskip-1.2cm    \sum_{j=1}^2 \Vert  A^\ast_{2,j}  ( \mathsf{x})    \Vert + \Vert  B^\ast_1 ( t, \mathsf{x}_1  )   \Vert  + \Vert  B_2   ( t,   \mathsf{x} ) \Vert \le  \tau (  \Vert  \mathsf{x}_1 \Vert  ) \label{F2}  
  \end{eqnarray}
for  $\tau   \in {\mathcal K}_+   $ and for all  $t \ge 0$ and $\mathsf{x}  \in  {\mathbb R}^n $.  We see how to meet Assumption SINE. Using \eqref{F0}-\eqref{F1}, the system  $\Sigma  (t) = 0$    can be written as
\begin{eqnarray} 
&&  \hskip-1cm  D_t x (t) =    \begin{pmatrix}  {\textbf 0}_{n \times n} \cr B_2 ( t ,  x  (t) ) \end{pmatrix} d(t)     + K_L  y  (t)  \label{sknew}  \\
&&  \hskip-1cm  +  \left (A_L  -K_L C_L  +   \begin{pmatrix}  B^\ast_1 (t , x_1 (t) )  & {\textbf 0}_{n \times n} \cr A^\ast_{2,1}  (t ,  x (t)  ) &  A^\ast_{2,2}  (t,  x (t)  )  \end{pmatrix} \right )  x   (t) \nonumber
\end{eqnarray}
where  $A_L \delequal \begin{pmatrix}  {\textbf 0}_{n \times n} & B_1 (0,0)  \cr {\textbf 0}_{n \times n} & {\textbf 0}_{n \times n}   \end{pmatrix}  $,  $C_L  \delequal \begin{pmatrix} {\textbf I}_n  & {\textbf 0}_{n \times n} \end{pmatrix} $ and  $K_L \in {\mathbb R}^{2n\times n}$. But $(C_L,A_L)$ is observable since $B_1(0,0) \in {\mathbb {GL}} (n)$, hence there exist $K_L \in {\mathbb R}^{2n \times n}$  and  $Q  \in {\mathbb S}_+ (2n)$  such that   
  \begin{eqnarray}
 &&\hskip-1cm   ( A_L  -K  C_L )^\top Q  + Q ( A_L  -K C_L )  + {\textbf  I}_{2n} \in {\mathbb S}_+ (2n)   \label{cond_matrix}  
  \end{eqnarray}
with $ {\bf \underline{q}} {\textbf I}_n \le Q  \le {\bf \overline{q}} {\textbf I}_n $  for some  ${\bf \underline{q}},  {\bf \overline{q}}  > 0$.  Using \eqref{F2}   with  $W ( \mathsf{x} ) \delequal \mathsf{x}^\top Q \mathsf{x}$,   we also  find $\ell_1 ,\ell_2   > 0$ such that 
\begin{eqnarray} 
&&  \hskip-.9cm D_t W ( x(t)) \vert_{\Sigma(t)=0} \le  - \ell_1   W ( x (t)  ) \nonumber \\
&&  \hskip-.9cm+ \ell_2 (  1 +   \tau ( \Vert   y(t)  \Vert ) ) (  W ( x (t)  )  +  \Vert d(t) \Vert^2    +    \Vert y (t) \Vert^2  )  \label{lyap-dissip}  
\end{eqnarray}
 for $t \ge 0$. Hence,  if  
 \begin{eqnarray}
&& V  (\mathsf{x})  \delequal \ln ( 1  +  W  (\mathsf{x}) ) ,  \label{vv} \\
 &&\alpha (s) \delequal \frac{\ell_1  s}{1+s} ,  \Phi (s) \delequal \ell_2 (1+\tau ( s ) )( 1 +   s^2  ) ,
  \label{rr}
 \end{eqnarray}
 we get \footnote{Here, we used the inequality $  \frac{e^{ s }-1}{e^{s} } \ge    \frac{s}{1+s}$ for all $s \ge 0$.} 
 $ D_t V  ( x (t)  ) \vert_{\Sigma(t)=0}  \le  - \alpha   ( V  ( x (t)  ) +  \Phi (  \Vert ( y(t) , d(t) ) \Vert )  $ for all $t \ge 0$.  Furthermore, since $V (\mathsf{x})  \ge \ln ( 1  +  {\bf \underline{q}} \Vert \mathsf{x} \Vert^2  ) $ for all $\mathsf{x} \in {\mathbb R}^n$ then $ \Vert  x (t)  \Vert  \le  \beta_1   (V  ( x (t)  ) )$ for all $t \ge 0$ with 
\begin{eqnarray} 
&&  \beta_1 (s) \delequal  \frac{ e^s  -1}{{\bf \underline{q}}}   .\label{rrr}
 \end{eqnarray}
 Finally, $ \Phi ( \Vert (  y(t) , d(t) ) \Vert )  \vert_{ \Sigma (t)  = 0  }  \le  \eta  (  \Vert   ( x (t)  , d(t) ) \Vert  )  
 $ for all $t \ge 0$ 
\begin{eqnarray} 
&&\hskip-1.4cm  \eta (s) \delequal   \ell_2 (1+\tau(0)) ( (1+\tau ( s )) ( 1 + s^2  ) -  \tau(0)  )  \in {\mathcal K}_+ .  \label{rrrr} 
\end{eqnarray}
 The functions $V  $ in \eqref{vv},  $\alpha \in {\mathcal K}$ and  $\Phi \in {\mathcal K}_+$ in \eqref{rr}, $\beta_1  \in {\mathcal K}_\infty$ in \eqref{rrr} and  $ \eta    \in {\mathcal K}_+$ in \eqref{rrrr} satisfy Assumption SINE (with $\beta_0=0$). 
%

\textbf{\emph{Example}} \emph{\ref{ex_0}} \textbf{\emph{(cont'ed)}}.  
Consider the system map $\Sigma $ associated to   \eqref{sys_0b}. Going through similar calculations from  \eqref{sknew} to  \eqref{lyap-dissip},  with $V  (\mathsf{x})  \delequal \ln ( 1  +  W  (\mathsf{x}) )$, $W  (\mathsf{x}) \delequal \begin{pmatrix} \mathsf{x}_1  &  e^{-\mathsf{x}_1 } \mathsf{x}_2 \end{pmatrix}  Q  \begin{pmatrix} \mathsf{x}_1 & e^{-\mathsf{x}_1 } \mathsf{x}_2 \end{pmatrix}^\top$ and $Q \in {\mathbb S}_+ (2)$ such that $ {\bf \underline{q}} {\textbf I}_2 \le Q  \le {\bf \overline{q}} {\textbf I}_2 $  for some  ${\bf \underline{q}},  {\bf \overline{q}}  > 0$, we find $D_t V  ( x (t)  ) \vert_{\Sigma (t) = 0} \le  - \alpha   ( V  ( x (t)  ) +  \Phi (  \Vert ( y(t) , d(t) ) \Vert ) $ and $\Vert  x (t)  \Vert  \le  \beta_1   (V  ( x (t)  )$  with
\begin{eqnarray}
&&\alpha(s) \delequal \frac{\ell_1 s}{1+s} , \Phi (s) \delequal \ell_2 (1+e^{s+1} )( 1 +   s^2  )   , \label{n1} \\
&& \beta_1 (s) \delequal  \frac{ e^s  -1}{ {\bf \underline{q}} }  ( 1+  e^\frac{ e^s  -1}{ {\bf \underline{q}} } ) , \ell_1, \ell_2 > 0  . \hfill \blacktriangleleft \label{n2}
\end{eqnarray} 
In the remaining part of this section, we provide the necessary details for designing SI-norm estimators under Assumption SINE according to a  procedure taking inspiration from  Proposition  4.1 of \cite{Battilotti2022} but modified so as to take into account the relaxed condition $\alpha \in {\mathcal K}$. Since  we restrict  our attention to   bounded inputs $d(t)$, we assume   that $\sup_{t \ge 0} \Vert d(t) \Vert \le N < +\infty$. Using the fact that $\alpha \in {\mathcal K}$, let  $\omega_1 > 0$,  $\lambda_1 \in ( 0  ,   1)$ and  $\lambda_0  \in (0,  \alpha  ( \omega_1 ) )$ be such that  
\begin{eqnarray}
 \alpha  ( s+\omega_1 )   \ge  \lambda_1  \alpha  ( s ) + \lambda_0   \label{alpha^Lambda}
 \end{eqnarray}
  for  all $s \ge  0$. Define the filter
\begin{eqnarray}
 &&\hskip-1cm D_t \widehat{V} (t)  =  -\lambda_1 \alpha (  \widehat{V}  (t)  ) +  \Phi (   \Vert y  (t) \Vert + N)   ,  \; \widehat{V} (0) \ge  0 ,  \label{secondo}
 \end{eqnarray}
and let $L  (V,\widehat{V}) \delequal   {\max}^2 \{ V   -\widehat{V} -\omega_1    , 0 \}$.  By \eqref{SN1} and using \eqref{alpha^Lambda},   we have $D_t L (V,\widehat{V}) \vert_{\Sigma (t) = 0}   \le   - 2\lambda_0 \sqrt{L (V,\widehat{V})}$  (for almost all $t \ge 0$) so that 
 \begin{eqnarray} 
 V (t, x(t) )  \le   \widehat{V} (t) +\omega_1 ,    \forall t \ge T_{x (0) }  :=\frac{V (0, x(0))  +  \omega_1   }{\lambda_0} . \label{pt}
 \end{eqnarray}
Also,  on account of \eqref{SN2},  $  \Vert   x (t)  \Vert   \le    \beta_1 ( \widehat{V} (t) +  \omega_1  ) + \beta_0$ for all $t \ge T_{x (0)}$, hence  
${\mathfrak p} \delequal  p(t) \delequal  \sigma (  \omega_0 + N +\beta    ( \widehat{V} (t) +  \omega_1  )   )  $  satisfies \eqref{widesigma} for all  $ \omega_0  \ge \beta_0$ and  $t \ge T_{x (0) }$ and provides  a SI-norm  estimator  for $t \ge T_{x (0) }$ as long as $\sup_{t \ge 0} \Vert d(t) \Vert \le N $. For avoiding  largely oscillatory  behaviours of $\widehat{V} (t)$, in place of the filter \eqref{secondo} we may use as well non-negative filters 
\[
D_t \widehat{V} (t)  = \max \{ -\lambda_1 \alpha (  \widehat{V}  (t)  ) +  \Phi (   \Vert y  (t) \Vert + N)    , 0\} .
\]
  
    \subsubsection{Global symmetry-based observers with infinite-time convergence}\label{S3.2.2}
        
 Global observers are the result of a suitable combination of semiglobal observers with SI-norm estimators.  Below, we sum up the main assumptions. First of all, we  slightly  reinforce  Assumptions S.1 and S.4 as follows.  
     
{\textbf{\emph{Assumption GS.1}}:  $ \Psi_{\mathfrak p}$   is a LS  of  $\Sigma (t)  =  0$ with ${\rm Dom} ( \Psi_{\mathfrak p} ) = {\mathcal M}$ for ${\mathfrak p} > 0$, it is SI-contracting with  contracting maps $(\sigma , \mu , \Gamma_{\mathfrak p}^\mathsf{x} \times  \Gamma_{\mathfrak p}^\mathsf{d}  )$ and UU-transforming. Moreover,  the infinitesimal generator of $\Psi^{t, {\mathsf x}}_{\mathfrak p}$  is   $  {\textbf g}^{t, {\mathsf x}} \vert_{ ( t, {\mathsf x} ) }  =   g^t  (t )  \frac{\partial}{\partial t} + \sum_{j=1}^n  (  g^{\mathsf{x}}  (t , {\mathsf x})   )_j \frac{\partial}{\partial {\mathsf x}_j} 
$  with 
\begin{eqnarray}
  g^t (t)   \le h t  \label{gtgx0}
  \end{eqnarray}
     for all $  t \ge 0$ and for some $h \ge  0$,  $  \sup_{t \ge 0} \left \Vert   \frac{\partial  g^{\mathsf{x}} (t ,{\mathsf x})}{\partial   {\mathsf x}  }\Big \vert_{{\mathsf x}=0} \right \Vert  < +\infty $ and for some $\kappa \in {\mathcal K}$ and  for all  $  {\mathsf x} \in {\mathbb R}^n $
     \begin{eqnarray}  
  && \sup_{t \ge 0}   \left \Vert    \frac{\partial  g^{\mathsf{x}} (t , {\mathsf x})}{\partial   {\mathsf x}  }  -\frac{\partial  g^{\mathsf{x}} (t ,{\mathsf x})}{\partial   {\mathsf x}  }\Big \vert_{{\mathsf x}=0} \right \Vert   \le \kappa  (\Vert {\mathsf x}\Vert )  . \label{gtgx2} \hfill \blacktriangleleft 
\end{eqnarray} 
Let $A(t), B(t), C(t)$ and $D(t)$ be as in \eqref{abcd} with the following slightly stronger detectability assumption.

{\textbf{\emph{Assumption GS.4}}:  $ \sup_{t \ge 0}  \Vert B (t) \Vert 
 < +\infty$, $ \sup_{t \ge 0} \Vert D (t) \Vert 
 < +\infty$  and there exist $K   : {\mathbb R}_+^0  \to  {\mathbb R}^{n \times p} $,  $P  \in  {\mathbb S}_+ (n)$ and  $ {\bf \underline{p}}  ,  {\bf \overline{p}}    > 0$ such that  $ \sup_{t \ge 0}  \Vert K (t) \Vert  
 < +\infty$ and  $  P  (A (t )   -K  (t)  C   (t))    + (A  (t )   -K  (t)  C   (t))^\top (t)  P   + 2{\textbf I}_n   \in {\mathbb  S}_- (n)   $  
  for all $t \ge 0$ with  
${\bf {\bf \underline{p}}}  {\textbf I}_n \le P  \le {\bf {\bf \overline{p}}} {\textbf I}_n $.$\hfill \blacktriangleleft$ 
  
 When ${\mathfrak p}=p(t) $ is time-varying and $ \Psi_{\mathfrak p}$ is UU-transforming for each ${\mathfrak p} > 0$,  we have to introduce some additional conditions for guaranteeing that also $\theta (t) \delequal \Psi^t_{\mathfrak p} (t) \vert_{{\mathfrak p} = p(t)} $ is an unbounded time scale. A  clarifying example is $ \Psi^t_{\mathfrak p} (t) = e^{- \mathfrak p} t$ with $p(t) \ge  \ln (1+t^k) $ for $t \ge 0$ and some $k \ge 1$,  which implies  $\theta (t) \le \frac{t}{1+t^k} \le 1$ for $t \ge 0$. Hence, although   $\lim_{t \to +\infty}  \Psi^t_{\mathfrak p}  (t) = +\infty$ and $\Psi_{\mathfrak p}^t  $ is strictly increasing on ${\mathbb R}^0_+$  for each   ${\mathfrak p} > 0$,  however  $\lim_{t \to +\infty}  \theta (t)  \le 1$ and  if $k > 1$ then $\theta $ is not even strictly  increasing on ${\mathbb R}^0_+$. 
 
Let  $\pi_2 \in {\mathcal L}$ be  as in Assumption S.2  and  $ \beta  \in {\mathcal K}_\infty $ and  $  \eta  \in {\mathcal K}_+$ as in Assumption SINE. For $ t, \omega \ge 0$, $N > 0$ and $s \ge N$ define 
 \begin{eqnarray}
 &&\hskip-.5cm {\mathcal V} ( t,s)   \delequal    \frac{ 1+ \vert \min \{0 , g^t (\Psi^t_s (t) )   \} \vert + \vert \frac{d \ln \pi_2  (s) }{ds}   \vert    }{   \frac{\partial \Psi^t_s (t)}{\partial t} }    ,  \nonumber \\
 &&\hskip-.5cm  {\mathcal E}  (s,\omega) \delequal  \frac{d\sigma   (r) }{dr} \Big \vert_{r=s+\omega}  \frac{d\beta  (r) }{dr} \Big \vert_{r=\beta^{-1} ( s-N) }    \eta (s) . \label{VE}
 \end{eqnarray}
 
{\textbf{\emph{Assumption GS.6}}: For each $N > 0$ there exists $k_0 \in {\mathcal L}$, $k_1 \in {\mathcal L}$ and $k_2 > 0$ such that  for all $\omega  \ge 0$
 \begin{eqnarray}
 &&\hskip-.5cm     \sup_{ t  \ge  0 , s \ge N  }{\mathcal E}  (s,\omega )  {\mathcal V} ( t,\sigma (\omega  +s))  \le  k_0 ( \omega)        \label{I_glob}  \\
 &&\hskip-1.5cm  \inf_{t \ge 0  , s \ge  N} \left \{ \frac{\partial \Psi^t_r (t)}{\partial t} \Big \vert_{r= \sigma (\omega  +s)}( 1 + k_2 t )  \right \}  \ge k_1 (\omega ) . \label{II_glob} \hfill \blacktriangleleft
\end{eqnarray}  
}
\begin{example}\label{por}
Consider  the system map $\Sigma $ associated to   \eqref{triang_ex} and  its global symmetry  \eqref{triang_symm}  with  infinitesimal generator \eqref{triang_gen}. As already seen, \eqref{triang_symm}  is SI-contracting with contracting maps \eqref{ex5.3.2.b0} and UU-transforming for   ${\mathfrak p}>0$. Moreover,  \eqref{triang_ex}  satisfies \eqref{F0}-\eqref{F2} with $\tau (s) \delequal 2(1+  s^k)$, $k \ge 2$.   We show how to meet Assumption GS.6. Collect $\sigma \in {\mathcal K}_\infty$ from  \eqref{ex5.3.2.b0} (with any $\varphi \in {\mathcal K}_\infty$ such that $\varphi(s) \ge s$ for all $s \ge 0$),  $\pi_2 \in {\mathcal L}$ from \eqref{pi2.1}, $Q\in {\mathbb S}_+$ with its upper and lower bounds ${\bf \overline{q}} , {\bf \underline{q}} > 0$ from \eqref{cond_matrix},  $\beta \in {\mathcal K}_\infty$ from \eqref{rrr} and $\eta \in  {\mathcal K}_+$ from \eqref{rrrr}. 
Pick $N > 0$.  We  choose  $  \varphi (s) \delequal s (1+\tau ( s )) ( 1 + s   )  + \tau(s)- \tau(0)   $ and  $\delta \in (0,1)$   such that $\frac{1}{1-\delta }  \le   ( 1+ N)^{\frac{1}{1-\delta }}$ and 
 \begin{eqnarray} 
  &&\hskip-.8cm    \frac{d\varphi (s) }{ds}     <  \frac{  {\bf \underline{q}}  ( 1+\varphi (s))^{\frac{k-1}{1-\delta }}   }{  \lambda_2   (1+\tau(0))   ( 1 + {\bf \underline{q}} )(1+   4(1+k ))      }    \frac{1}{1+s}    \label{phisup}
\end{eqnarray}
for all $s \ge N$. From \eqref{triang_gen}-\eqref{triang_symm}  we recover   $ \Psi_{\mathfrak p}^t (t)  = e^{ {\mathfrak p}k } t$ and its infinitesimal generator ${\textbf g}^t  \vert_{t}   =g^t(t)  \frac{\partial}{\partial t}  $ with $g^t(t) = kt$. Hence, from definition \eqref{VE}
and since   $\eta (s) \le  \lambda_2 (1+\tau(0)) ( 1 + \varphi(s ))$ and $ \frac{1}{1-\delta}  \le  (1+s)^ \frac{1}{1-\delta}  $  for all $s \ge N$, 
      \begin{eqnarray}
 &&\hskip-1cm     {\mathcal E}  (s,\omega )  {\mathcal V} ( t,\sigma (\omega  +s)) \le   \lambda_2  (1+\tau(0))  \frac{   (1+   4(1+k))  }{  {\bf \underline{q}} } \times \nonumber \\
  &&\hskip-1cm   \times  \frac{ 1 + {\bf \underline{q}} }{ ( 1+\varphi (s+\omega ))^{ \frac{k  -1}{1-\delta}  }}    \frac{d\varphi}{dr} \vert_{r=s+\omega}  \nonumber   
\end{eqnarray}
which, on account of \eqref{phisup}, implies  \eqref{I_glob}  of  Assumption GS.6  with $k_0 (s) = \frac{1}{1+s} \in {\mathcal L}$. Furthermore, for all $\omega \ge 0$
      \begin{eqnarray}
  &&\hskip-1cm      \inf_{t \ge 0 , s \ge N} \left \{ \frac{\partial \Psi^t_r (t)}{\partial t} \Big \vert_{r= \sigma (\omega +s)} ( 1 +  t ) \right \}    \ge  e^{ -\omega } \delequal k_ 1 ( \omega)  \in {\mathcal L} \nonumber 
  \end{eqnarray}
which implies  \eqref{II_glob} of  Assumption GS.6 with $k_2=1$. $\hfill \blacktriangleleft$
\end{example}
Consider the  dynamic output filter   
     \begin{eqnarray}
 &&\hskip-1.2cm\begin{pmatrix}  D_t \chi (t) \cr \zeta(t) \end{pmatrix}  = \begin{pmatrix} \frac{\partial \Psi^t_{\mathfrak p} (t) }{\partial t} {\textbf I}_{n \times n} & {\textbf 0}_{n \times p} \cr {\textbf 0}_{p \times n} & {\textbf I}_{p \times p} \end{pmatrix}  \times \nonumber \\
  &&\hskip-1.2cm \times \Bigg (  \begin{pmatrix} A ( \Psi^t_{\mathfrak p} (t) )  +     \frac{\partial  g^{\mathsf x} (\Psi^t_{\mathfrak p} (t) , \mathsf{x}  )  }{\partial \mathsf{x}} \Big \vert_{\mathsf{x}=0}  (  \frac{\partial \Psi^t_{\mathfrak p} (t) }{\partial t}  )^{-1}  D_t  p (t)       \cr  C  (\Psi^t_{\mathfrak p} (t))    \end{pmatrix}  \chi(t)   \nonumber \\
  &&\hskip-1.2cm  + (\Delta \Sigma_L)   (\Psi^t_{\mathfrak p} (t) ,  \chi^{\rm sat} (t)  , 0 )   \nonumber \\
  &&\hskip-1.2cm  + \begin{pmatrix} g^{\mathsf x} (\Psi^t_{\mathfrak p} (t)   , \chi^{\rm sat} (t)    )   -  \frac{\partial  g^{\mathsf x}  ( t , \mathsf{x} )  }{\partial \mathsf{x}} \Big \vert_{\mathsf{x}=0} \chi^{\rm sat} (t)   \cr {\textbf 0}_{p \times 1} \end{pmatrix}   \frac{D_t  p (t)  }{ \frac{\partial \Psi^t_{\mathfrak p} (t) }{\partial t}  } \nonumber \\
  &&\hskip-1.2cm  +   \begin{pmatrix}        K (\Psi^t_{\mathfrak p} (t)) ( \Psi^{\mathsf y}_{\mathfrak p} (t,y(t))   - \zeta (t)   ) \cr   {\textbf 0}_{p \times 1}   \end{pmatrix}   \Bigg )  , \label{observer_4.-2}  \\
 &&\hskip-1.2cm {\mathfrak p} = p(t)  \delequal   \sigma (  \omega_0 + N +\beta    ( \widehat{V} (t) +  \omega_1  )   ) ,\label{observer_4.-1} \\
  &&\hskip-1.2cm D_t \widehat{V} (t)  = \max \{  -\lambda_1 \alpha (  \widehat{V}  (t)  ) +   \Phi (  t, \Vert y  (t) \Vert + N)    , 0\} , 
\label{observer_4} \end{eqnarray} 
with   $\Delta \Sigma_L$ as in \eqref{Delta_2}  and  $\chi^{\rm sat} (t) = \Gamma_{- \mathfrak p}^{\mathsf x} ( {\rm sat}_{\mu ({\mathfrak p} ) } ( \Gamma_{ \mathfrak p}^{\mathsf x} ( \chi (t) ) ) $. The proof of the following result is reported in  the Appendix. 
 \begin{theorem}\label{TH3}
 Under Assumptions SINE, GS.1, S.2, S.3, GS.4, S.5 and GS.6  consider the solutions $z(t)$ of $\Sigma (t) = 0 $  for which $\sup_{t \ge 0} \Vert     d(t)    \Vert \le N$ together with the filter \eqref{observer_4.-2}-\eqref{observer_4}, its design parameters $\lambda_1 ,  \omega_0, \omega_1   >0$ and its output $\widehat{x} (t) \delequal   \Psi_{-{\mathfrak p}}^{\mathsf{x}}  (\Psi_{{\mathfrak p}}^t (t) , \chi^{\rm sat} (t)   )$. For each $\varepsilon > 0$ there exist  $\lambda_1 ,  \omega_0, \omega_1   >0$ such that \eqref{error}  holds true. $\hfill \blacktriangleleft$
 \end{theorem}
\begin{remark}
The design parameters  $\lambda_1 ,  \omega_0, \omega_1   >0$ claimed in Theorem \ref{TH3} are tuned as follows:  the parameters $\lambda_1 ,   \omega_1 >0$ are numbers satisfying  \eqref{alpha^Lambda},  while the parameter $ \omega_0 \ge \beta_0$ ($\beta_0 \ge 0$ from Assumption SINE)  is tuned   as the corresponding parameter $\omega > 0$ in the proof of Theorem \ref{TH1}, taking into account the additional conditions \eqref{I_glob}-\eqref{II_glob}.
\end{remark}
 \begin{example}\label{ex_5}
 {\sl (Global symmetry-based observers versus others)}. Consider the system map  $\Sigma $ associated to   
    \begin{eqnarray} 
 (F  , H )  \delequal  \left ( \begin{pmatrix} A_1  ( \mathsf{x}_1) +  \mathsf{x}_2      \cr    A_2 ( \mathsf{x})  +  \mathsf{d}    \end{pmatrix}   ,  \mathsf{x}_1  \right )  ,\label{triang2}
      \end{eqnarray}
 $\mathsf{x} \delequal (\mathsf{x}_1, \mathsf{x}_2 )$ and $\mathsf{x}_1 , \mathsf{x}_2  , \mathsf{d}  \in {\mathbb R} $, and 
    \begin{eqnarray} 
&&\hskip-1.5cm  A_1  ( \mathsf{x}_1 ) =  a_1 \mathsf{x}_1^{\frac{\gamma + r_1}{r_1}}  ,   A_2  ( \mathsf{x} ) = a_{2,1}  \mathsf{x}_1^{\frac{\gamma + r_2}{r_1}} + a_{2,2}  \mathsf{x}_2^{\frac{\gamma + r_2}{r_2}}     \label{triang4}
      \end{eqnarray}
  for some $a_1, a_{2,1}, a_{2,2} \in {\mathbb R}$, $ r_1  > 0$,  $\gamma \ge  0$ with  $r_2 \delequal r_1 +\gamma$. Hence, $F$ and $H$ are  homogeneous with weights $(r_1, r_2)$ and degrees $\gamma$ and $r_1$, respectively. It is possible to design a global homogeneous observer for $\Sigma (t) = 0$ (see for instance \cite{Andrieu_Praly_Astolfi_2008})   if  
    \begin{eqnarray} 
 &&\hskip-1.5cm   \vert  A_1  ( \mathsf{x}_1^a) -  A_1  ( \mathsf{x}_1^b) \vert \le c_1    \vert \mathsf{x}_1^a - \mathsf{x}_1^b \vert^{\frac{\gamma + r_1}{r_1}} ,  \nonumber \\
&&\hskip-1.5cm  \vert  A_2  ( \mathsf{x}^a) -  A_2  ( \mathsf{x}^b) \vert \le c_2 (   \vert \mathsf{x}_1^a - \mathsf{x}_1^b \vert^{\frac{\gamma + r_2}{r_1}} +  \vert \mathsf{x}_2^a - \mathsf{x}_2^b \vert^{\frac{\gamma + r_2}{r_2}}  )      \label{triang3}
      \end{eqnarray}
      for all $  \mathsf{x}^a , \mathsf{x}^b \in {\mathbb R}^2$  and for some $c_1, c_2   > 0$, which  holds only if  $\gamma = 0$ (the case $\gamma < 0$ may be considered but this would imply $F \notin C^1$). It is not even 
 possible to design a sliding-mode  observer for $\Sigma (t) = 0$ if $\sup_{t \ge 0}  \Vert (x(t) , d(t) \Vert = +\infty$.  On the other hand, as already discussed in the case of \eqref{triang} in Example \ref{ex_3}, we find that $\Psi_{\mathfrak p} (t,{\mathsf z}) = ( e^{{\mathfrak p} \gamma} t, e^{- {\mathfrak p}r_1 } {\mathsf x}_1  , e^{- {\mathfrak p} r_2 } {\mathsf x}_2 ,e^{- {\mathfrak p} (\gamma + r_2)  } {\mathsf d} , e^{- {\mathfrak p}r_1 } {\mathsf y}   )$  is a GS of  $\Sigma (t) = 0$, SI-contracting  with contracting maps $( \frac{1}{\delta} \ln (1+\varphi(s) ) ,  e^{-s(1-\delta)} ,  \mathbbm{1}_{\mathfrak p}^2  \times  \mathbbm{1}_{\mathfrak p} )$ for any $\delta \in (0,1)$ and $\varphi \in {\mathcal K}_\infty$ such that $\varphi (s) \ge s$ for all $s \ge 0$.  Furthermore,   
   from   \cite{Andrieu_Praly_Astolfi_2008}  we   borrow  a $C^1$ function   $V :   {\mathbb R}^n  \to {\mathbb R}_+^0 $ of the form  $V (\mathsf{x} ) =   \int_{\mathsf{x}_2^{\frac{r_1}{r_2}}}^{d_1 \mathsf{x}_1} (s^{\frac{d_2-r_1}{r_1}} - \mathsf{x}_2^{\frac{d_2-r_1}{r_2}} ) ds + \vert \mathsf{x}_2 \vert^{\frac{d_2}{r_2}}$   such that  $ D_t V (x(t)) \vert_{\Sigma(t)=0}  \le - \ell_1 V^{ \frac{d_2+\gamma}{r_2} } (x(t)) + \ell_2 \vert y(t) \vert^{\frac{d_2+\gamma}{r_1}} + \ell_3 ( \sup_{t \ge 0} \Vert d(t) \Vert )^{\frac{d_2+\gamma}{r_2+\gamma}}$ for $t \ge 0$ and  for some $\ell_1 , \ell_2, \ell_3  > 0$ and  sufficiently large $ d_1 > 0$ and $d_2 > \max \{ r_1, r_2 \}$. Moreover,  $ \Vert x(t)   \Vert \le b_1   V^{\frac{r_2}{d_2}} + b_0$ for all $t \ge 0$ and for some $b_1, b_0  > 0$.   Hence, Assumption SINE is satisfied  with $\alpha(s) = \ell_1 s^{ \frac{d_2+\gamma}{r_2} }$, $\Phi (s) = \ell_2 s^{\frac{d_2+\gamma}{r_1}} + \ell_3 s^{\frac{d_2+\gamma}{r_2+\gamma}}$,   $\beta_1 (s) = b_1 s^{\frac{r_2}{d_2}} $ and $\beta_0=b_0$. Assumption GS.6, as well as the remaining assumptions of Theorem \ref{TH3},  can be satisfied as in Example \ref{por}. In conclusion,  while for a system $\Sigma (t) = 0$ with $(F,H)$ in \eqref{triang2}-\eqref{triang4} we can design a global  symmetry-based observer, we cannot use any other existing observer unless $\Sigma (t) = 0$ satisfies either  \eqref{triang3} with $\gamma=0$ or $\sup_{t \ge 0}  \Vert (x(t) , d(t) \Vert < +\infty$. 
  $\hfill \blacktriangleleft$
 \end{example}
 Analogue results to Theorem \ref{TH3} can be stated for global symmetry-based observers when considering PSI-contracting symmetries and using Assumptions PS.1, PS.2 and PS.5 instead of Assumptions  S.1, S.2 and S.5. 
\begin{example}\label{ex_6} 
Consider  the system map  $\Sigma $ associated 
to  \eqref{sys_0b} and  its global symmetry \eqref{Psi_0_alt}. Notice that \eqref{sys_0b} does not fall in the class  \eqref{triang2}-\eqref{triang4} for any choice of $ r_1  > 0$ and $\gamma \ge  0$.   As already seen, \eqref{sys_0b} is PSI-contracting with contracting maps  $( \sigma , \mu , \Gamma_{\mathfrak p}^{\mathsf{x}_2} \times  \Gamma_{\mathfrak p}^\mathsf{d}  )  = ( \frac{1}{\delta} \ln (1+\varphi(s)) ,   e^{-  s (1-\delta) }   ,   \mathbbm{1}_{\mathfrak p}  \times  \mathbbm{1}_{\mathfrak p}  )$ for any $\delta \in (0,1)$ and $\varphi \in {\mathcal K}_\infty$ such that $\varphi (s) \ge s$ for all $s \ge 0$. Furthermore, a SI-norm estimator for $\Sigma (t) = 0$ is designed from   \eqref{n1}-\eqref{n2} as well as  Assumption GS.6 is
satisfied by some  $\varphi \in {\mathcal K}_\infty$ and $\delta \in (0,1)$ selected as in Example \ref{por}. A global symmetry-based observer  for \eqref{sys_0b}  is 
\begin{eqnarray}
&&\hskip-.7cm D_t \chi (t)    = e^{p(t)}  \Big \{  \begin{pmatrix}  \chi_2(t)    \cr  {\rm sat}^2_{   e^{-p(t) (1-\delta)} }    (\chi_2 (t))  -  e^{- p(t)}  D_t  p (t)     \chi_2(t)  \end{pmatrix}    \nonumber \\
&&\hskip-.7cm  +  \begin{pmatrix}  k_1  Y(t)    \cr   k_2 Y(t)   \end{pmatrix}  \Big \} ,  \;   \widehat{x} (t)  = \begin{pmatrix}  
y (t)  \cr  e^{ p(t) }   {\rm sat}_{   e^{-p(t) (1-\delta)} }    (\chi_2 (t))  \end{pmatrix} , \nonumber
\end{eqnarray}
with $Y(t) \delequal y(t)    - \chi_1 (t)$, the gain matrix $K=\begin{pmatrix}  
k_1  &  k_2 \end{pmatrix}^\top$ such that $A-KC$ is Hurwitz and   $p (t)$ as in \eqref{observer_4.-1}-\eqref{observer_4}. To the best of our knowledge,   this is the first type of global observer proposed for the system $\Sigma (t)  =0 $.  $\hfill \blacktriangleleft$
\end{example}

\subsubsection{Global symmetry-based observers with prescribed finite-time convergence}\label{S3.2.3}
    
Theorems \ref{TH1}, \ref{TH1.1} and \ref{TH3} make  use of UU-transforming symmetries for designing   observers with asymptotic error convergence.   The following result invoke  BU-transforming and, in particular, CBU-transforming symmetries  (Definition \ref{time_scale}) for designing global observers with prescribed finite-time error convergence. For keeping the exposition as self-contained as possible, consider a   system map  $\Sigma $ with $(F  , H )\delequal(A  (t) {\mathsf x} +  B  (t) {\mathsf d}, C  (t) {\mathsf x} +  D  (t) {\mathsf d} ) $ (i.e. linear and time-varying) and the following assumptions replacing Assumptions S.1-S.3:
 
{\textbf{\emph{Assumption LTS.1}}: $ \Psi_{\mathfrak p}$   is a LS  of  $\Sigma (t)  =  0$ with ${\rm Dom} ( \Psi_{\mathfrak p} ) = \{  (t , {\mathsf z} ) \in {\mathcal M} :  t \in  {\rm Dom} ( \Psi_{\mathfrak p}^t ) \}$   and  CBU-transforming for   ${\mathfrak p} > 0$.$\hfill \blacktriangleleft$
}

{\textbf{\emph{Assumption LTS.2}}: There exist   $\psi \in {\mathcal L}$ and  $\pi_2 : {\mathbb R}_+^0 \times {\mathbb R}_+^0  \to {\mathbb R}_+$     such that for all $ {\mathfrak p} > 0$, ${\mathsf x} \in {\mathbb R}^n $ and $ t \in [   \Psi^t_{- \mathfrak p} ( \tau ) ,    {\rm Dom}^+  \Psi_{\mathfrak p}^t  ) $, with    $ \tau \in  {\rm Dom} (   \Psi^t_{- \mathfrak p} ) \cap {\mathbb R}_+^0$,  we have   $({\mathsf t}_{\mathfrak p}  ,  \mathsf{x} )  \in  {\rm Dom} (   \Psi^{ \mathsf{x}}_{- \mathfrak p} ) $,  
$  \left \Vert \frac{\partial 
\Psi^{\mathsf{x}}_{-\mathfrak p} ({\mathsf t}_{\mathfrak p}  ,  \mathsf{x} ) }{\partial  \mathsf{x} }   \right \Vert   \le \frac{1 }{ \pi_2 ({\mathsf t}_{\mathfrak p} , {\mathfrak p})   }  $ and $    \left \vert   \frac{\partial \ln\pi_2 ( s ,  {\mathfrak p} )}{ \partial s }      \right \vert_{s={\mathsf t}_{\mathfrak p}}  \le \psi ( {\mathsf t}_{\mathfrak p} )  $. $\hfill   \blacktriangleleft $ 
}
 
{\textbf{\emph{Assumption LTS.3}}: There exist  $\pi_3 \ge  0 $  such that $ \Vert  {\mathsf d}_{\mathfrak p}    \Vert   \le \pi_2 ({\mathsf t}_{\mathfrak p} , {\mathfrak p} )   \pi_3  \Vert  \mathsf{d} \Vert  $   
for all $ {\mathfrak p}  > 0$,  $ ( \mathsf{x},  \mathsf{d} ) \in {\mathbb R}^n \times {\mathbb R}^m$ and  $ t \in [   \Psi^t_{- \mathfrak p} ( \tau ) ,    {\rm Dom}^+  \Psi_{\mathfrak p}^t  )$. $\hfill \blacktriangleleft$  
}  

    The proof of the following result  is similar to the proof of Theorem \ref{TH3}, except for few issues on the time scale which we omit for brevity.
       \begin{theorem}\label{TH1b}
 Under Assumptions LTS.1-LTS.3 and  S.4  consider the solutions   $z(t)$ of $\Sigma (t) = 0 $ for which $\sup_{t \ge 0} \Vert d(t) \Vert    < +\infty  $ with the filter 
   \begin{eqnarray}
 &&\hskip-1cm D_t \chi (t)   =  \frac{\partial \Psi^t_{\mathfrak p}  (t) }{\partial t} \Big  (  (  A   (  \Psi^t_{\mathfrak p}  (t)   )  - K (  \Psi^t_{\mathfrak p}  (t)  )   C (  \Psi^t_{\mathfrak p}  (t)  )    )  \chi(t) \nonumber \\
  &&\hskip-1cm+       K  (  \Psi^t_{\mathfrak p}  (t)  )   \Psi^{\mathsf y} _{\mathfrak p}  (t,y (  t)  )  \Big  ) ,   \label{observer_12}  
   \end{eqnarray} 
   its design parameter ${\mathfrak p} >0$ and its output  $ \widehat{x} (t)  \delequal  \Psi_{-{\mathfrak p}}^{\mathsf{x}}  (\Psi_{{\mathfrak p}}^t (t) ,   \chi  (t) )$. For each   $\varepsilon , T > 0$   there exists  $ {\mathfrak p} >0$   such that  $0 <   {\rm Dom}^+  \Psi^t_{\mathfrak p}  \le  T$ and 
  \begin{eqnarray}
 && \hskip-.8cm  \limsup_{t \nearrow    {\rm Dom}^+  \Psi^t_{\mathfrak p}  }  \Vert x (t) - \widehat{x}  (t)   \Vert   \le  \{ \varepsilon +  8{\bf \overline{p}}  \sqrt\frac{2{\bf \overline{p}}}{{\bf \underline{p}}}   \times \label{error_0}\\
  && \hskip-.8cm \times   (1+ \sup_{ t \ge 0 }   \Vert K ( t ) \Vert  )  (  \sup_{ t \ge 0  }   \Vert B (  t ) \Vert  +   \sup_{t \ge  0 }   \Vert D ( t ) \Vert  )   \}   \pi_3  \sup_{t \ge 0} \Vert d(t) \Vert   \nonumber   \hfill \blacktriangleleft
   \end{eqnarray}   
 \end{theorem} 
 \begin{remark}  
Since we are considering a linear time-varying system $\Sigma(t) = 0$,   the  group $\Psi_{{\mathfrak p}} $ is not required to be SI-contracting and, therefore, a SI-norm estimator is not needed.  The estimate $\widehat{x}  (t)$ converges to $x(t) $ (up to an error which depends on $\sup_{t \ge 0} \Vert d(t) \Vert < +\infty$) even when $\sup_{t \ge 0} \Vert x(t) \Vert = +\infty$. 
\end{remark}

\textbf{\emph{Example}} \emph{\ref{ex_12}} \textbf{\emph{(cont'ed)}}.    Consider the system map $\Sigma  $ associated to  \eqref{infinite}. We have seen that 
\eqref{locgr}  is a LS of $\Sigma (t) = 0 $, with ${\rm Dom} ( \Psi_{\mathfrak p} ) = \{  (t , {\mathsf z} ) \in {\mathcal M} :  t \ne \frac{1}{\mathfrak p} \}$ and CBU-transforming for   ${\mathfrak p} > 0$ (Assumption  LTS.1). Assumptions  LTS.2-LTS.3  are met with  $\tau:=1$,  $\psi (t ) \delequal \frac{6}{ 1+ t }$, $\pi_2 (t,{\mathfrak p}) \delequal \frac{1}{ (1+{\mathfrak p} t)^3}$ and $\pi_3 :=1$. By Theorem \ref{TH1b}  an observer for $\Sigma (t) = 0$ with convergence  within prescribed  finite-time  $T$  is provided by \eqref{observer_12} as 
\begin{eqnarray}
&&D_t \chi (t)    =\frac{1}{  (  1-{\mathfrak p}t )^2  } \Big ( \begin{pmatrix} \chi_2 (t) \cr 0\end{pmatrix} + K    (  \frac{ y(t) }{1- {\mathfrak p} t}   -   \chi_1 (t)          )  \Big )   , \nonumber \\
&& \widehat{x} (t) =  \begin{pmatrix} 1- {\mathfrak p} t & 0 \cr  -{\mathfrak p}  &  \frac{1}{1- {\mathfrak p} t } \end{pmatrix}  \chi (t)  \label{pw}
\end{eqnarray}
     with $K \in {\mathbb R}^2$ such that $ A-KC$ is Hurwitz. 
     
     We can compare the symmetry-base observer \eqref{pw} (more generally, the results of Theorem \ref{TH1b}) with    existing ones for linear systems such as the (homogeneous) one proposed in \cite{Pol}, although to our best knowledge none of them takes into account disturbance-insensitivity issues with global error convergence, or the SLO's proposed for instance in \cite{R_2022}. Homogeneous observers and SLO's are {\sl time-invariant  fractional-power or even discontinuous} function of the output, while our observers are {\sl  time-varying continuously differentiable}  (actually smooth) function of the   output. Our observers are more resemblant to the ones used in \cite{SWK} and, as  also noticed in \cite{SWK}, a time-varying gain-based finite-time observer is built upon smooth   observers, not on fractional-power or discontinuous observers, thus resulting in a smooth estimation process. Moreover,  the finite-time observer is characterized with uniformly pre-specifiable convergence time that is independent of initial condition and any other design parameters and can be preassigned as needed within the physically allowable range.  

\section{Asymptotic symmetries}\label{as_symm}
 
As it is apparent from  \eqref{error},  a favorable situation for attenuating the effect of unknown  exogenous inputs $d(t)$ on the   estimation  error is when   the linearization of $\Sigma (t) = 0$  around the origin is input-insensitive (i.e. $B(t)\equiv 0$ and $D(t) \equiv0$). Indeed, in this case the error bound resulting from \eqref{error} is the arbitrarily small number $ \varepsilon \pi_3 (N)$.   When the linearization of  $\Sigma (t) = 0$  is not  input-insensitive, we are motivated to introduce relaxed notions of 
 symmetries $\Psi_{\mathfrak p}$ for which  $     \Sigma_{\mathfrak p} \vert_{\Sigma  = 0}  \ne 0$ (i.e. $\Psi_{\mathfrak p}$ is not a LS of $\Sigma (t)=0$)  but   there exists a system map $\Sigma_\infty  \ne \Sigma $  such that  $     ( \Sigma_\infty )_{\mathfrak p} \vert_{\Sigma  = 0}    \to 0$ as ${\mathfrak p} \to+\infty$ 
 \footnote{According to the Notational Remark II, $ ( \Sigma_\infty )_{\mathfrak p}$ denotes  the  transformed system map  $    \Sigma_\infty $   under the action of $\Psi_{\mathfrak p}$.} 
  (i.e.   $\Psi_{\mathfrak p}$   maps $\Sigma =0$ into  $( \Sigma_\infty )_{\mathfrak p} =0$  as ${\mathfrak p} \to+\infty$).  The ``asymptotic'' system $\Sigma_\infty (t) = 0$ is designed so as to have an input-insensitive linearization or more specific and simple structures (linear, triangular, etc etc). 
   
 \begin{definition}\label{2}
 \emph{(Asymptotic symmetries)}. A LGT (resp. GGT)   $  \Psi_{\mathfrak p}  $  is \emph{a local (resp. global)  $C^\infty$ asymptotic   symmetry (LAS, resp. GAS)}  of the   system  $   \Sigma (t) = 0$  with asymptotic system map $\Sigma_\infty  $  and contracting maps $(\sigma,  \mu,  \Gamma_{\mathfrak p}^{\mathsf x} \times \Gamma_{\mathfrak p}^{\mathsf d}  )$  if  it is   SI-contracting with contracting maps $(\sigma,  \mu,  \Gamma_{\mathfrak p}^{\mathsf x} \times \Gamma_{\mathfrak p}^{\mathsf d}  )$ and  there exist a   system  map $\Sigma_\infty \ne \Sigma$ (associated to a pair $(  {\mathsf F}_\infty (t,{\mathsf x}  , {\mathsf d}  ) ,  {\mathsf H}_\infty (t,{\mathsf x}  , {\mathsf d}  ) ) \ne (  {\mathsf F}  (t,{\mathsf x}  , {\mathsf d}  ) ,  {\mathsf H}  (t,{\mathsf x}  , {\mathsf d}  ) )$)     and $\nu   \in {\mathcal{K_+ L}}  $   such that for all ${\mathfrak p} \ge 0$ and  $(   t, {\mathsf z} ) \in {\rm Dom} (\Psi_{\mathfrak p} )  $ 
\begin{eqnarray} 
&&  \hskip-1cm    \Vert  \Delta \Sigma_\infty  ( {\mathfrak p} ,  {\mathsf t}_{\mathfrak p}  ,   {\mathsf x}_{\mathfrak p}    ,  {\mathsf d}_{\mathfrak p}  )  \Vert    \le  \nu      (      \Vert (\Gamma_{\mathfrak p}^{\mathsf x} \times \Gamma_{\mathfrak p}^{\mathsf d}   ) (   {\mathsf x}_{\mathfrak p}    ,  {\mathsf d}_{\mathfrak p}  )  \Vert     , {\mathfrak p}     )     \Vert ( {\mathsf x}_{\mathfrak p}    ,  {\mathsf d}_{\mathfrak p} ) \Vert \nonumber \\
&&  \hskip-1cm  \label{approximation}  
\end{eqnarray}  
where $\Delta \Sigma_\infty  ({\mathfrak p} ,  \overline{t} ,   \overline{\mathsf  x}  ,   \overline{\mathsf  u} ) \delequal   [  ( \Sigma_\infty )_{\mathfrak p}    \vert_{    \Sigma =  0    } ] \vert_{    ( t   ,  {\mathsf  x} , {\mathsf  u}   ) = [ \Psi_{\mathfrak p}^{t, \mathsf{x}, \mathsf{d}} ]^{-1} (  \overline{t} ,   \overline{\mathsf  x}  ,   \overline{\mathsf  u}   ) }  $. $ \hfill \blacktriangleleft $  
\end{definition} 
Notice that $ \Delta \Sigma_\infty  ( {\mathfrak p} ,  {\mathsf t}_{\mathfrak p}  ,   {\mathsf x}_{\mathfrak p}    ,  {\mathsf d}_{\mathfrak p}  ) = ( \Sigma_\infty )_{\mathfrak p}    \vert_{    \Sigma =  0    }$ for all ${\mathfrak p} \ge 0$ and  $(   t, {\mathsf z} ) \in {\rm Dom} (\Psi_{\mathfrak p} )  $, since $ ( t   ,  {\mathsf  x} , {\mathsf  u}   ) = [ \Psi_{\mathfrak p}^{t, \mathsf{x}, \mathsf{d}} ]^{-1} (  {\mathsf t}_{\mathfrak p}  ,   {\mathsf x}_{\mathfrak p}    ,  {\mathsf d}_{\mathfrak p}  )$. Whenever ${\rm Dom} ( \Psi_{\mathfrak p} ) = {\mathcal M}$ for all ${\mathfrak p} > 0$,  for each $ (   t, {\mathsf z} ) \in  {\mathcal M}$ and as ${\mathfrak p} \to +\infty$ we have $ \Vert (\Gamma_{\mathfrak p}^{\mathsf x} , \Gamma_{\mathfrak p}^{\mathsf d} ) (   {\mathsf x}_{\mathfrak p}    ,  {\mathsf d}_{\mathfrak p}  )  \Vert  \to 0$ by the SI-contraction property and  finally   $  ( \Sigma_\infty )_{\mathfrak p}    \vert_{    \Sigma =  0    }   \to 0 $ by \eqref{approximation}.   Furthermore, since any LS (GS) of  $   \Sigma (t) = 0$ is such that $ \Sigma_{\mathfrak p}  \vert_{\Sigma = 0}  =0$ (Definition \ref{1d}), if it is also SI-contracting   then it  can be considered as the limit case of  a LAS (GAS)  of   $   \Sigma (t) = 0$  with   asymptotic system map $\Sigma_\infty =  \Sigma$. 
\begin{example}\label{ex_1bb}
 {\sl (Weighted homogeneity in the $\infty$-limit)}. As for the symmetry condition \eqref{1b},    also the asymptotic symmetry condition \eqref{approximation}  can be given  a geometric interpretation and, as a particular case, we    recover the notion of {\sl homogeneity in the $\infty$-limit} (\!\!\cite{Andrieu_Praly_Astolfi_2008}).   Consider the same  system map  $\Sigma   $  and  GGT $\Psi_{\mathfrak p}$ as Section \ref{geometric}. On account of the prolongation formula \eqref{EPRW2} for computing $\Psi_{\mathfrak p}^{{\mathsf x}^{(1)}}$ and the asymptotic symmetry condition \eqref{approximation},  $ \Psi_{\mathfrak p} $ is a GAS of  $\Sigma (t) = 0$ with asymptotic system map  $\Sigma_\infty  $   associated to  $(F_\infty ( t, \mathsf{x},\mathsf{d} ), H_\infty ( t, \mathsf{x},\mathsf{d} ) ) = (A_\infty (\mathsf{x}) ,C_\infty (\mathsf{x}) ) \ne (A (\mathsf{x}) ,C  (\mathsf{x}) )  $   if (and only if) the vector field $A$,  resp. the mapping $C$,  is homogeneous in the $\infty$-limit with weights  $(-g^{\mathsf{x}_1}   , \cdots , -  g^{\mathsf{x}_n}   )$,  degree  $g^t$, resp. degrees $( g^{\mathsf{y}_1} , \cdots , g^{\mathsf{y}_p} )$, and $\infty$-limit vector field $A_\infty $, resp.$\infty$-limit map $C_\infty $: in formulas, for each $\mathsf{x}$ it holds that  
 $   \Vert  {\rm diag} \{   e^{  {\mathfrak p} ( g^{{\mathsf x}_i}  -   g^t )  } \}  A (   \mathsf{x}    )  - A_\infty (  e^{  {\mathfrak p} g^{{\mathsf x}_i}  }  {\mathsf x} )   \Vert  \to  0$ and $  \Vert   e^{  {\mathfrak p}   g^{{\mathsf y}_i}   }   C  (   \mathsf{x}    )  - C_\infty (  e^{  {\mathfrak p}  g^{{\mathsf x}_i}     } {\mathsf x} )     \Vert \to 0$   as ${\mathfrak p}\to +\infty$. 
$\hfill\blacktriangleleft$
 \end{example} 
 \begin{example}\label{E8} 
Back to the system map $\Sigma $ associated to   \eqref{triang_ex},  notice that the linearization of $\Sigma (t) = 0$ at $(\mathsf{x}, \mathsf{d} )=(0,0)$  is not input-insensitive. Consider  the GGT  $\Psi_{\mathfrak p} (\mathsf{t} ,\mathsf{z} )= ( \mathsf{t}_{\mathfrak p}, \mathsf{x}_{\mathfrak p},\mathsf{d}_{\mathfrak p},\mathsf{y}_{\mathfrak p})$   
 \begin{eqnarray}
  &&\hskip-.8cm \mathsf{t}_{\mathfrak p} \delequal  e^{    {\mathfrak p} g^t }   ,   \mathsf{x}_{\mathfrak p} \delequal  (    e^{- \mathfrak p}    \mathsf{x}_1 ,   e^{-{\mathfrak p} (1+g^t)}    \mathsf{x}_2 ) ,  
\nonumber   \\ 
  &&\hskip-.8cm  \mathsf{d}_{\mathfrak p} \delequal  
  e^{-{\mathfrak p} (1+\frac{2}{3} g^t ) } \mathsf{d},  \mathsf{y}_{\mathfrak p} \delequal     e^{-\mathfrak p}  \mathsf{y}    \label{wp2b.2asy}
\end{eqnarray}
with any  $g^t > 0$, which   is    SI-contracting with contracting maps    
     \begin{eqnarray}
&&\hskip-1cm ( \sigma , \mu ,  \Gamma_{\mathfrak p}^\mathsf{x} \times  \Gamma_{\mathfrak p}^\mathsf{d}  ) \nonumber \\
&&\hskip-1cm = ( \frac{1}{1 - \delta}  \ln (\varphi (s)  +1),  e^{-\delta s } ,   {\rm diag} \{   1 ,  e^{{\mathfrak p} g^t  }  \} \times e^{ \frac{2}{3}  {\mathfrak p} g^t }     )  \label{ex5.3.2asy}
\end{eqnarray}           
for any $\delta \in (0,1)  $ and $\varphi \in {\mathcal K}_\infty$ such that $\varphi(s) \ge s$ for all $s \ge 0$. Also, consider a candidate asymptotic system map $\Sigma_\infty  \ne \Sigma$,   associated to 
\begin{eqnarray}
    (F_\infty( t, \mathsf{x}, \mathsf{d} )  , H_\infty ( t, \mathsf{x}, \mathsf{d} ) ) = \left ( \begin{pmatrix}  \mathsf{x}_2 \cr 0   \end{pmatrix} ,   \mathsf{x}_1  \right ) \label{as_map} 
\end{eqnarray}
which is input-insensitive.  By direct calculations
  \begin{eqnarray}
&&\hskip-.9cm  \Vert (  \Sigma_\infty) _{\mathfrak  p}  \vert_{\Sigma = 0}  \Vert   =   e^{-{\mathfrak p} ( 1 +2  g^t  ) }  \vert  \mathsf{x}_1^k  \mathsf{x}_2  +  ( 1+  \mathsf{x}_1^2   ) \mathsf{d}     \vert , \nonumber \\
&&\hskip-.9cm    \Vert  \Delta \Sigma_\infty  (  {\mathfrak p}  ,  {\mathsf t}_{\mathfrak p}  ,   {\mathsf x}_{\mathfrak p}    ,  {\mathsf d}_{\mathfrak p}  )  \Vert    \le   e^{-2 {\mathfrak p}    g^t    }  \times  \nonumber \\
&&\hskip-.9cm   \times  ( e^{ {\mathfrak p} k } \Vert \Gamma_{\mathfrak p}^\mathsf{x} \mathsf{x}_{\mathfrak p} \Vert^k  \Vert  \mathsf{x}_{\mathfrak p}  \Vert  +  ( 1+ e^{ 2{\mathfrak p}  }  \Vert \Gamma_{\mathfrak p}^\mathsf{x} \mathsf{x}_{\mathfrak p} \Vert^2   ) e^{  \frac{2}{3}  {\mathfrak p}      g^t  ) }   \vert  \mathsf{d}_{\mathfrak p} \vert  )   
\nonumber    
\end{eqnarray} 
for all ${\mathfrak p} \ge 0$ and  $(t, \mathsf{z}) \in {\mathcal M}$. If  $g^t \ge \max \{6, k  \} $ then we obtain \eqref{approximation} 
for all ${\mathfrak p} \ge 0$ and  $(t, \mathsf{z}) \in {\mathcal M}$,   with  $  \nu ( r,s   ) = e^{- s g^t } (r^k+ 1 +   r^2 ) \in {\mathcal {K_+L}}$. Hence, the GGT 
 \eqref{wp2b.2asy}, for any $g^t \ge \max \{6, k  \} $,    is a GAS of  $\Sigma(t)=0$ with asymptotic system  map $\Sigma_\infty$  (associated to    \eqref{as_map})  and contracting maps \eqref{ex5.3.2asy}. $\hfill \blacktriangleleft$ 
\end{example}
A LAS is conceived in such a way to disentangle the transformed system under the action of a LGT  from the system   itself, in the sense that   the transformed system is different from the original one (as ${\mathfrak p} \to +\infty$ when this makes sense). As  pointed out earlier,  this may be beneficial in terms of input-sensitivity if we design the observer on the transformed system.  However,  condition \eqref{approximation} which characterizes a LAS  is not yet sufficient to design a convergent observer. An analogue condition  \eqref{approximation}   has to be  introduced on  the ``variation'' of  the function $  \Delta \Sigma_\infty  ( {\mathfrak p} ,  {\mathsf t}_{\mathfrak p}  ,   \cdot   ,  \cdot ) $ leading to the following definition of ``variational'' LAS.   
 \begin{definition}\label{3}
 A LGT (resp. GGT)   $  \Psi_{\mathfrak p}  $  is \emph{a local (resp. global) $C^\infty$ asymptotic variational symmetry (LAVS, resp. GAVS)} of $   \Sigma (t) = 0$  with asymptotic system map $\Sigma_\infty   $  and   contracting maps $(\sigma,  \mu,   \Gamma_{\mathfrak p}^{\mathsf x} \times \Gamma_{\mathfrak p}^{\mathsf d}  )$  if  it is a LAS (resp. GAS) of   $   \Sigma (t) = 0$  with asymptotic system map $\Sigma_\infty  $ and  contracting maps $(\sigma,  \mu,   \Gamma_{\mathfrak p}^{\mathsf x} \times \Gamma_{\mathfrak p}^{\mathsf d} )$ 
 and, in addition, there exists $\lambda \in {\mathcal {K_+ L}}$ such that  for all ${\mathfrak p} \ge 0$ and  $(   t, {\mathsf z} ) \in {\rm Dom} (\Psi_{\mathfrak p} )  $ 
\begin{eqnarray}  
&&  \hskip-1.2cm   \Vert   \frac{\partial  \Delta \Sigma_\infty   ( {\mathfrak p} , {\mathsf t}_{\mathfrak p}  ,   \overline{\mathsf  x}  ,   \overline{\mathsf  u}  )     }{ \partial    (   \overline{\mathsf  x}  ,   \overline{\mathsf  u}   )     }   \Vert_{   (  \overline{\mathsf  x}  ,   \overline{\mathsf  u}   ) \atop  =  (     {\mathsf x}_{\mathfrak p}    ,  {\mathsf d}_{\mathfrak p}  ) }    \le  \lambda (      \Vert (\Gamma_{\mathfrak p}^{\mathsf x} \times \Gamma_{\mathfrak p}^{\mathsf d}  ) (   {\mathsf x}_{\mathfrak p}    ,  {\mathsf d}_{\mathfrak p}  )   \Vert    , {\mathfrak p}     )      \label{approximation2}  
\end{eqnarray}  
where  $\Delta \Sigma_\infty$  is as in \eqref{approximation}. $\hfill \blacktriangleleft$
\end{definition}  

Since any LS (GS) of  $   \Sigma (t) = 0$ is such that   $ \Sigma_{\mathfrak p}  \vert_{\Sigma = 0}  =0$, if it is also  SI-contracting   then it can be considered as the limit case of a   LAVS (resp. GAVS)  of   $   \Sigma (t) = 0$  with   asymptotic system map $\Sigma_\infty = \Sigma$.

\textbf{\emph{Example}} \emph{\ref{E8}} \textbf{\emph{(cont'ed)}}.  
Once more consider the system map $\Sigma  $ associated to   \eqref{triang_ex} with the GGT \eqref{wp2b.2asy}, SI-contracting with  contracting maps  \eqref{ex5.3.2asy}. Also, let $(F_\infty ( t, \mathsf{x}, \mathsf{d} ) ,H_\infty ( t, \mathsf{x}, \mathsf{d} ) )$ be as  in \eqref{as_map}. By direct calculations
  \begin{eqnarray}
&&\hskip-.9cm    \Vert   \frac{\partial  \Delta \Sigma_\infty   ( {\mathfrak p},  {\mathsf t}_{\mathfrak p}  ,    \overline{\mathsf  x}  ,   \overline{\mathsf  u}   )     }{ \partial    (   \overline{\mathsf  x}  ,   \overline{\mathsf  u}   )     }   \Vert_{   (  \overline{\mathsf  x}  ,   \overline{\mathsf  u}   ) \atop   =  (    {\mathsf x}_{\mathfrak p}    ,  {\mathsf d}_{\mathfrak p}  ) } 
\le   e^{-  {\mathfrak p}   g^t    }  \Vert \Gamma_{\mathfrak p}^\mathsf{x} \mathsf{x}_{\mathfrak p} \Vert  \times  \nonumber \\
&&\hskip-.9cm \times (   e^{ {\mathfrak p} k } \Vert \Gamma_{\mathfrak p}^\mathsf{x} \mathsf{x}_{\mathfrak p} \Vert^{k-1}  (k+  1 )  + ( 1+    \Vert \Gamma_{\mathfrak p}^\mathsf{x} \mathsf{x}_{\mathfrak p}  \Vert +   2  \Vert \Gamma_{\mathfrak p}^\mathsf{d} \mathsf{d}_{\mathfrak p} \Vert  ) e^{  {\mathfrak p}   }   )   \nonumber
\end{eqnarray} 
for all ${\mathfrak p} \ge 0$ and  $(t, \mathsf{z}) \in {\mathcal M}$. If  $g^t \ge \max \{2, 2k  \} $ then we obtain \eqref{approximation2} 
for all ${\mathfrak p} \ge 0$ and  $(t, \mathsf{z}) \in {\mathcal M}$,   with   $  \lambda ( r,s   ) = e^{-\frac{1}{2} s g^t } ( 1 +3r + r^{k-1} (k+1)  )  \in {\mathcal {K_+L}}$. Therefore, since   
 \eqref{wp2b.2asy} is for any $g^t \ge \max \{6, k  \}$  a GAS of  $\Sigma(t)=0$ with asymptotic system  map $\Sigma_\infty$ (associated to   \eqref{as_map})   and contracting maps \eqref{ex5.3.2asy},   it  is for any $g^t \ge \max \{6, 2k  \} $  a GAVS of  $\Sigma(t)=0$ with asymptotic system  map $\Sigma_\infty $  (associated to   \eqref{as_map})   and contracting maps \eqref{ex5.3.2asy}.  

\section{Variational asymptotic symmetry-based observers}\label{S4}

For designing variational asymptotic symmetry-based observers,  we modify Assumptions S.1, S.4 and S.5  coherently with the new definition of  LAVS: 

{\textbf{\emph{Assumption AVS.1}}: $ \Psi_{\mathfrak p}$   is a LAVS  of  $\Sigma (t)  =  0$, with asymptotic system  map  $\Sigma_\infty $,  contracting maps $(\sigma , \mu , \Gamma_{\mathfrak p}^\mathsf{x} \times  \Gamma_{\mathfrak p}^\mathsf{d}  )$  and  UU-transforming and 
 ${\rm Dom} ( \Psi_{\mathfrak p} ) = {\mathcal M}$ for ${\mathfrak p} > 0$. $\hfill \blacktriangleleft$ 
 
Let   
\begin{eqnarray}
&&\hskip-.7cm A_\infty (t)  \delequal   \frac{\partial    F_\infty   }{\partial  \mathsf{x}} \Big \vert_{(\mathsf{x},\mathsf{d})=0} , B_\infty (t) \delequal   \frac{\partial    F_\infty  }{\partial  \mathsf{d}} \Big \vert_{(\mathsf{x},\mathsf{d})=0} , \nonumber \\
&&\hskip-.7cm  C_\infty (t) \delequal   \frac{\partial  H_\infty   }{\partial \mathsf{x} } \Big \vert_{(\mathsf{x},\mathsf{d})=0} ,  D_\infty (t) \delequal   \frac{\partial   H_\infty   }{\partial  \mathsf{d}} \Big \vert_{(\mathsf{x},\mathsf{d})=0}  , \nonumber \\
&&\hskip-.7cm ( \Delta \Sigma_{\infty , L}  ) ( t, \mathsf{x} , \mathsf{d})    \delequal  \begin{pmatrix}  F_\infty   ( t,  \mathsf{x} , \mathsf{d}) \cr  H_\infty  ( t,   \mathsf{x} , \mathsf{d})  \end{pmatrix}   -   \begin{pmatrix} A_\infty  (t) &  B_\infty  (t) \cr  C_\infty  (t)& D_\infty  (t)\end{pmatrix} \begin{pmatrix}  \mathsf{x}   \cr   \mathsf{d} \end{pmatrix}   \nonumber 
\end{eqnarray}

{\textbf{\emph{Assumption AVS.4}}: $ \sup_{t \ge 0}  \Vert B_\infty (t) \Vert 
 < +\infty$, $ \sup_{t \ge 0} \Vert D_\infty (t) \Vert 
 < +\infty$  and there exist $K_\infty   : {\mathbb R}_+^0  \to  {\mathbb R}^{n \times p} $,  $P_\infty  : {\mathbb R}_+^0  \to    {\mathbb S}_+ (n)$ and  $ {\bf \underline{p}}_\infty  ,  {\bf \overline{p}}_\infty    > 0$ such that  $ \sup_{t \ge 0}  \Vert K_\infty (t) \Vert  
 < +\infty$ and $ D_t P_\infty (t) + P_\infty (t) (A_\infty (t)   -K_\infty  (t)  C_\infty   (t))   + (A_\infty (t)   -K_\infty  (t)  C_\infty   (t))^\top (t)  P_\infty (t)  + 2{\textbf I}_n   \in {\mathbb  S}_- (n)   $ with  
${\bf {\bf \underline{p}}}_\infty  {\textbf I}_n \le P_\infty (t) \le {\bf {\bf \overline{p}}}_\infty {\textbf I}_n $  for all $t \ge 0$.$\hfill \blacktriangleleft$ 

{\textbf{\emph{Assumption AVS.5}}: There exist   $\xi  \in {\mathcal K}$ such that $   \left \Vert   \frac{\partial   ( \Delta \Sigma_{\infty , L}  ) (t , {\mathsf x} , {\mathsf d}  )  }{\partial ( {\mathsf x} , {\mathsf d}   )   }   \right  \Vert \le  \xi  (   \Vert  ( {\mathsf x} , {\mathsf d}  ) \Vert )    $ for all $t \ge 0$ and $ ( {\mathsf x} , {\mathsf d}  ) \in {\mathbb R}^n \times {\mathbb R}^m $. $\hfill \blacktriangleleft$ 
   
 Define the   filter 
  \begin{eqnarray}
 &&\hskip-1.2cm\begin{pmatrix}  D_t \chi (t) \cr \zeta (t)   \end{pmatrix}  = \begin{pmatrix} \frac{\partial\Psi_{\mathfrak p}^t (t) }{\partial t} {\textbf I}_n & {\textbf 0}_{n \times p} \cr {\textbf 0}_{p \times n} & {\textbf I}_p \end{pmatrix}   \Bigg (   \begin{pmatrix} A_\infty  (\Psi_{\mathfrak p}^t (t)  )   \cr  C_\infty ( \Psi_{\mathfrak p}^t (t) )   \end{pmatrix}   \chi (t)     \nonumber \\
  &&\hskip-1.2cm  +  ( \Delta \Sigma_\infty )  (\Psi_{\mathfrak p}^t (t) ,    \chi^{\rm sat} (t) ,0 )  +  ( \Delta \Sigma_{\infty , L}  )  (\Psi_{\mathfrak p}^t (t) ,    \chi^{\rm sat} (t) , 0 )  \nonumber \\
  &&\hskip-1.2cm    +    \begin{pmatrix}        K  (\Psi_{\mathfrak p}^t (t))   (  \Psi^{\mathsf y}_{\mathfrak p} (t,y(t))   - \zeta(t)) \cr   {\textbf 0}_{p \times 1}   \end{pmatrix}   \Bigg )  \label{observer_1_AS2}  
   \end{eqnarray}
with  $   \Delta \Sigma_\infty  $ as in \eqref{approximation}   and   $  \chi^{\rm sat} (t)   \delequal  
    \Gamma^{\mathsf x}_{-\mathfrak p}    {\rm sat}_{\mu ( {\mathfrak p}  )}   (  \Gamma^{\mathsf x}_{\mathfrak p}  (  \chi  (t)  )      $. The proof of the following result follows closely the proof of Theorem \ref{TH1} and it is omitted. 
      \begin{theorem}\label{TH1_AS}
 Under Assumptions AVS.1, 
 S.2,  S.3, AVS.4 and 
 AVS.5 and for each $N>0$  consider the solutions $z(t)$ of $\Sigma (t) = 0 $    for which $\sup_{t \ge 0}  \Vert    (   x  (t) , d (t) )  \Vert  \le N$ together with the filter  \eqref{observer_1_AS2}, its parameter design ${\mathfrak p} > 0$ and its output $ \widehat{x} (t)  \equiv  \Psi_{-{\mathfrak p}}^{\mathsf{x}}  (\Psi_{{\mathfrak p}}^t (t) ,   \mathsf{\chi}^{\rm sat} (t) )$. For each $\varepsilon > 0$ there exists   $ {\mathfrak p} >0$   such that 
  \begin{eqnarray}
 && \hskip-1cm   \limsup_{t \rightarrow +\infty}  \Vert x (t) - \widehat{x}  (t)   \Vert    \le  \{  \varepsilon +  8{\bf \overline{p}}  \sqrt\frac{2{\bf \overline{p}}}{{\bf \underline{p}}}     (1+ \sup_{ {\mathsf t} \ge 0  }   \Vert K_\infty (  t ) \Vert  ) \times \nonumber \\
  && \hskip-1cm \times (  \sup_{ {\mathsf t} \ge 0  }   \Vert B_\infty (  t  ) \Vert  +   \sup_{ {\mathsf t} \ge 0 }   \Vert D_\infty (  t  ) \Vert  ) \pi_1(0)  \}  \pi_3 (\sup_{t \ge 0} \Vert d(t) \Vert  )  .    \label{error_AS} \hfill \blacktriangleleft 
   \end{eqnarray} 
 \end{theorem} 
\begin{remark}\label{fin_rem}
If the linearization of  $\Sigma_\infty (t) =0$ is   input-insensitive (i.e. $B_\infty(t) \equiv 0$ and $D_\infty(t)\equiv0$)   Theorem  \ref{TH1_AS} certifies that $\widehat{x}  (t) $ is an asymptotic estimate (up to an arbitrarily small error $\varepsilon \pi_3 (N ) $) of any solution $x(t)$ of $\Sigma (t) = 0 $ for which $\sup_{t \ge 0} \Vert (x(t), d(t) ) \Vert \le N$.  Hence, in order to  attenuate the effect of exogenous inputs on the error estimation it is important to look for asymptotic system maps $\Sigma_\infty$ with   input-insensitive linearization.    
\end{remark}

\textbf{\emph{Example}} \emph{\ref{E8}} \textbf{\emph{(cont'ed)}}.       
Consider the system map $\Sigma $ associated to  \eqref{triang_ex} with  the GGT \eqref{wp2b.2asy}  
which is for any $g^t \ge \max \{6, k  \}$ a GAVS of  $\Sigma(t)=0$  with asymptotic systems map  $\Sigma_\infty  $, where $(F_\infty (t,\mathsf{x} , \mathsf{d} ) , H_\infty (t,\mathsf{x} , \mathsf{d} )  ) $ is as in \eqref{as_map}, and contracting maps \eqref{ex5.3.2asy}. %
 Theorem \ref{TH1_AS} applies to $\Sigma(t)=0$  providing the  variational asymptotic  symmetry-based observer 
 \begin{eqnarray}
&& \hskip-.8cm D_t \chi (t)  = e^{{\mathfrak p}g^t }  \Big \{ 
 \begin{pmatrix}      \chi_2 (t)     \cr  e^{ {\mathfrak p}( k - 2g^t)}  {\rm sat}^k_{    e^{- {\mathfrak p}\delta  } }  ( \chi_1 (t)    ) {\rm sat}_{     e^{- {\mathfrak p}\delta  }    }    (   e^{  {\mathfrak p} g^t   }  \chi_2  (t)    )    \end{pmatrix}    \nonumber  \\
&& \hskip-.8cm  + 
 \begin{pmatrix}    k_1  Y(t)   \cr     k_2 Y(t)   \end{pmatrix}    \Big \}   ,  \;  \widehat{x} (t) = e^{   \mathfrak p } \begin{pmatrix}   {\rm sat}_{    e^{- {\mathfrak p}\delta  } }  ( \chi_1 (t)    ) \cr    {\rm sat}_{     e^{- {\mathfrak p}\delta  }    }    (   e^{  {\mathfrak p} g^t   }  \chi_2  (t)    )  \end{pmatrix}   ,  \nonumber 
 \end{eqnarray}
   with $Y(t):=e^{-  {\mathfrak p} } y (t)       -   \chi_1 (t)$,   $g^t \ge \max \{6, 2k  \} $ and    
 $K = (k_1,k_2)^\top$   such that $A-KC$ is Hurwitz, which can be easily compared with the   symmetry-based observer \eqref{obs1}. The error bound provided by Theorem \ref{TH1_AS} is the arbitrarily small number $\varepsilon \pi_3 (N) $, since in this case $B_\infty (t) \equiv 0$ and $D_\infty (t) \equiv 0$. Hence, we recover the same performances of the semiglobal   HGO  \eqref{due} (SLOs perform better yet,  achieving exact asymptotic  convergence).  $\hfill \blacktriangleleft$
%
  
\section{Conclusions}

Different types of symmetries   have been introduced for ODE systems with motivating and illustrative examples. Based on these symmetries,  we have  designed a variety of  semiglobal and global   observers, in some cases recovering the same performances of semiglobal classical  HGOs  and in other cases  obtaining novel  global  observers  where existing design techniques cannot provide any. 

\appendix
   
\section{Proofs of main  results}\label{main_proofs}

\emph{Proof of Theorem \ref{TH1}}.  Since $\Psi_{\mathfrak p}$ is UU-transforming for ${\mathfrak p} > 0$, $ \Psi_{\mathfrak p}^t $   is for each ${\mathfrak p}>0$  a  monotonically increasing function of $t  \ge 0$ with $\lim_{t \to +\infty} \Psi_{\mathfrak p}^t (t) = +\infty$.   Choose any  $\varepsilon , N > 0$ and  consider all the solutions $z(t)$ of $\Sigma (t) = 0$ for which $\sup_{t \ge 0}  \Vert  ( x (t)  , d(t)  )  \Vert \le  N$. Define the parameter design ${\mathfrak p}$ of the observer \eqref{observer_1} as  $
 {\mathfrak p} \delequal \sigma  ( \omega   ) $  
where $  \omega > \max \{ \sigma^{-1} ({\mathfrak p}_0 )  , N \} $ is  selected as follows with ${\mathfrak p}_0$ introduced in Assumption S.2.  By  Assumption S.1  $ \Psi_{\mathfrak p}$   is  SI-contracting with contracting maps $(\sigma , \mu , \Gamma_{\mathfrak p}^{\mathsf x} \times \Gamma_{\mathfrak p}^{\mathsf d}  )$.   Since $\sigma \in {\mathcal K}_\infty$ it  follows $     \sup_{t \ge 0}   \sigma (       \Vert  ( x (t)  , d(t)  )  \Vert     ) \le \sigma (N) \le  {\mathfrak p}  $. Hence,     by the SI-contracting property of $\Psi_{\mathfrak p}$
   \begin{eqnarray}
&& \hskip-1cm  \sup_{t \ge 0}   \Vert   ( \Gamma_{\mathfrak p}^{\mathsf x} \times \Gamma_{\mathfrak p}^{\mathsf d} ) (   x_{\mathfrak p} (t)  ,      d_{\mathfrak p}  (t)    )  \Vert \le      \mu (  {\mathfrak p} ) .  \label{alternate6Q_b2}  
  \end{eqnarray}
In what follows we consider $t \ge 0$. Let
\begin{eqnarray}
&&\hskip-1cm ( x_{\mathfrak p}^{\rm sat} (t) ,d_{\mathfrak p}^{\rm sat} (t) ) \nonumber \\
&&\hskip-1cm \delequal  
      (   \Gamma_{- \mathfrak p}^{\mathsf x} \times  \Gamma_{- \mathfrak p}^{\mathsf d} )  (  {\rm sat}_{ \mu ({ \mathfrak p})} (  (  \Gamma_{ \mathfrak p}^{\mathsf x} \times    \Gamma_{ \mathfrak p}^{\mathsf d}   )   ( x_{\mathfrak p} (t)  ,  d_{\mathfrak p} (t)    )  ) )  \label{def_x_sat}  
  \end{eqnarray}
 and notice that,  by  the properties of  the saturation functions (Section \ref{note}), for any integer $r \ge 1$ 
 \begin{eqnarray} 
 &&\hskip-1.1cm \left \Vert  \underset{  i=1,\dots , r}{\rm diag}     \{ \frac{{\rm sat}_\mu (\zeta^a_i ) -{\rm sat}_\mu (\zeta^b_i )}{\zeta^a_i-\zeta^b_i}  \} \right \Vert \le 1 ,   \label{prop_sat_2} \\
&&\hskip-1.1cm  \Vert   {\rm sat}_{\mu  } (   \zeta )   \Vert ,   \Vert  \theta {\rm sat}_{\mu  } (   \zeta^a ) + (1-\theta) {\rm sat}_{\mu  } (   \zeta^b )   \Vert    \le r \mu     \label{prop_sat_4} 
 \end{eqnarray}
  for all  $\zeta,  \zeta^a  ,  \zeta^b   \in {\mathbb R}^r$, $\mu > 0$ and $\theta \in [0,1]$. 
 By \eqref{alternate6Q_b2}, \eqref{prop_sat_2} and since     $\Gamma_{ \mathfrak p}^{\mathsf x} \times    \Gamma_{ \mathfrak p}^{\mathsf d} $ is a group of linear and diagonal transformations,  
    \begin{eqnarray}
&&\hskip-.8cm  x_{\mathfrak p}  (t)     -  \chi^{\rm sat}  (t)  =  x_{\mathfrak p}^{\rm sat}  (t)       -   \chi^{\rm sat}  (t)   =  \Lambda_{\mathfrak p}^{\mathsf x}    (  t ) (x_{\mathfrak p}  (t)   - \chi (t) )  , \nonumber \\
 &&\hskip-.8cm    d_{\mathfrak p}  (t)       = d_{\mathfrak p}^{\rm sat}  (t)  =  \Lambda_{\mathfrak p}^{\mathsf d}    ( t  )  d_{\mathfrak p} (t)     \label{XI2.1}
 \end{eqnarray}
  with $ \Lambda_{\mathfrak p}^{\mathsf x}    ( t ) \in {\mathbb R}^n$  and $ \Lambda_{\mathfrak p}^{\mathsf d}    (  t ) \in {\mathbb R}^m$ such that $  \Vert  \Lambda_{\mathfrak p}^{\mathsf x}    ( t )  \Vert  $ and $  \Vert  \Lambda_{\mathfrak p}^{\mathsf d}    ( t )  \Vert \le 1  $.  Using the mean value theorem, Assumption   S.5, \eqref{prop_sat_4}, \eqref{XI2.1} and   the  non-contracting property \eqref{mono} of  $\Gamma_{\mathfrak p}^\mathsf{x} \times \Gamma_{\mathfrak p}^\mathsf{d}$,  we write
     \begin{eqnarray}
 &&\hskip-1cm   ( \Delta \Sigma_L )   ( t ,    x_{\mathfrak p}^{\rm sat}  (t)      , d_{\mathfrak p}^{\rm sat}  (t)       )      -   ( \Delta \Sigma_L )  ( t ,   \chi^{\rm sat}    (t)   ,  0 )   \nonumber \\
    &&\hskip-1cm =  \Xi_{\mathfrak p}    ( t )    \begin{pmatrix}   x_{\mathfrak p}  (t)   - \chi (t)   \cr  d_{\mathfrak p}  (t)  \end{pmatrix}  \label{XI2}
 \end{eqnarray}
where $ \Xi_{\mathfrak p} (t) \in {\mathbb R}^{(n+p)\times (n+m)}$ is such that for some $\xi \in {\mathcal K}$
\begin{eqnarray}
 &&\hskip-1.3cm \Vert   \Xi_{\mathfrak p}   (t)  \Vert  \le  \xi ( 2( n+m    ) \mu ( {\mathfrak p} )    )   .   \label{norm_bound2}  
 \end{eqnarray}
 Using the fact that $\xi  \in {\mathcal K}$, $\pi_1 \in {\mathcal K}_+$ (from Assumption S.2)  and $ \mu  \in {\mathcal L}$ (from the contracting maps),   select  $  \omega > \max \{ \sigma^{-1} ({\mathfrak p}_0 )  , N \} $ and, therefore, ${\mathfrak p} = \sigma(\omega)$  such that if $ \phi ( {\mathfrak p} ) \delequal \left (   1+  \Vert K   \Vert_\infty  )  )      \xi (  ( n+m    ) \mu ( {\mathfrak p} )        \right )$  
 \begin{eqnarray}
&&\hskip-.8cm    \frac{8{\bf \overline{p}^2}      }{{\bf \underline{p}}}  \phi ( {\mathfrak p} )  \le  1  ,
\label{fine1.0} \\ 
&&\hskip-.8cm \pi_1 (  n\mu ({\mathfrak p})  ) \le 2 \pi_1 (0) ,
\label{fine1.1} \\
&&\hskip-.8cm (   1+  \Vert K  \Vert_\infty  )    \{ ( \Vert B  \Vert_\infty    +  \Vert D   \Vert_\infty    +  \phi ( {\mathfrak p} )    )^2 -  (    \Vert B  \Vert_\infty     +    \Vert D   \Vert_\infty     )^2 \}  \nonumber \\
&&\hskip-.8cm  +  \Big (  4{\bf \overline{p}}   \sqrt\frac{2{\bf \overline{p}}}{{\bf \underline{p}}} \Big )^{-2}
  \frac{\varepsilon^2}{ 4\pi_1^2 (0 )}       \label{fine1}  
    \end{eqnarray} 
where for brevity we use the notation $\Vert f \Vert _\infty \delequal \sup_{ t \ge 0  } \Vert f (t) \Vert$.   The rest of the proof is provided in the non-negative unbounded time scale ${\mathsf t}_{\mathfrak p} = \Psi_{\mathfrak p}^t (t)$ with  $ {\mathsf t}_{\mathfrak p} \ge 0$. By Assumption S.1  $ \Psi_{\mathfrak p}$   is a  LS  of  $ \Sigma (t) = 0$, hence  $ \Psi_{\mathfrak p}$  maps    $\Sigma (t) = 0$ into $\widetilde{\Sigma}_{\mathfrak p}  ({\mathsf t}_{\mathfrak p}) = 0$ which reads as  \footnote{The variables  $\widetilde{x}_{\mathfrak p}   , \widetilde{x}_{\mathfrak p}^{\rm sat}  , \widetilde{u}_{\mathfrak p}  , \widetilde{u}_{\mathfrak p}^{\rm sat} $ and $     \widetilde{y}_{\mathfrak p} $ denote $x_{\mathfrak p}   , x_{\mathfrak p}^{\rm sat}  ,d_{\mathfrak p}  , d_{\mathfrak p}^{\rm sat} $ and $     y_{\mathfrak p} $  in the time scale ${\mathsf t}_{\mathfrak p}$, i.e. $x_{\mathfrak p} \circ \Psi^t_{-\mathfrak p}   , x_{\mathfrak p}^{\rm sat} \circ \Psi^t_{-\mathfrak p}  ,d_{\mathfrak p}  \circ \Psi^t_{-\mathfrak p}  , d_{\mathfrak p}^{\rm sat} \circ \Psi^t_{-\mathfrak p} $ and $     y_{\mathfrak p}\circ \Psi^t_{-\mathfrak p}  $ (see also Notational Remark II).}
 \begin{eqnarray}
 &&\hskip-1.5cm \begin{pmatrix}  D_{\mathsf{t}_{\mathfrak p}} \widetilde{x}_{\mathfrak p} ({\mathsf t}_{\mathfrak p}) \cr \widetilde{y}_{\mathfrak p}({\mathsf t}_{\mathfrak p}) \end{pmatrix} =  \begin{pmatrix} A   ({\mathsf t}_{\mathfrak p})  &B  ({\mathsf t}_{\mathfrak p})  \cr C ({\mathsf t}_{\mathfrak p})  & D  ({\mathsf t}_{\mathfrak p})  \end{pmatrix}    \begin{pmatrix}  \widetilde{x}_{\mathfrak p}  ({\mathsf t}_{\mathfrak p}) \cr  \widetilde{u}_{\mathfrak p} ({\mathsf t}_{\mathfrak p}) \end{pmatrix} \nonumber \\
  &&\hskip-1.5cm +  \Delta \Sigma_L   ({\mathsf t}_{\mathfrak p} , \widetilde{x}_{\mathfrak p}({\mathsf t}_{\mathfrak p})   , \widetilde{u}_{\mathfrak p} ({\mathsf t}_{\mathfrak p}) )   \label{sys_new_coord} 
 \end{eqnarray} 
    where  $ \Delta \Sigma_L $ is defined in \eqref{Delta_2} and satisfies $\Vert \frac{ \partial ( \Delta \Sigma_L  ) (t ,  {\mathsf  x}  ,   {\mathsf  u}  ) }{\partial (  {\mathsf  x}  ,   {\mathsf  u}   ) } \Vert \le  
 \xi (  \Vert ( {\mathsf  x}  ,   {\mathsf  u} ) \Vert ) $ for some $\xi \in {\mathcal K}$ and for all $t \ge 0$ and $  (  {\mathsf  x}  ,   {\mathsf  u}  ) \in {\mathbb R}^n \times {\mathbb R}^m $ by   Assumption S.5.  
   On the other hand, the filter   \eqref{observer_1}  in the time scale ${\mathsf t}_{\mathfrak p}$ is \footnote{We denote by $\widehat\chi     , \widehat\chi^{\rm sat}    $ and $\widehat\zeta$  the variables $ \chi  ,  \chi^{\rm sat}   $ and  $\zeta $ in the time scale ${\mathsf t}_{\mathfrak p}$, i.e. $ \chi \circ  \Psi_{-\mathfrak p}^t  ,  \chi^{\rm sat} \circ \Psi_{-\mathfrak p}^t    $ and  $\zeta  \circ \Psi_{-\mathfrak p}^t  $.}  
\begin{eqnarray}
 &&\hskip-.8cm\begin{pmatrix}  D_{ {\mathsf t}_{\mathfrak p} }\widehat\chi ({\mathsf t}_{\mathfrak p})  \cr \widehat\zeta ({\mathsf t}_{\mathfrak p}) \end{pmatrix}  =  \begin{pmatrix} A   ({\mathsf t}_{\mathfrak p})     \cr  C ({\mathsf t}_{\mathfrak p})    \end{pmatrix}      \widehat\chi ({\mathsf t}_{\mathfrak p})   +  (   \Delta \Sigma_L ) ( {\mathsf t}_{\mathfrak p} ,   \widehat\chi^{\rm sat}  ({\mathsf t}_{\mathfrak p})     ,  0  )     \nonumber  \\
  &&\hskip-.8cm +   \begin{pmatrix}   K  ({\mathsf t}_{\mathfrak p})  \cr {\textbf 0}_{p \times 1}  \end{pmatrix} 
 \Big (  C  ({\mathsf t}_{\mathfrak p}) ( \widetilde{x}_{\mathfrak p} ({\mathsf t}_{\mathfrak p})   -\widehat\chi ({\mathsf t}_{\mathfrak p})) +   D  ({\mathsf t}_{\mathfrak p})  \widetilde{u}_{\mathfrak p} ({\mathsf t}_{\mathfrak p})  \Big   )  \nonumber  \\
  &&\hskip-.8cm   +    \begin{pmatrix} {\textbf 0}_{n \times n} &   K  ({\mathsf t}_{\mathfrak p}) \cr {\textbf 0}_{p \times n}  & {\textbf 0}_{p \times p} \end{pmatrix}   \Big (   (   \Delta \Sigma_L ) ( {\mathsf t}_{\mathfrak p} ,   \widetilde{x}_{\mathfrak p}  ({\mathsf t}_{\mathfrak p})    ,  \widetilde{u}_{\mathfrak p} ({\mathsf t}_{\mathfrak p})  )      \nonumber  \\
  &&\hskip-.8cm  -  (   \Delta \Sigma_L ) ( {\mathsf t}_{\mathfrak p} ,  \widehat\chi^{\rm sat}   ({\mathsf t}_{\mathfrak p})   , 0 )  \Big  ) .\label{sys_1_err} 
  \end{eqnarray} 
On account of Assumption   S.3,   \eqref{alternate6Q_b2} and \eqref{def_x_sat}
 \begin{eqnarray}
 &&\hskip-1.3cm ( \widetilde{x}^{\rm sat}_{\mathfrak p}({\mathsf t}_{\mathfrak p}) , \widetilde{u}^{\rm sat}_{\mathfrak p}({\mathsf t}_{\mathfrak p}) ) = ( \widetilde{x}_{\mathfrak p}({\mathsf t}_{\mathfrak p}) , \widetilde{u}_{\mathfrak p}({\mathsf t}_{\mathfrak p}) ) ,    \label{norm_bound4.0}  \\
 &&\hskip-1.3cm \Vert   \widetilde{u}_{\mathfrak p}  ({\mathsf t}_{\mathfrak p})   \Vert  \le  \pi_2 ({\mathfrak p})   \pi_3 ( \Vert d (t)  \Vert_{t =  \Psi^{t}_{-\mathfrak p} ( {\mathsf t}_{\mathfrak p}  ) ) }  .  \label{norm_bound4}  
 \end{eqnarray}
If  $ \varrho  \delequal \widetilde{x}_{\mathfrak p}  -\widehat\chi  $ and collecting \eqref{XI2} and  \eqref{norm_bound4.0}, 
\[
D_{\mathsf{t}_{\mathfrak p}}\varrho  ({\mathsf t}_{\mathfrak p})  =    (     \widetilde{A} ({\mathsf t}_{\mathfrak p})   + \Delta     \widetilde{A} ({\mathsf t}_{\mathfrak p})   )  \varrho   ({\mathsf t}_{\mathfrak p})   +   (   \widetilde{B}  ({\mathsf t}_{\mathfrak p})  +  \Delta \widetilde{B}  ({\mathsf t}_{\mathfrak p})  )  \widetilde{u}_{\mathfrak p}  ({\mathsf t}_{\mathfrak p}) 
\]   
 where  $ \widetilde{A}   \delequal A   - K  C  $,  $   \widetilde{B}   \delequal B   -   K D  $, $  \Delta     \widetilde{A}  \delequal  \begin{pmatrix} {\textbf I}_n  &-K    \end{pmatrix}     \Xi_{\mathfrak p}  \begin{pmatrix} {\textbf I}_n \cr  {\textbf 0}_{p \times n} \end{pmatrix} $ and $  \Delta \widetilde{B}   \delequal  \begin{pmatrix} {\textbf I}_n  &-K  \end{pmatrix} \Xi_{\mathfrak p} \begin{pmatrix} {\textbf 0}_{n \times p} \cr     {\textbf I}_p \end{pmatrix}   $.  
 Furthermore,  on account of \eqref{norm_bound2}   
     \begin{eqnarray}
  &&\hskip-1.2cm  \Vert     \widetilde{B}  ({\mathsf t}_{\mathfrak p})  \Vert \le   (1+   \Vert K  \Vert_\infty )     (  \Vert B   \Vert_\infty    +   \Vert D   \Vert_\infty  )         ,\nonumber  \\
  &&\hskip-1.2cm  \Vert  \Delta     \widetilde{B}   ({\mathsf t}_{\mathfrak p})  \Vert \le   (1+  \Vert K   \Vert_\infty  )           \xi (  ( n+m    ) \mu ( {\mathfrak p} )    )  ,\label{norm_bound4b}   \\
  &&\hskip-1.2cm \Vert   \Delta     \widetilde{A}   ({\mathsf t}_{\mathfrak p}) \Vert \le   (1+  \Vert K   \Vert_\infty  )  )      \xi (  ( n+m    ) \mu ( {\mathfrak p} )    )    . \nonumber  
   \end{eqnarray}  
Consider a candidate Lyapunov function $V (\mathsf{t}_{\mathfrak p}, \varrho    ) =  \varrho^\top   P (\mathsf{t}_{\mathfrak p})  \varrho $ with $P (\mathsf{t}_{\mathfrak p}) $ from Assumption S.4. We have  
  \begin{eqnarray}
   &&  \hskip-.8cm  D_{\mathsf{t}_{\mathfrak p}}  V (\mathsf{t}_{\mathfrak p}, \varrho  (t_{\mathfrak p}) )   \le  -  \Big (  \frac{ 1 }{2{\bf \overline{p}}} -  \frac{2{\bf \overline{p}}      }{{\bf \underline{p}}}   \phi ( {\mathfrak p} )     \Big ) V (\mathsf{t}_{\mathfrak p}, \varrho   ({\mathsf t}_{\mathfrak p})  )   \nonumber \\
   &&\hskip-.8cm   +  2{\bf \overline{p}}^2    \{ (1+ \Vert K   \Vert_\infty     ) ( \Vert B   \Vert_\infty      +   \Vert D  \Vert_\infty  )   +  \phi   (  {\mathfrak p} )    \}^2  \Vert   \widetilde{u}_{\mathfrak p}  ({\mathsf t}_{\mathfrak p})  \Vert^2   \nonumber  
  \end{eqnarray}
 From \eqref{fine1.0}, \eqref{fine1},  \eqref{norm_bound4}  and (to cite one) Lemma 5 of \cite{Bernard_Praly_Andrieu}  it follows that 
     \begin{eqnarray}
  &&\hskip-.8cm V ({\mathsf t}_{\mathfrak p}, \varrho  ({\mathsf t}_{\mathfrak p}) )  \le   e^{   -  \frac{  {\bf \underline{q}}  }{8 {\bf \overline{p}}}  {\mathsf t}_{\mathfrak p}   } V (0 , \varrho  (0)  ) +  16 {\bf \overline{p}}^3 \{  (1+  \Vert K   \Vert_\infty  )^2 \times \nonumber \\
   &&\hskip-.8cm  \times (   \Vert B  \Vert_\infty   +    \Vert D \Vert_\infty   )^2  + \Big (  4{\bf \overline{p}} \sqrt\frac{2{\bf \overline{p}}}{{\bf \underline{p}}} \Big )^{-2} \frac{\varepsilon^2}{ 4\pi_1^2 (0)} \}   \pi_3^2  ( \Vert u   \Vert_\infty  )   \pi_2^2 ({\mathfrak p})        \nonumber  
  \end{eqnarray}
  and finally  $ \Vert  \varrho  ({\mathsf t}_{\mathfrak p}) \Vert  \le \sqrt\frac{ {\bf \overline{p}}}{ {\bf \underline{p}}}  e^{   -  \frac{  {\bf \underline{q}}  }{16 {\bf \overline{p}}}    {\mathsf t}_{\mathfrak p}    } \Vert  \varrho  (0) \Vert     +   4{\bf \overline{p}}  \sqrt\frac{2{\bf \overline{p}}}{{\bf \underline{p}}}  $$  [   (1+    \Vert K  \Vert_\infty  )  (  \Vert B   \Vert_\infty   +    \Vert D  \Vert_\infty   )  +  \Big (  4{\bf \overline{p}} \sqrt\frac{2{\bf \overline{p}}}{{\bf \underline{p}}} \Big )^{-1}  \frac{\varepsilon }{ 2\pi_1  (0)}  ]  \pi_3 (  \Vert u   \Vert_\infty  ) \pi_2  ({\mathfrak p})   $.   
Since $ {\mathfrak p}   \ge  {\mathfrak p}_0$   and $\Vert \widehat\chi^{\rm sat} ({\mathsf t}_{\mathfrak p} ) \Vert  \le  n \mu ( {  \mathfrak p} )$,  on account of  Assumption  S.2 $( {\mathsf t}_{\mathfrak p} ,\widehat\chi^{\rm sat} ({\mathsf t}_{\mathfrak p} )) \in {\rm Dom} ( \Psi^{ t,{\mathsf x} }_{-\mathfrak p} ) $ so that $  \Psi^{ \mathsf x  }_{-\mathfrak p}({\mathsf t}_{\mathfrak p} ,\widehat\chi^{\rm sat} ({\mathsf t}_{\mathfrak p} )) $ is well-defined and, in addition,  from the non-contracting property \eqref{mono} of  $\Gamma_{\mathfrak p}^\mathsf{x}$ and  \eqref{prop_sat_4},   we  have  for all $\theta \in [0,1]$ 
 \begin{eqnarray}
&&\hskip-1.5cm \left \Vert     \frac{\partial  \Psi_{-{\mathfrak p} }^{\mathsf{x}} (\mathsf{t}_{\mathfrak p},\mathsf{x}) }{\partial \mathsf{x}}\right \Vert_{\mathsf{x}=\theta {\widetilde{x}}_{\mathfrak p}^{\rm sat} ({\mathsf t}_{\mathfrak p} )    + (1-\theta)  \widehat\chi^{\rm sat} ({\mathsf t}_{\mathfrak p} )   }   
 \le \frac{ \pi_1  ( n \mu ({\mathfrak p} ) ) }{ \pi_2 ({\mathfrak p})   } . \label{fff}
\end{eqnarray}   
By Assumption  S.2 with \eqref{fine1.1} and \eqref{fff},  the mean value theorem and since $  \pi_1 \in {\mathcal K}_+
$ and $\mu   \in {\mathcal L}$ 
   \begin{eqnarray}
  && \hskip-1cm \limsup_{t \rightarrow +\infty}  \Vert x (t) -\widehat{x} (t) \Vert   \nonumber \\
  && \hskip-1cm = \limsup_{{\mathsf t}_{\mathfrak p} \rightarrow +\infty}   \Vert    \Psi_{-{\mathfrak p} }^{\mathsf{x}} ({\mathsf t}_{\mathfrak p}  ,   {\widetilde{x}}_{\mathfrak p}^{\rm sat} ({\mathsf t}_{\mathfrak p} ) ) -  \Psi_{-{\mathfrak p} }^{\mathsf{x}} ({\mathsf t}_{\mathfrak p}  , \widehat\chi^{\rm sat} ({\mathsf t}_{\mathfrak p} ) )\Vert  \nonumber \\
  && \hskip-1cm  \le  \limsup_{{\mathsf t}_{\mathfrak p} \rightarrow +\infty} \int_0^1 \left \Vert     \frac{\partial  \Psi_{-{\mathfrak p} }^{\mathsf{x}} ({\mathsf t}_{\mathfrak p},\mathsf{x}) }{\partial \mathsf{x}}\right \Vert_{\mathsf{x}=\theta {\widetilde{x}}_{\mathfrak p}^{\rm sat} ({\mathsf t}_{\mathfrak p} )     + (1-\theta)  \widehat\chi^{\rm sat} ({\mathsf t}_{\mathfrak p} )   } \hskip-.5cm  d\theta \times \nonumber \\
  && \hskip-1cm   \times   \limsup_{t_{\mathfrak p} \rightarrow +\infty} \Vert  \varrho  ({\mathsf t}_{\mathfrak p}) \Vert \le \{    \varepsilon +  8 {\bf \overline{p}}  \sqrt\frac{2{\bf \overline{p}}}{{\bf \underline{p}}}     (1+  \Vert K   \Vert_\infty  ) \times \nonumber \\
  && \hskip-1cm \times (    \Vert B   \Vert_\infty  +     \Vert D \Vert_\infty  )  \pi_1 (0)  \} \pi_3 ( \Vert u   \Vert_\infty  )    . \nonumber  
   \end{eqnarray}  
  
\emph{Proof of Theorem \ref{TH3}}. In this case, we set ${\mathfrak p}= p(t) \delequal \sigma (\omega_0 + N + \beta_1 ( \widehat{V} (t) + \omega_1))$, where $\widehat{V} (t)$ is the output of the SI-norm estimator $D_t \widehat{V} (t) = \max \{ -\lambda_1 \widehat{V} (t) + \Phi (   \Vert y(t) \Vert +N ) ,  0 \}$,  the parameters $ \lambda_1 , \omega_1 >0$  are selected as in  \eqref{alpha^Lambda}, 
 $\omega_0 \ge \beta_0$ is selected as the corresponding parameter $\omega > 0$ in the proof of Theorem \ref{TH1} and $\beta_1 \in {\mathcal K}_\infty$ comes from Assumption SINE. The proof of Theorem follows the proof of Theorem \ref{TH1}, except for the fact that we have to prove that $\theta (t) \delequal \Psi_{\mathfrak p}^t   (t) \vert_{{\mathfrak p} = p(t)}  $ is an unbounded  time scale.  To this aim, we prove that, no matter the values of the design parameters $ \lambda_1 , \omega_1 > 0$ and $  \omega_0 \ge \beta_0 $ are,   there exists $T > 0$ such that $\theta (t)   $  is a monotonically increasing function of $t \ge T$ and $\lim_{t \to +\infty} \theta (t) = +\infty  $.  First,  notice that $D_t \theta (t) =  \frac{\partial \Psi_{\mathfrak p}^t  (t) }{\partial t}  +   g^t (\theta (t) ) D_t p (t) $ for $t \ge 0$.  Moreover,  the time   $T \delequal T_{x(0)} $ (defined in \eqref{pt}) is such that  $ V (t, x(t) )  \le   \widehat{V} (t) +\omega_1$ for all $ t \ge T $.  On account of  Assumption GS.6 and the equation  for $D_t p(t)$,   there exist  $k_0 \in {\mathcal L}$, $k_2 > 0$ and $k_1 \in {\mathcal K}_+$ such that for all $  t \ge T$  and $\omega_0 \ge  \beta_0 $ for which $k_0 ( \omega_0) < 1$ we have 
   \begin{eqnarray}
 &&\hskip-.8cm  D_t \theta (t)  \ge   \frac{\partial \Psi_{\mathfrak p}^t   (t)}{\partial t} \Big (  1 +  \Big ( \frac{\partial \Psi_{\mathfrak p}^t   (t) }{\partial t} \Big )^{-1} \min \{ 0 ,  g^t ( \theta (t) ) \}  D_t  p (t)  \Big )  \nonumber \\
  &&\hskip-.8cm   \ge \frac{\partial \Psi_{\mathfrak p}^t  (t) }{\partial t} (1-k_0 ( \omega_0) ) \ge \frac{(1-k_0 ( \omega_0))k_1 (\omega_0)}{1 +k_2t}  > 0  . \label{tempo1b} 
  \end{eqnarray}
Hence,  $D_t \theta (t) \ge \frac{(1-a)k_1 ( \omega_0)}{1 +k_2t} > 0 $  for $t  \ge T $ by \eqref{tempo1b} which implies  $ \theta (t)$   monotonically increasing for $t \ge T$.  Integrating w.r.t. $t$ both members of the above differential inequality, $ \theta (t) > \frac{(1-a)k_1 ( \omega_0)}{k_2}\ln \frac{ 1 +k_2t }{1+ k_2 T} $ for $t  \ge T $, which implies $\lim_{t \to +\infty} \theta (t) =+\infty$. This concludes the proof that  $\theta(t)$ is an unbounded time scale.   

At this point, we follow the proof of Theorem \ref{TH1} and use the Lyapunov function   $
 V ( \varrho    ) = \frac{1}{\pi_2^2 ( {\mathsf t}_{\mathfrak p} , {\mathfrak p} ) }  \varrho^\top   P    \varrho   $ with $P  $ from Assumption GS.4. 
 \end{document}